\documentclass[aoas]{imsart}

\RequirePackage{amsthm,amsmath,amsfonts,amssymb}
\RequirePackage[authoryear]{natbib}
\RequirePackage{graphicx}
\usepackage{mathtools}
\usepackage{amsmath}
\usepackage{bm}
\usepackage{subcaption}
\usepackage{url}

\usepackage{xcolor}

\startlocaldefs

\endlocaldefs

\begin{document}

\begin{frontmatter}
\title{Estimation of Time-Varying Treatment Effects in a Joint Model for Longitudinal and Recurrent Event Outcomes in Mobile Health Data}
\runtitle{Time-Varying Treatment Effects in a Joint Model}

\begin{aug}
\author[A]{\fnms{Madeline R}~\snm{Abbott}\ead[label=e1]{mrabbott@umich.edu}\orcid{0000-0002-5344-3732}},
\author[A]{\fnms{Jeremy M G}~\snm{Taylor}\ead[label=e2]{jmgt@umich.edu}}
\author[B]{\fnms{Inbal}~\snm{Nahum-Shani}\ead[label=e3]{inbal@umich.edu}}
\author[C]{\fnms{Lindsey N}~\snm{Potter}\ead[label=e4]{lindsey.potter@hci.utah.edu}}
\author[C]{\fnms{David W}~\snm{Wetter}\ead[label=e5]{david.wetter@hci.utah.edu}}
\author[C]{\fnms{Cho Y}~\snm{Lam}\ead[label=e6]{cho.lam@hci.utah.edu}}
\and
\author[A]{\fnms{Walter}~\snm{Dempsey}\ead[label=e7]{wdem@umich.edu}}

\address[A]{Department of Biostatistics, University of Michigan\printead[presep={,\ }]{e1,e2,e7}}

\address[B]{Institute for Social Research, University of Michigan\printead[presep={,\ }]{e3}}

\address[C]{Department of Population Health Sciences and Huntsman Cancer Institute, University of Utah\printead[presep={,\ }]{e4,e5,e6}}
\end{aug}

\begin{abstract}
Not only does mobile health technology enable researchers to track changes in multiple longitudinal outcomes of interest and to record the occurrence of health-related events over time, but it also allows for the delivery of repeated low-cost treatments directly to individuals in real time. We present a model-based approach for estimating the effect of repeatedly delivered treatments in a micro-randomized trial (MRT) via an extension of a joint longitudinal-survival model. We discuss different ways that these repeated treatment effects can be incorporated into the joint model; these different model specifications correspond to different mechanisms by which treatment is assumed to impact the longitudinal and event processes. Taking a Bayesian approach to inference, we model the association between repeated treatments, multiple longitudinally measured outcomes, and recurrent events. We also demonstrate how to calculate information criteria for model selection and present goodness-of-fit plots for assessing survival submodel calibration. We then illustrate the performance of our method via simulations and analysis of data collected in an MRT of substance use.
\end{abstract}


\end{frontmatter}


\doublespacing

\section{Introduction}

Mobile health (mHealth) technology enables frequent measurement of multiple outcomes over time. The resulting intensive longitudinal data (ILD) can provide insight into underlying behavioral, psychological, or other states that are indirectly measured through these many observed outcomes. Beyond data collection, mHealth tools also support the delivery of repeated, low-cost treatments directly to individuals, often via just-in-time adaptive interventions (JITAIs) \citep{nahum-shani_2018}. JITAIs are typically delivered multiple times per day and are tailored---in timing, type, and intensity---to individual’s context and state. They can be developed in micro-randomized trials (MRTs) in which individuals are randomized---possibly multiple times per day---to be sent or not sent a treatment.

Since their proposal \citep{klasnja_2015, liao_2015}, MRTs have been used to inform development of effective JITAIs in settings ranging from promoting physical activity among sedentary adults \citep{klasnja_2019} to increasing engagement with mHealth tools for substance use \citep{rabbi_2018}. In MRTs, researchers are often interested in understanding the effect of treatment on an outcome measured shortly after randomization; e.g., the effect of smartphone notifications on substance use within the next 12 hours.

The treatment effects of interest---i.e., of repeatedly delivered treatments on repeatedly measured outcomes---may account for the time-varying nature of treatments, covariates, and repeatedly measured outcomes. \cite{boruvka_2018} propose an estimator of a causal excursion effect that incorporates the time-varying nature of the treatment, outcome, and potential moderators. This estimator, which can be conditional on all past treatments or on a selected set of moderators (e.g., past engagement with notifications), captures the effect of treatment on a future outcome under different treatment scenarios and is marginal over all variables except the potential moderators on which the effect is conditioned. These MRT treatment effects are commonly estimated using generalized estimating equation methods, such as weighted and centered least squares \citep{boruvka_2018}, with extensions in \cite{qian_2020} and \cite{shi_2023}. Although these estimating equation approaches reduce parametric assumptions, they cannot directly model possible treatment effects on multiple outcome types---e.g., a longitudinal and a time-to-event outcome---simultaneously.

Here, we propose a method for obtaining model-based estimates of treatment effects. Such estimators are common in the joint longitudinal–survival model setting where understanding the hazard of an event as a function of time-varying predictors is of interest. Modeling longitudinal and survival outcomes jointly, rather than separately, can reduce bias and increase efficiency by accounting for dependencies between the longitudinal and survival processes \citep{ibrahim_2010}. Joint models typically treat longitudinal outcomes as intermediate variables through which treatment affects event risk. Much of this work is motivated by observational data (e.g., \cite{yu_2004, taylor_2014, rizopoulos_2024}), where a longitudinal outcome determines treatment, thus making it a “time-dependent confounder” that requires additional consideration (\cite{taylor_2014, rizopoulos_2024}). In MRTs, however, the timing of treatment delivery is random, so the longitudinal outcome is not a time-dependent confounder. Instead, the key challenge is that treatment can be delivered multiple times per day, rather than at a single decision point.

In the MRT context, a primary advantage of our model-based approach is that it can disentangle treatment effects on a latent longitudinal process measured via multiple longitudinal outcomes and on the hazard of recurrent events. Our work is motivated by an MRT aiming to use behavioral strategies to promote smoking cessation among current smokers attempting to quit. Throughout the study, prompts (i.e., treatments) to engage in behavioral strategies are delivered multiple times per day via app-based notifications, while information on emotions (i.e., ILD) and repeated instances of substance use (i.e., recurrent events) is also reported multiple times per day. We incorporate these repeated treatments into a joint longitudinal–recurrent event model that links latent psychological states with the hazard of recurrent substance-use events. While this approach has similarities to mediation analysis, we view our contribution as developing a framework that is useful for exploratory analyses of associations of interest. Modeling the time-varying effects of interventions on both the recurrent-event risk and the underlying latent process has the potential to provide insights into how JITAIs influence events of interest and thus inform the design of improved JITAIs.

The main contribution of this work is a model-based approach for estimating treatment effects---on both the risk of recurrent events and the trajectory of the latent factors---in an MRT. We consider two different mechanisms by which the treatments can impact the longitudinal latent process, which we model using a continuous-time multivariate stochastic process. Specifically, we allow treatment to impact the latent process via an additive shift to its mean, or we allow treatment to impact the underlying dynamics of the latent process by altering the rate at which it reverts towards the mean (i.e., as time-varying drift). Useful consequences of this model-based approach are threefold: we can (1) disentangle the impact of treatment directly on the event outcome and via the latent factors; (2) use the fitted model to predict the risk of an event within a fixed interval of time, given the trajectory of the latent factors and past treatment; and (3) use the fitted model to inform the development of improved JITAIs or other treatments through increased understanding of the pathways by which treatment may impact the recurrent event outcome. When analyzing the motivating MRT data, we also demonstrate how information criteria and posterior predictive checks can be used to examine the fit of models making different assumptions about the relationship between the recurrent events, longitudinal latent process, and repeated treatment effects.

The rest of the paper is organized as follows: we describe the motivating MRT in Section \ref{s:motivating_data}, we present the joint longitudinal-recurrent event models with time-varying treatment effects in Section \ref{s:methods}, we demonstrate the statistical properties of our method via simulation in Section \ref{s:sim_study}, we analyze the MRT data in Section \ref{s:case_study}, and we provide a discussion in Section \ref{s:discussion}.

\section{Motivating Data}\label{s:motivating_data}

\begin{figure}
\centering
\includegraphics[width=10cm]{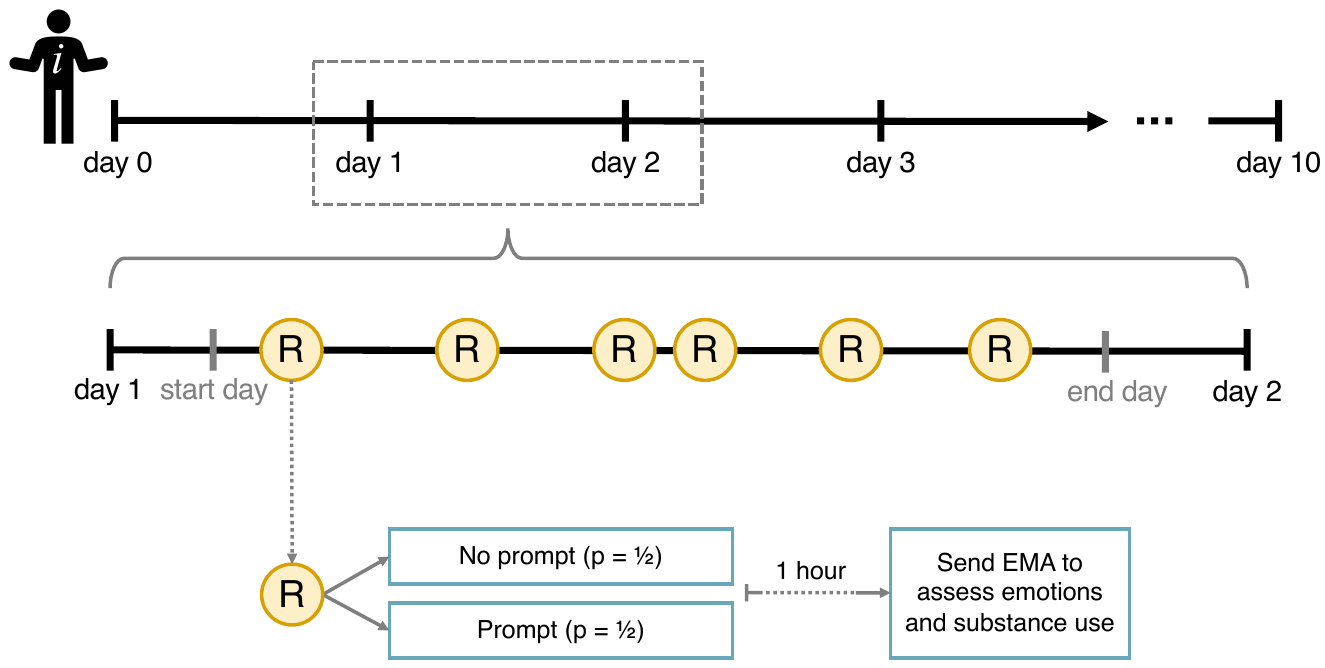}
\caption[Diagram of the MRT design.]{Simplified diagram of the MRT. Each day, a participant is randomized up to six times to potentially receive an app-based prompt. About one hour after randomization, an EMA is sent to the participant's smartphone to assess the current intensity of multiple emotions and substance use since the prior EMA. Figure is adapted from \cite{nahum-shani_2021}.}
\label{fig:mrt_timeline}
\end{figure}

This work is motivated by data from the Affective Science MRT. At the time of drafting this manuscript, the study was still ongoing and thus we only include data available from enrolled participants at that time (N = 72). The design of this study is identical to that described in \cite{nahum-shani_2021} and uses ecological momentary assessments (EMAs) to collect the self-reported intensity of 15 emotions, along with recent substance use (smoking cigarettes, vaping, or using marijuana). Up to six times per day over a ten day period, study participants are randomized with equal probability to be sent or not sent a prompt to engage with interventions delivered via smartphones. These prompts are aimed at improving their engagement in behavioral and self-regulatory activities associated with decreased vulnerability to substance use \citep{nahum-shani_2021}. Approximately one hour after randomization, the participants are sent an EMA on their smartphone. A simplified diagram of the MRT design is provided in Figure \ref{fig:mrt_timeline}.

Our longitudinal outcome is a set of 15 emotions---5 positive and 10 negative---measured via self-report on a 5-point Likert scale at each EMA. Emotions are: happy, proud, relaxed, grateful, enthusiastic, angry, ashamed, irritable, guilty, lonely, anxious, sad, restless, bored, and hopeless. On average, emotions are measured 28 times per person (min. = 3, max. = 53).

Our recurrent event is poly-substance use, which is also self-reported in the EMAs. To define an instance of poly-substance use (i.e., using marijuana, vaping, or smoking cigarettes), we use a set of deterministic rules that consolidate information across EMA questions assessing the timing and type of substance use (see Web Appendix A.1 for details).

The within-individual average EMA completion rate is 47\%, but this rate varies widely across individuals (range: 5\% - 88\%) according to a combination of non-response to delivered EMAs and failure to deliver EMAs. While many individuals respond frequently to the EMAs, some individuals go for multiple days without responding to an EMA. The event submodel assumes that individuals are at risk of experiencing an event (up until censoring), so to help reduce the impact of non-response on our event submodel, we censor individuals at the time of their most recent longitudinal measurement if they fail to respond to an EMA for a period of more than 48 hours. Otherwise, we censor individuals at the time of their final completed EMA. We also exclude any individuals who fail to respond to an EMA within the first 48 hours of the study, resulting in an analytic sample size of $N = 64$ individuals.

The observed data, which consist of the longitudinal outcomes, recurrent events, and treatment timings, are plotted in Appendix Web Figure 4 for a subset of individuals in the motivating MRT. Across all individuals, we observe 1,162 events of poly-substance use, which corresponds to an average of 19 events per person spread over a maximum of 10 days. Additional plots detailing the timing of events are given in Web Appendix A.1.

\section{Methods}\label{s:methods}

To model the relationship between repeated treatments, a multivariate longitudinal latent process, and the risk of recurrent events, we use a joint longitudinal-recurrent event model. The model extends the joint longitudinal-survival model previously presented in \cite{abbott_2024} to allow for recurrent events and to include a model for the effect of treatment on the latent process and the hazard of events.  The longitudinal submodel---without treatment---was originally presented in \cite{abbott_2023}.

Here, we first introduce our joint longitudinal-recurrent event model in a setting without treatment effect models. Let $i = 1, ..., N$ index the independent individuals in our dataset, and let $j = 1, ..., n_i$ index the longitudinal measurement occasions for individual $i$. This joint model consists of the following three submodels:

\begin{itemize}
    \item[] \textbf{Measurement submodel:} The measured $k$ longitudinal outcomes, represented by $k$-length vector $\bm{Y}_i(t_{ij})$, are assumed to be noisy observations of a $p$-dimensional latent process $\bm{\eta}_i(t_{ij})$, where $p < k$. We model the measured longitudinal outcomes using a dynamic factor model, \begin{align}\label{eq:meas_submod}
        \bm{Y}_i(t_{ij}) = \bm{\Lambda} \bm{\eta}_i(t_{ij}) + \bm{u}_i + \bm{\epsilon}_i(t_{ij}).
    \end{align} $\bm{\Lambda}$ is a $k \times p$ time-invariant loadings matrix that captures the association between the measured longitudinal outcomes and the time-varying latent process. $\bm{u}_i$ is a $k$-length vector of outcome-specific random intercepts that account for baseline differences in the measured longitudinal outcomes across individuals; we assume $\bm{u}_i \sim N_k(\bm{0}, \bm{\Sigma}_u)$ where $\bm{\Sigma}_u$ is a diagonal matrix. $\bm{\epsilon}_i(t_{ij})$ captures measurement error; we assume $\bm{\epsilon}_i(t_{ij}) \sim N_k(0, \bm{\Sigma}_{\epsilon})$ and that $\bm{\Sigma}_{\epsilon}$ is diagonal. In the motivating data, $\bm{Y}_i(t_{ij})$ is the set of 15 emotions measured at time $t_{ij}$ for individual $i$.
    \item[] \textbf{Structural submodel:} The $p$-dimensional latent process $\bm{\eta}_i(t)$ is assumed to be a multivariate Ornstein-Uhlenbeck (OU) stochastic process. This process can be thought of as a continuous-time version of a vector autoregressive process and captures correlated change within and between multiple latent factors over time. The stochastic differential equation definition of the OU process is \begin{align}\label{eq:ou_sde}
        d\bm{\eta}_i(t) = \left[ \bm{\mu} -\bm{\theta} \bm{\eta}_i(t) \right] dt + \bm{\sigma} dW_i(t).
    \end{align} $\bm{\theta}$ is a $p \times p$ parameter matrix that captures the speed at which the latent process reverts towards a constant mean $\bm{\mu}$, and $\bm{\sigma}$ is a $p \times p$ parameter matrix that captures the volatility of the process. $W_i(t)$ is a Wiener process. Often, the constant mean is assumed to be 0 ($\bm{\mu} = 0$). Later, when we introduce interventions, we allow $\bm{\mu}$ to depend on time. For more discussion of the OU process, see \cite{brigo_2007}. In the motivating data, we summarize the 15 measured emotions as a two time-varying latent factors using a bivariate ($p = 2$) OU process. These latent factors can be interpreted as capturing different aspects of psychological state, namely positive affect and negative affect.
    \item[] \textbf{Event-time submodel:} The hazard of the $r^\text{th}$ event for individual $i$ can depend on the history of the latent process $\mathcal{H}_i(t) = \{\bm{\eta}_i(s), 0 \leq s \leq t \}$, the history of recurrent events $\mathcal{R}_i(t) = \{T_{ir} < t, r = 1, 2, ...\}$, and possibly baseline covariates $\bm{X}_i$. Very generally, we define the hazard for recurrent event $r$ as \begin{align}
        h_{ir} \{t | \mathcal{H}_i(t), \mathcal{R}_i(t), X_i\} = h_0(t) \exp\big\{f(\mathcal{H}_i(t); \bm{\beta}_H) + g(\mathcal{R}_i(t); \bm{\beta}_R) + \bm{X}_i^{\top} \bm{\gamma} \big\}.
    \end{align} $h_0(t)$ is the baseline hazard, which may be time-varying. We take a clock-reset approach in which we use the hazard to model the time between each recurrent event and thus reset the clock for the baseline hazard (but not the time-dependent predictors) to 0 after the occurrence of an event. We then account for associations between repeated events through $g(\mathcal{R}_i(t); \bm{\beta}_R)$; for example, $g(\mathcal{R}_i(t); \bm{\beta}_R)$ could be some transformation of the time since the most recent prior event, $T_{i, r-1}$. $\bm{X}_i$, which is a vector containing baseline covariates, could adjust for pre-quit substance use or other time-independent risk factors. We discuss more concrete forms of $f$ and $g$ later in the simulation study.
\end{itemize}

Due to the complexity of our model and the MRT data, we could potentially adapt our joint model to capture the effect of these randomized treatments in up to three different ways. Each treatment could be associated with changes in the measured longitudinal outcome, changes in the latent process, or modifications directly to the risk of a recurrent event. We summarize these possible pathways for treatment effect in Appendix Web Figure 11. In the remainder of this paper, we focus on the scenarios in which treatment directly impacts (a) the latent process and (b) the risk of a recurrent event. We could potentially also allow the treatment to directly impact the measured longitudinal outcome, but this modeling decision would imply that the treatment changes the observed longitudinal outcome without modifying the latent process that the longitudinal outcome is assumed to measure. This mechanism for treatment effect is less scientifically relevant than (a) or (b) because we view the observed longitudinal outcomes simply as noisy measurements of the latent factors of interest. Our measurement error perspective implies that any effect of treatment on the measured outcomes should result from changes to the latent factors themselves. Thus, our main contribution is the development of treatment effect models for (a) and (b).

\subsection{Modeling the Impact of Treatment on the Latent Process}\label{ss:ou_tx_effect}

We now build on the models defined in the previous section to model treatment effects. Let $a_i(t_{ij})$ denote the decision to treat individual $i$ at time $t_{ij}$ and $\mathcal{A}_i(t) := \{a_i(s), 0 \leq s \leq t \}$ be their treatment history.

We can model the impact of treatment directly on the latent process by incorporating a model for the treatment effect into our structural submodel. The prompts in the Affective Science MRT target engagement in behavioral activities posited to decrease vulnerability to smoking and so, via this model formulation, we assume that these vulnerability-related psychological states are represented by the low-dimensional latent process. This assumption about the treatment effect implies that treatment could also impact the risk of a recurrent event and the measured longitudinal outcomes via changes in the trajectory of the latent process.

\begin{figure}
 \begin{subfigure}[t]{7cm}
     \includegraphics[width=7cm]{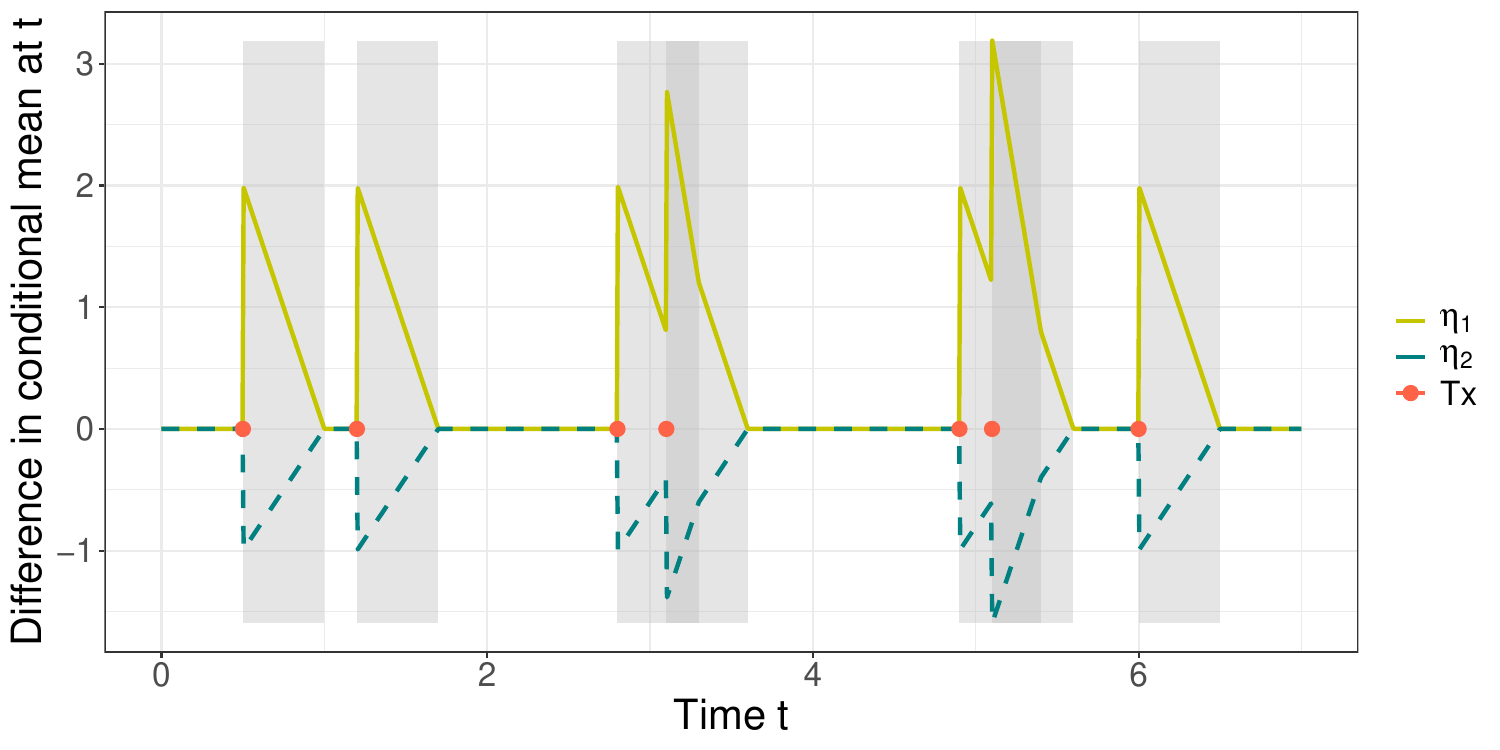}
     \caption{The difference in means at time $t$ between a latent process with treatment effect modeled as an \textbf{additive} shift to the mean and a latent process without a treatment effect is $\bm{\mu}_i(t) - e^{-\bm{\theta} (t - s)} \bm{\mu}_i(s)$, where $s = 0$.}
     \label{fig:linear_mu_setting1}
 \end{subfigure}
 \hfill
 \begin{subfigure}[t]{7cm}
     \includegraphics[width=7cm]{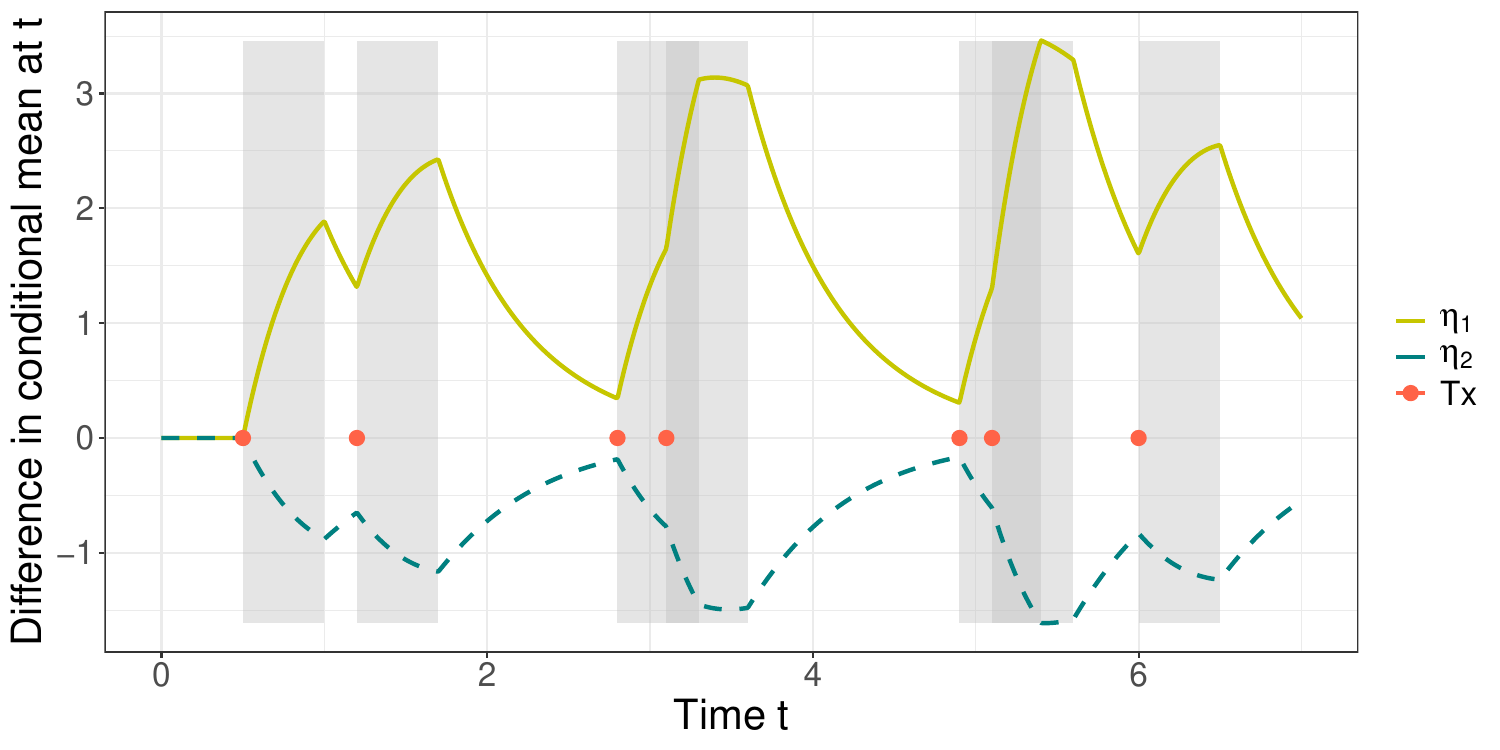}
     \caption{The difference in means at time $t$ between a latent process with treatment effect modeled on the derivative scale as time-varying \textbf{drift} and a latent process without a treatment effect is $\int_s^t e^{-\bm{\theta}(t - u)} \bm{\mu}_i(u) du$, where $s = 0$.}
     \label{fig:hw_integral_setting1}
 \end{subfigure}
 \caption[Treatment effect models for the longitudinal process.]{Treatment models for the longitudinal process. The plots show the difference between the latent process with a treatment effect model---either additive or drift---and a latent process with a constant mean of 0. We plot this difference across time $t$, with $s$ fixed at 0, if treatments were to be sent as indicated by the red dots. The shaded grey bars highlight the (potentially overlapping) regions over which treatments may have some non-zero effect.}\label{fig:a:tx_terms}
\end{figure}

We consider two different approaches to modeling the impact of treatment on the latent process trajectory, which make different assumptions about how treatment might affect the latent process. In this first approach, we assume that the impact of treatment is additive and directly shifts the OU process away from the constant mean for a short window of time after the treatment. In the motivating MRT, this approach would assume that sending a prompt to an individual results in a shift in the level of their psychological state (e.g., positive and negative affect) on the scale of the state itself. Let $\bm{\eta}^*(t)$ denote the OU process without treatment effect (i.e., Eq.\ref{eq:ou_sde} with $\bm{\mu} = 0$). Then, the impact of treatment is modeled as
\begin{align}\label{eq:tx_additive}
    \bm{\eta}_i(t) = \bm{\eta}^*_i(t) + \bm{\mu}_i(t)
\end{align}

\noindent where $\bm{\mu}_i(t)$ is a simple function that describes the short-term impact of treatment. Here, we assume that, for each individual $i$, $\bm{\mu}_i(t)$ is a simple deterministic function:
\begin{align}\label{eq:mu}
    \bm{\mu}_i(t) = \sum_{t_{ia} \in \mathcal{A}_i(t)} \bm{\tau} \left(1 - \frac{t - t_{ia}}{\delta_a} \right)_+
\end{align}

\noindent where $\mathcal{A}_i(t)$ contains the time $t_{ia}$ of all treatments sent to individual $i$ prior to time $t$, $\bm{\tau}$ is a length-$p$ vector that captures the maximum impact of the treatment at the time at which it is delivered, and $\delta_a$ defines the window over which each treatment is active. We assume that $\delta_a$ is known, but that $\bm{\tau}$ is estimated. Note that the deterministic function $\bm{\mu}_i(t)$ implicitly conditions on the treatment history $\mathcal{A}_i(t)$. This form for $\bm{\mu}_i(t)$ assumes that the effect of treatment on the latent process is largest at the time of treatment; then, this effect decays linearly until no effect remains $\delta_a$ units of time after the treatment was delivered. If multiple treatments are delivered in rapid succession, the cumulative impact of the treatments is additive, as illustrated in Figure \ref{fig:linear_mu_setting1}. Due to the complexity of our joint model, we assume a fairly simple form for $\bm{\mu}_i(t)$ here, but one could specify a different form for $\bm{\mu}_i(t)$ in a different setting. 

Assuming this additive treatment effect in Eq. \ref{eq:tx_additive}, we can write the conditional distribution for our structural submodel as follows: if $\bm{\eta}_i(0) \sim N_p(\bm{0}, \bm{V})$ where $\bm{V} = vec^{-1}\big\{(\bm{\theta} \oplus \bm{\theta})^{-1} vec(\bm{\sigma} \bm{\sigma}^{\top}) \big\}$, then for times $t$ and $s$, $t > s$,
\begin{align}\label{eq:ou_add_dist}
    \bm{\eta}_i(t) | \bm{\eta}_i(s) \sim N_p \left(\bm{\mu}_i(t) + e^{-\bm{\theta} (t-s)} \left( \bm{\eta}_i(s) - \bm{\mu}_i(s) \right), \bm{V} - e^{- \bm{\theta} (t-s)} \bm{V} e^{- \bm{\theta}^{\top} (t-s)}\right)
\end{align}

\noindent where $e$ is the matrix exponential.

As an alternative to modeling the treatment effect as an additive shift to the mean of the OU process, we can instead model treatment as impacting the dynamics of the latent process through a time-varying drift term on the derivative scale. In the motivating MRT, this approach assumes that sending a prompt to an individual alters the rate at which their psychological states (e.g., positive and negative affect) change over time. For example, sending a prompt could increase the rate at which levels of negative affect revert towards their average level. The standard OU process assumes a constant value for $\bm{\mu}$, but the Hull-White process, which is often used in financial math applications, extends the OU process to allow for time-varying drift. For more discussion of the OU process and Hull-White model, see \cite{brigo_2007}. The stochastic differential equation (SDE) for the Hull-White model is
\begin{align}\label{eq:hw_sde}
        d \bm{\eta}_i(t) = \left[\bm{\mu}_i(t) -\bm{\theta} \bm{\eta}_i(t) \right] dt + \bm{\sigma} dW_i(t)
\end{align}

\noindent where $\bm{\mu}_i(t)$ is still a simple function that describes the short-term treatment effect. From this SDE, it follows that the conditional distribution of $\bm{\eta}_i(t)$ with time-dependent drift is
\begin{align}\label{eq:hw_dist}
    \bm{\eta}_i(t)|\bm{\eta}_i(s) \sim N_p\left( e^{-\bm{\theta} (t - s)} \bm{\eta}_i(s) + \int_s^t e^{-\bm{\theta} (t-u)} \bm{\mu}_i(u) du, \bm{V} - e^{- \bm{\theta} (t-s)} \bm{V} e^{- \bm{\theta}^{\top} (t-s)} \right)
\end{align}

\noindent The mean of this distribution requires integrating across $e^{-\bm{\theta} (t-u)} \bm{\mu}_i(u)$ as a function of $u$. Depending on the specific formulation of $\bm{\mu}_i(t)$, an analytic solution to this integral may or may not exist. If we assume that the treatment effect model takes the linear form given in Eq.\ref{eq:mu}, we can derive the analytic solution to the integral in the conditional mean in Eq.\ref{eq:hw_dist}. In a setting in which only a single treatment impacts the drift of this latent process, integration would be straight forward; in our setting, however, we must carefully account for overlapping active treatments. We use $\mathcal{A}_i(s - \delta_a, t)$ to denote the set of times at which treatments were sent to individual $i$ between time $s - \delta_a$ and time $t$; this set of treatment times corresponds to all treatments that are active between times $s$ and $t$. If we solve the integral in Eq.\ref{eq:hw_dist}, then we can re-write the distribution in an analytic form:

\scriptsize
\begin{equation}\label{eq:hw_dist_analytic}
    \begin{aligned}
    \bm{\eta}_i(t) &| \bm{\eta}_i(s)  \sim \\ & N_p \Bigg( e^{-\bm{\theta} (t - s)} \bm{\eta}_i(s) + \sum_{t_{ia} \in \mathcal{A}_i(s - \delta_a, t)} \left[ \left( 1 - \frac{u - t_{ia}}{\delta_a} \right) e^{-\bm{\theta} (t - u)} \theta^{-1} + \frac{1}{\delta_a} e^{-\bm{\theta} (t - u)} \right] \bm{\tau}\Bigg\rvert_{u = max(t_{ia}, s)}^{u = min(t, t_{ia} + \delta_a)}, \\ & \bm{V} - e^{- \bm{\theta} (t-s)} \bm{V} e^{- \bm{\theta}^{\top} (t-s)} \Bigg)
\end{aligned}
\end{equation}

\normalsize
\noindent More details on this derivation are given in the Appendix C.

Comparing the conditional distribution of $\bm{\eta}_i(t) | \bm{\eta}_i(s)$ when the treatment effect is modeled as an additive term (Eq. \ref{eq:ou_add_dist}) to the conditional distribution when treatment effect is modeled as impacting the dynamics of the latent process through drift (Eq. \ref{eq:hw_dist_analytic}), we see the variance terms are the same but the treatment function $\bm{\mu}_i(t)$ appears differently in the conditional means:

\textbf{Additive treatment effect:}
\footnotesize
\begin{align*}
    \mathbb{E}\left[ \bm{\eta}_i(t)|\bm{\eta}_i(s) \right] &= e^{-\bm{\theta} (t-s)}\bm{\eta}_i(s) + \bm{\mu}_i(t) - e^{-\bm{\theta} (t-s)} \bm{\mu}_i(s)
\end{align*}
\normalsize

\textbf{Drift treatment effect:} 
\footnotesize
\begin{align*}
    \mathbb{E}\left[ \bm{\eta}_i(t)|\bm{\eta}_i(s) \right] = & e^{-\bm{\theta} (t - s)} \bm{\eta}_i(s) \\ &+ \sum_{t_{ia} \in \mathcal{A}_i(s - \delta_a, t)} \left[ \left( 1 - \frac{u - t_{ia}}{\delta_a} \right) e^{-\bm{\theta} (t - u)} \bm{\theta}^{-1} + \frac{1}{\delta_a} e^{-\bm{\theta} (t - u)} \right] \bm{\tau}\Bigg\rvert_{u = max(t_{ia}, s)}^{u = min(t, t_{ia} + \delta_a)}
\end{align*}

\normalsize

\noindent In the additive version, the mean reversion parameter $\bm{\theta}$ and treatment effect function $\bm{\mu}_i(t)$ show up clearly in the terms in the sum. Using the definition of the latent process under additive treatment effect as given in Eq.\ref{eq:tx_additive}, we can re-write the conditional mean as $\mathbb{E}\left[ \bm{\eta}_i(t)|\bm{\eta}_i(s) \right] = e^{-\bm{\theta}(t - s)} \bm{\eta}_i^*(s) + \bm{\mu}(t)$. Writing the conditional mean in this format makes the additive impact of treatment quite clear. When treatment is incorporated into the drift of the latent process, the impact of treatment on the conditional expectation is still additive; however, the impacts of $\bm{\theta}$ and $\bm{\mu}_i(t)$ on the trajectory of the latent process are linked together in a much more complicated way (see Figure \ref{fig:hw_integral_setting1}).

\subsection{Modeling the Impact of Treatment on the Hazard of Recurrent Events}

In the previous section, we described how treatment can modify the trajectory of the latent process. Now, we discuss modeling the direct impact of treatment on the hazard of an event. Modeling the impact of treatment through a term included in the hazard model implies that the treatment alters the risk of a recurrent event through some mechanism that is not captured by the latent process. If the app-based notifications in the motivating study target engagement in certain behaviors, then, by using this model formulation, we assume that these states are different from those states captured by the latent factors. To model this treatment effect, we include a treatment-related term $\Tilde{\mu}_i(t)$ in the event-time submodel for individual $i$'s $r^{th}$ recurrent event:
\small
\begin{align}\label{eq:haz_mod}
    h_{ir} \{t | \mathcal{H}_i(t), \mathcal{R}_i(t), X_i\} = h_0(t) \exp\big\{f(\mathcal{H}_i(t); \bm{\beta}_H) + g(\mathcal{R}_i(t); \bm{\beta}_R) + \Tilde{\mu}_i(t) + \bm{X}_i^{\top} \bm{\gamma} \big\}
\end{align}

\normalsize

For simplicity, we assume that $\Tilde{\mu}_i(t)$ has a form similar to the treatment model for the latent process (Eq.\ref{eq:mu}). In a different setting, one may want to specify an alternative form for $\Tilde{\mu}_i(t)$, but here we use:
\begin{align}\label{eq:mu_for_haz}
    \Tilde{\mu}_i(t) = \sum_{t_{ia} \in \mathcal{A}_i(t)} \Tilde{\tau} \left(1 - \frac{t - t_{ia}}{\delta_b} \right)_+
\end{align}

\noindent where the impact of treatment on the hazard is captured through the unknown (scalar) parameter $\Tilde{\tau}$. As before, we assume that the duration of the treatment effect $\delta_b$ is known.

\subsection{Inference}\label{ss:inference}

We take a Bayesian approach to fit our joint longitudinal-recurrent event model. Combining the measurement submodel (Eq. \ref{eq:meas_submod}), hazard submodel (Eq. \ref{eq:haz_mod}), and structural submodel with the treatment effect modeled either as an additive shift (Eq. \ref{eq:ou_add_dist}) or as drift (Eq. \ref{eq:hw_dist}), our likelihood is
\footnotesize
\begin{equation}\label{eq:likelihood}
    \begin{aligned}
        p(\bm{T}, \bm{\delta}, \bm{Y}; \bm{\Theta}_R, \bm{\Theta}_M, \bm{\Theta}_S) = & \prod_{i = 1}^{N} \prod_{r = 1}^{R_i} \int \Bigg[ p(T_{ir}, \delta_{ir} | \mathcal{H}_i(T_{ir}), \mathcal{R}_i(T_{i,r-1}), \mathcal{A}_i(T_{ir}) ; \bm{\Theta}_R) \\ & \times \prod_{j = 1}^{n_i} p(Y_i(t_{ij}) | \mathcal{H}(T_{ir})_i, \mathcal{A}(T_{ir})_i ; \bm{\Theta}_M) p(\bm{\eta}_i; \bm{\Theta}_S) d \bm{\eta}_i \Bigg]
    \end{aligned}
\end{equation}
\normalsize

\noindent where $R_i$ is the number of events for individual $i$, $\bm{\Theta}_R = (\bm{\beta}_H, \bm{\beta}_R, \Tilde{\tau})$, $\bm{\Theta}_M = (\bm{\Lambda}, \bm{\Sigma}_u, \bm{\Sigma}_{\epsilon})$, and $\bm{\Theta_S} = (\bm{\theta}, \bm{\sigma}, \bm{\tau})$. Recall that $\mathcal{H}_i(t)$ is the history of the latent process until time $t$, $\mathcal{R}_i(t)$ is the history of recurrent events until time $t$, and $\mathcal{A}_i(t)$ is the treatment history until time $t$. The likelihood does depend on the treatment history, but we can factor out this term after conditioning on the observed treatment history and so we omit it from our likelihood definition above. We do not include baseline covariates in this definition of the likelihood, but incorporating them is straightforward. 

The likelihood contribution of the recurrent events is given by 
\small
\begin{equation*}
    p(T_{ir}, \delta_{ir} | \mathcal{H}_i(T_{ir}), \mathcal{R}_i(T_{i,r-1}), \mathcal{A}_i(T_{ir}) ; \bm{\Theta}_R) = h_{ir}(T_{ir} | \mathcal{H}_i(T_{ir}))^{\delta_{ir}} exp\left\{ - \int_{T_{i,r-1}}^{T_{ir}} h_{ir}(s) ds \right\}.
\end{equation*}
\normalsize
\noindent Note in this definition of the joint likelihood, we opt to write the distribution of the observed longitudinal outcome conditional only on the latent factors and integrate out the random intercepts in order to decrease the number of unknown parameters that must be sampled within the Bayesian algorithm. This distribution for the measured longitudinal outcome is $$ \bm{Y}_{i}(t_{ij}) | \mathcal{H}_i(T_{ir}), \mathcal{A}_i(T_{ir}) ; \bm{\Theta}_M \sim N(\bm{\Lambda} \eta_i(t_{ij}), \bm{\Sigma}_u + \bm{\Sigma}_{\epsilon}).$$ The distribution of latent process $p(\bm{\eta}_i ; \bm{\Theta}_S)$ is the product of $p$-dimensional conditional Gaussian distributions of either the form in Eq.\ref{eq:ou_add_dist} or \ref{eq:hw_dist}, depending on how treatment is assumed to impact the latent process.

When fitting this joint model, we must consider (i) how to ensure both the latent process parameters and the loadings matrix are identifiable and (ii) how to calculate the cumulative hazard, which requires integrating over the multivariate OU process. To address (i), we model the latent factors on the correlation scale (see \cite{tran_2021b}). By forcing the latent factors to have a stationary variance of 1, we fix the amount of variability in the process. The loadings matrix $\bm{\Lambda}$ then rescales the latent factors to capture their association with the measured longitudinal outcomes $\bm{Y}$. To implement this identifiability constraint, we follow the approach described in \cite{tran_2021b} and reparameterize the OU process: instead of estimating parameters $\bm{\theta}$ and $\bm{\sigma}$, we estimate $\bm{\theta}$ and the off-diagonal elements $\bm{\rho}$ of the stationary correlation matrix of the OU process $\bm{V}$ (previously defined in Section \ref{ss:ou_tx_effect} as $\bm{V} = vec^{-1}\big\{(\bm{\theta} \oplus \bm{\theta})^{-1} vec(\bm{\sigma} \bm{\sigma}^{\top}) \big\}$). For a bivariate OU process, we estimate $\bm{V} = \begin{bmatrix}
     1 & \rho \\ \rho & 1 
\end{bmatrix}$. We also require that $\bm{\theta}$ has eigenvalues with positive real parts (see \cite{tran_2021b} for details), and that $\bm{\Lambda}$ contains structural zeros with known locations. This constraint on $\bm{\Lambda}$ means that we know which of the observed longitudinal outcomes are measurements of which of the latent factors. In practice, domain knowledge can help inform the placement of these structural zeros. All non-zero elements of $\bm{\Lambda}$ must be positive.

To address (ii) and avoid the complex integration required to evaluate the cumulative hazard in Eq.\ref{eq:likelihood}, we take a discrete approximation based on a midpoint rule. Within the Bayesian algorithm, we generate values of the latent process on a fine grid and then sum the hazard across this fine grid using a midpoint rule that allows us to closely approximate the integral with a sum. We previously used this strategy in the context of a joint longitudinal-survival model (for single time-to-event outcomes) and found that, in the ILD setting, parameters' posterior distributions were not sensitive to the choice of grid width. This sensitivity analysis, along with a more detailed description of this midpoint approximation to the cumulative hazard, can be found in \cite{abbott_2024}.

We use Stan, a software that carries out a Hamiltonian Monte Carlo sampling algorithm, to fit our model \citep{carpenter_2017}. Priors are given in Appendix D.

After fitting the joint model, it may be of interest to compare models that specify, for example, different mechanisms for the effect of treatment or different structures for the loadings matrix and latent factors. To do so, deviance information criterion (DIC) and widely applicable information criterion (WAIC) \citep{watanabe_2010} can be used. \cite{gelman_2014} define these information criteria as consisting of two terms: one term for the log-likelihood and another term that captures the effective number of parameters. Letting $\bm{\Theta} = (\bm{\Theta}_R, \bm{\Theta}_M, \bm{\Theta}_S)$, DIC is defined as:
$$DIC = -2log p (Y, T, \delta | \hat{\bm{\Theta}} ) + 2p_{DIC}$$

\noindent where $\hat{\bm{\Theta}}$ is the posterior mean; $p_{\text{DIC}}$ is the effective number of parameters, with $p_{\text{DIC}} = 2 \Big( log p (Y, T, \delta | \hat{\bm{\Theta}}) - $ $\frac{1}{S} \sum_{s = 1}^S log p (Y, T, \delta | \hat{\bm{\Theta}}^s)  \Big)$; and $s$ indexes posterior samples. Similarly,
$$WAIC = -2 \widehat{\text{lppd}} + 2p_{\text{WAIC}}$$

\noindent where $\widehat{\text{lppd}} = \sum_{i = 1}^N log\left( \frac{1}{S} \sum_{s = 1}^S p(Y_i, T_i, \delta_i | \bm{\Theta}^s_R, \bm{\Theta}^s_M, \bm{\Theta}^s_S) \right)$ and the effective number of parameters is $p_{\text{WAIC}} = \sum_{i = 1}^N V_{s = 1}^S \left( log p (Y_i, T_i, \delta_i | \bm{\Theta}^s_R, \bm{\Theta}^s_M, \bm{\Theta}^s_S) \right)$, with $V_{s = 1}^S (a_s) = \frac{1}{S-1} \sum_{s = 1}^S (a_s - \Bar{a})^2$. Both of these information criteria rely on evaluating the log-likelihood, either at the posterior mean $\hat{\bm{\Theta}}$ or each of the posterior samples $\bm{\Theta}^s, s = 1, ..., S$. To fit the model in Stan, we use the conditional likelihood, $log p (Y_i, T_i, \delta_i | \mathcal{H}_i; \bm{\Theta})$, rather than a version of the likelihood that is marginalized over the latent process. The marginal version, however, is generally recommended when comparing fits of latent variable models \citep{rizopoulos_2023, merkle_2019}. To compute DIC and WAIC, we must integrate the conditional likelihood in Eq. \ref{eq:likelihood} over the continuous-time multivariate OU process in the longitudinal and event submodels. Specifically, we use a Monte Carlo-based approach to sample values of the latent process and approximate the integral (see Appendix F). This approach allows us to estimate the value of the marginal log-likelihood at both the posterior mean $\hat{\bm{\Theta}}$ and each of the posterior samples $\bm{\Theta}^s$, which we then use to calculate DIC and WAIC.

We can also examine model fit using a posterior predictive check; specifically, we assess the form of the hazard model, which is a function of the time-varying treatment effect and latent variables that capture affective states. Generally, our approach involves comparing the observed number of events to the posterior expected number of events over different windows of time, conditional on event history prior to the start of the window. Treatment history and the posterior draws of the latent process are assumed known when generating these posterior predictions.  We use mean cumulative functions (MCFs) to compare the posterior predictions to the observed events. Details on this approach are given in Appendix G.

\section{Simulation Study}\label{s:sim_study}

The goal of the simulation study is to assess the statistical properties of correctly specified models fit to simulated datasets informed by other mHealth studies of smoking cessation. When generating data, we use two sets of true parameter values---called setting 1 and setting 2---which are informed by parameter estimates from observational mHealth studies similar to the motivating mHealth MRT. Our simulated data, however, are slightly simpler than the data from these mHealth studies to modulate the computational cost of conducting the simulation study. 

\subsection{Data Generation}

A single simulated dataset consists of $N = 100$ individuals who are followed for 14 days. At four random times each day, we observe longitudinal outcomes $\bm{Y}$, which consists of $k = 4$ observed outcomes that measure $p = 2$ latent factors. Treatments are sent randomly once per day and the effect of each treatment lasts for half a day ($\delta_a = 0.5, \delta_b = 0.5$). Depending on the exact time at which each treatment is delivered, two treatments may be active at the same time. As discussed earlier, we do not consider the setting in which treatment directly impacts the longitudinal outcome, so the measurement submodel always takes the form described in Eq.\ref{eq:meas_submod}. For the structural submodel, we assume that the treatment impacts the latent process as either an additive shift to the mean (Eq. \ref{eq:tx_additive}) or through a drift term (Eq. \ref{eq:hw_sde}). The treatment function $\bm{\mu}_i(t)$ is defined as in Eq.\ref{eq:mu}. For each version of the structural submodel, we consider two different event-time submodels:

\begin{enumerate}
    \item Treatment impacts the hazard of event $r$ through the latent process and treatment modifies the hazard directly: $h_{ir}(t) = h_0 \exp\big\{\beta_1 \eta_{1i}(t) + \beta_2 \eta_{2i}(t) + \Tilde{\mu}_i(t) \big\}.$
    \item Treatment impacts the hazard of event $r$ through the latent process and treatment modifies the hazard directly. The hazard also depends on the time since the most recent prior $(r-1)^{th}$ event, where experiencing an event (i.e., engaging in substance use) temporarily increases the risk of a subsequent event, but this temporary increase decays to 0 after a certain amount of time: $h_{ir}(t) = h_0 \exp\big\{\beta_1 \eta_{1i}(t) + \beta_2 \eta_{2i}(t) + \beta_3 g(t - t_{i,r-1}) + \Tilde{\mu}_i(t) \big\}.$
\end{enumerate}

\noindent For setting 1, $g(x) = \frac{1}{1 + \exp\{4(x - 2)\}}$ and for setting 2, $g(x) = \frac{1}{1 + \exp\{1.5(x - 2)\}}$. We assume that these functions are known when fitting the model. $\Tilde{\mu}_i(t)$ is given in Eq.\ref{eq:mu_for_haz} with $\delta_b = 0.5$. In both hazard models, we assume that the baseline hazard is constant, $h_0 = \exp(\beta_0)$.

When simulating the data, we assume that treatment has a positive impact on one of the latent factors and a negative impact on the other. In the motivating MRT, we might expect that receiving a prompt would temporarily increase positive affect (via $\tau_1 = 2$) while decreasing negative affect (via $\tau_2 = -1$). These effects are interpreted as either occurring on the scale of the mean (if the treatment effect is modeled additively) or on the scale of the derivative (if the treatment effect is modeled as drift). We also assume that receiving a prompt decreases the hazard of an event (via $\Tilde{\tau} = -0.8$), which is what prompts in the MRT aim to do by encouraging engagement in certain behavioral strategies. True values for other model parameters, which are informed by previous mHealth smoking cessation studies, are in Appendix E.

In Appendix E, we provide some plots of the simulated data. The number of observed events ranges from an average of 1.9 events per person in setting 2 when the treatment effect modeled through the drift term and the hazard takes the form of model 2, to an average of 4.7 events per person in setting 2 when the treatment effect modeled as an additive shift to the latent process and the hazard takes the form of model 1.

\subsection{Results}

For each parameterization of the treatment effect on the latent process (additive or drift) and for each hazard model (1 or 2), we simulate 100 datasets and fit the model. We repeat this process for the true parameter values corresponding to both setting 1 and setting 2. When fitting the joint model, we initialize parameters at values with approximately the correct order of magnitude and with the correct sign. In practice, a two-stage approach could be used to determine reasonable starting values when true values are not known. For each dataset, we run 1 chain for 2000 iterations and discard the first 1000 as burn-in.

\begin{figure}
    \centering
    \captionsetup{width=14cm}
    \includegraphics[width=14cm]{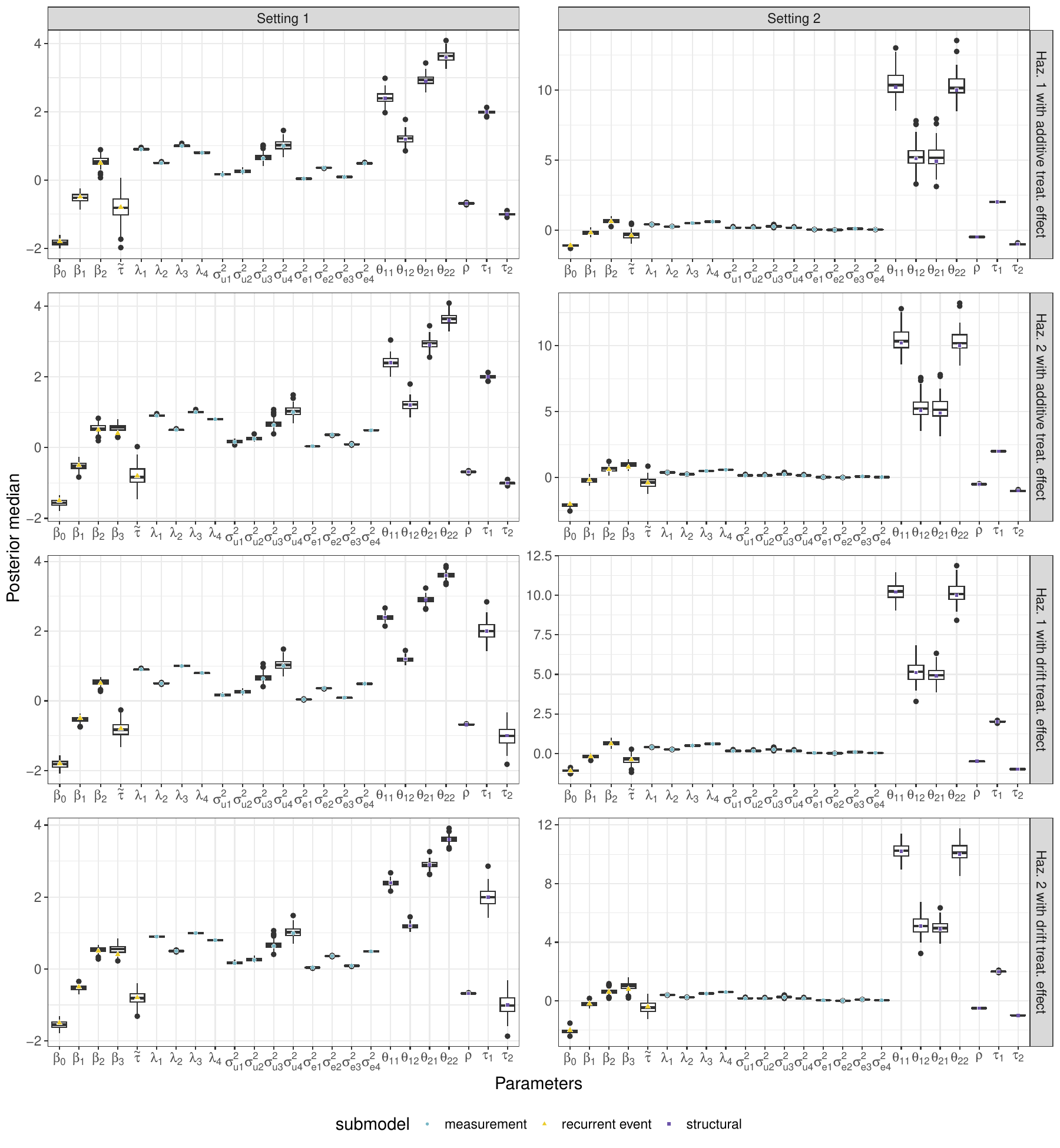}
    \caption[For data generated under settings 1 and 2 with different hazards and treatment effect models, we use box plots to summarize the distribution of the posterior medians for all parameters across the 100 simulated datasets.]{Box plots summarize the distribution of the \textbf{posterior medians for parameters} across the 100 datasets generated under each simulation scenario. When fitting the model, we assume that the grid used in the midpoint approximation of the cumulative hazard function has a width of 0.5 days. True parameter values are indicated with colored dots.} \label{fig:post_medians}
\end{figure}

\begin{figure}
    \centering
    \captionsetup{width=14cm}
    \includegraphics[width=14cm]{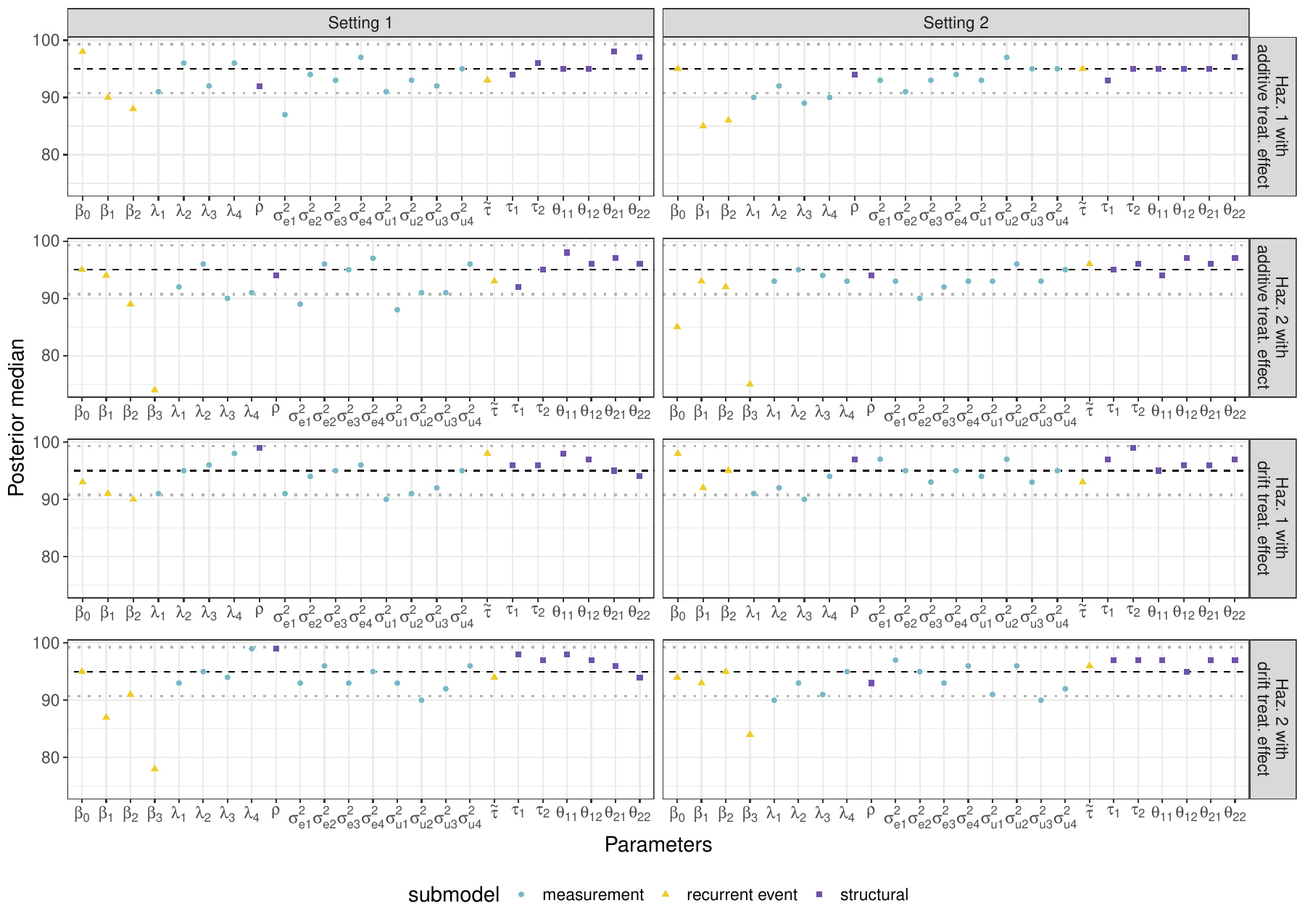}
    \caption[For data generated under settings 1 and 2 with different hazards and treatment effect models, we summarize the coverage rate of 95\% credible intervals across the 100 simulated datasets with the colored dots.]{For data generated under settings 1 and 2 with different hazards and treatment effect models, we summarize the \textbf{coverage rate of 95\% credible intervals} across the 100 simulated datasets with the colored dots. The black horizontal dashed lines indicate target coverage and the dotted grey lines corresponds to the upper and lower bounds of a 95\% binomial proportion confidence interval for a probability of 0.95.} \label{fig:post_cov}
\end{figure}

To assess bias and variance, we compare posterior medians to true values (Figure \ref{fig:post_medians}) and evaluate the coverage of 95\% credible intervals (Figure \ref{fig:post_cov}). We find that the posterior medians are close to the true values and that the posterior distributions have close-to-nominal coverage. In hazard model 2, point estimates for $\beta_3$---the parameter that captures the association between the hazard and a known function of time since the most recent event---show some bias and, as a result of this bias, lower-than-nominal coverage. We further investigate the bias in this parameter with some additional simulations, which we present in more detail in Appendix E.1. These supplemental simulations suggest that this bias is likely related to fitting a rather complicated model to a fairly small dataset; they also suggest that the higher temporal correlation in the latent process in setting 1, compared to setting 2, makes recovering unbiased estimates of $\beta_3$ more difficult. Overall, however, the magnitude of the bias in our point estimates for $\beta_3$ is still small and reasonable given the complexity of this joint model.

\section{Analysis of the MRT Data}\label{s:case_study}

We use our joint longitudinal recurrent event model to analyze data from the motivating MRT. We use the longitudinal submodel to summarize the observed 15 emotions as two latent factors that represent positive affect and negative affect. We consider two models for the repeated treatments (i.e., the app-based prompts): one in which treatment has an additive impact directly on the mean of the latent factors and another in which treatment impacts the dynamics of the latent factors on the derivative scale through the time-varying drift term. In the recurrent event submodel, we model the hazard of recurrent poly-substance use events using the time-varying predictors of positive affect, negative affect, and repeated treatment effects:
$h_{ir}(t) = h_0(t) \exp\left[ \beta_1 \eta_{1i}(t) + \beta_2 \eta_{2i}(t) + \Tilde{\mu}_i(t) \right]$.

We assume that the treatment effect duration is 8 hours for both submodels ($\delta_a = \delta_b = 8 \text{ hours}$). Alternative durations could be compared in a sensitivity analysis. 
We specify a log-normal baseline hazard and use a clock-reset approach. In the MRT, participants are asked to quit smoking on day 4; we account for potential changes in substance use before and after day 4 by specifying a separate baseline hazard for the pre- and post-quit periods. When modeling the impact of treatment directly on the hazard (i.e., when specifying $\Tilde{\mu}_i(t)$), we consider one version that assumes the treatment parameter $\Tilde{\tau}$ is constant across the entire study and another other version that allows this treatment parameter to differ between the pre- and post-quit periods (i.e., we have $\Tilde{\tau}_{\text{pre}}$ and $\Tilde{\tau}_{\text{post}}$). Further details are in Appendix A.3.

One additional advantage of using a log-normal baseline hazard is that it is potentially flexible enough to allow the underlying risk of an event to change as a function of the time since the prior event. In an attempt to avoid specifying an overly complex model given our limited sample size, we opt to account for potential changes in the hazard of an event as a function of time since the previous event through this log-normal baseline hazard with the clock reset, rather than by modeling the hazard as a separate function of time since the most recent event (i.e., by including a $\beta_3 g(\cdot)$ term in our hazard model).

When fitting the model, we rescale time so that the 10-day interval has time units in the range of 0 to 1 (this approach is suggested in \cite{tran_2021b} to help deal with potentially oscillating OU processes). Rescaling of the time interval impacts the interpretation of the $\bm{\theta}$ parameter in the structural submodel and the log-normal parameters in the baseline hazard.

To initialize parameters, we use a two-stage approach and first fit the longitudinal submodel, then a simple version of the hazard submodel. To fit the joint model, we use 4 chains with 3,000 samples, discarding the first 1,000 as burn-in. We find that fitting a joint model with treatment impacting the latent process in an additive way, rather on the derivative scale, is easier for a dataset of this size. When modeling treatment on the derivative scale and allowing separate pre- and post-quite treatment parameters in the hazard model, we increase the burn-in to 1,500 iterations but find that fitting this model is still somewhat challenging given a dataset of this size. More details on model convergence are in Section A.3.  

We compare model fit using DIC and WAIC. WAIC is given in Table \ref{tab:waic}. As WAIC is generally preferred over DIC \citep{gelman_2014b}, DIC is provided in the Appendix (Web Table 3). WAIC indicates that a time-varying drift treatment effect model for the latent process and two treatment-related parameters in the hazard model fit our data the best; DIC agrees. The WAIC for this model, however, is only very slighly better than WAIC for the other models. Due to the sampling-based approach used to calculate the marginal log-likelihood, some uncertainty exists in the exact value of WAIC and so we emphasize that our main motivation for presenting two different approaches for modeling the impact of treatment on the latent process is that these approaches correspond to different scientific beliefs about the mechanisms by which treatment impacts the latent process. Given the smaller sample size of this MRT data ($N = 64$ individuals), it may be difficult to assess which model is preferred.  

\begin{figure}
    \centering
    \captionsetup{width=14cm}
    \includegraphics[width=14cm]{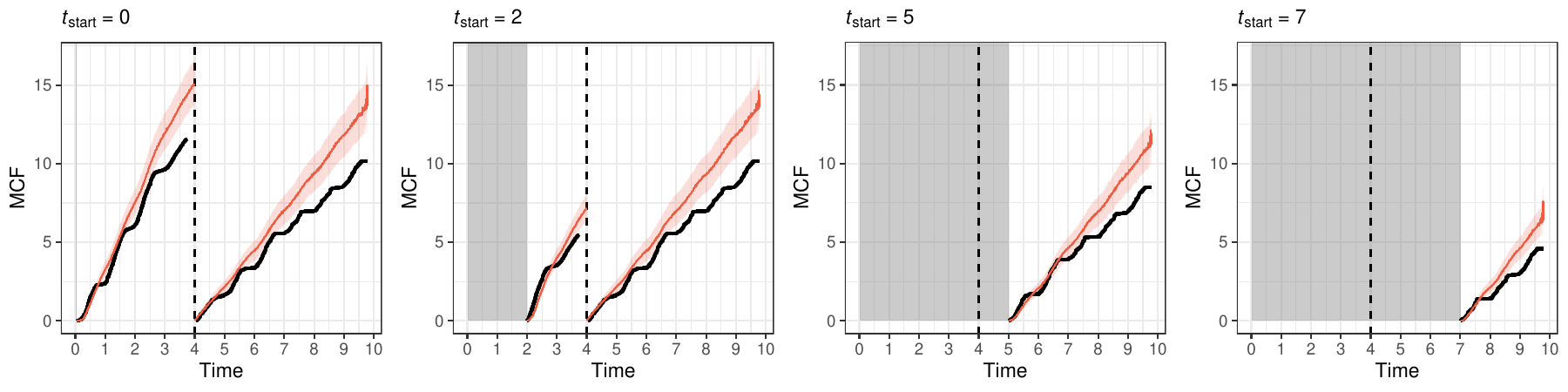}
    \caption{Median and (5, 95)$^{th}$ percentiles of the mean cumulative function (MCF) for posterior predicted events (red line with shaded ribbon) based on conditional posterior predictions from the joint model when the model for treatment effect on the latent process assumed the drift form and the hazard had two treatment-related coefficients. The black lines show the MCF for observed events.}
    \label{fig:ppc_realdata_h0lognormal}
\end{figure}

In addition to comparing models using information criteria, we examine model fit using a posterior predictive check. We consider posterior predictions of events across day 0--10, day 2--10, day 5--10, and day 7--10, conditioning on the event history prior to the start of the window and assuming the treatment history and posterior draws of the latent process are known.  Comparisons of MCFs based on posterior predictions and observed data are shown in Figure \ref{fig:ppc_realdata_h0lognormal} for the model with a time-varying additive treatment effect model for the latent process and one treatment-related parameter in the hazard model (i.e., the best model according to WAIC and DIC). The figure shows that the MCF for posterior predictions and for observed data are in close agreement shortly after the start of the prediction window, but error accumulates over the course of the window and so the difference between the predicted and observed MCFs increases. Plots for posterior predictions from the other three models show similar patterns (see Appendix Web Figure 22).

\begin{figure}
    \centering
    \captionsetup{width=14cm}
    \includegraphics[width=14cm]{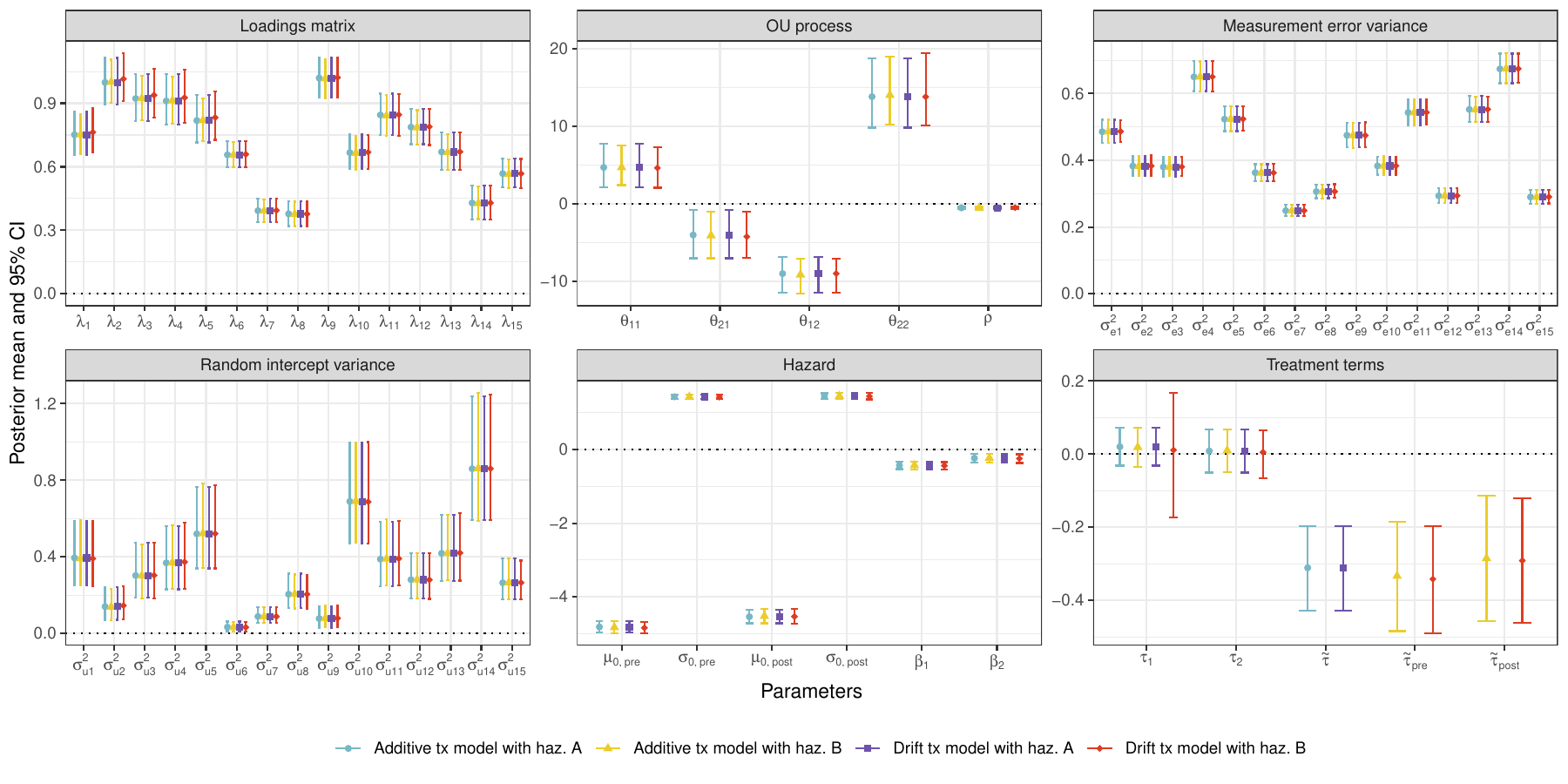}
    \caption[Posterior means and 95\% credible intervals for parameters in the joint models fit to the motivating MRT data.]{Posterior means and 95\% credible intervals for parameters in the  joint models fit to the motivating MRT data. Hazard model A has a single treatment (tx)-related parameter; hazard model B has separate pre- and post-quit treatment parameters. Emotions: 1 = grateful, 2 = happy, 3 = proud, 4 = relaxed, 5 = enthusiastic, 6 = angry, 7 = ashamed, 8 = guilty, 9 = irritable, 10 = lonely, 11 = anxious, 12 = sad, 13 = restless, 14 = bored, 15 = hopeless.} 
    \label{fig:case_study_forest_plt}
\end{figure}

Posterior medians and 95\% credible intervals for all four models are in Figure \ref{fig:case_study_forest_plt}. The estimated correlation ($\rho$) between the latent factors for positive and negative affect is negative, as expected. Across all models, the posterior mean of $\rho$ ranges from -0.53 to -0.50. Three of the negative emotions (angry, ashamed, and irritable) have smaller variance estimates for their item-specific random intercepts than the other negative emotions and the positive emotions. When examining the data directly, we see that the empirical variability in the individual-specific averages for these three negative emotions tends to be slightly smaller than for the other emotions, supporting this result (see Appendix A.3.3).

Across all joint models, we find that the estimated effect of treatment ($\bm{\tau}$) in the treatment model for the latent process is near zero. This result is the same whether we model the impact of treatment on the latent process as an additive shift to the mean or on the derivative scale. When examining posterior parameter estimates for the treatment terms in the latent process, we see similar protective effects of sending a prompt to an individual after they have attempted to quit ($\Tilde{\tau}_{\text{post}}$), compared to before quit ($\Tilde{\tau}_{\text{pre}}$).

\begin{table}
\resizebox{10cm}{!}{
\begin{tabular}{r|cc}
\textbf{WAIC} & \multicolumn{2}{c}{\textbf{Impact of treatment on latent process}} \\ \hline
\textbf{Hazard model} & \multicolumn{1}{c}{Additive} & \multicolumn{1}{c}{Drift} \\ \hline
Single treatment parameter & 193,664.3 & 195,059.2 \\ \hline
Separate pre- and post-quit treatment parameters & 194,316.9 & \textbf{193,297.4}
\end{tabular}}
\caption[WAIC for the joint models fit to the motivating MRT data.]{WAIC for the joint models fit to the motivating MRT data. When calculating the marginal log-likelihood, we subsampled every 5th iteration of the final 1,000 posterior samples. The lowest WAIC is bolded.}\label{tab:waic}
\end{table}

\section{Discussion}\label{s:discussion}

We proposed a method for jointly modeling multivariate ILD, a recurrent event outcome, and the effect of repeatedly delivered treatments, as often found in data from MRTs. In the longitudinal submodel, we only consider modeling the impact of treatments on the latent process. In addition to allowing treatments to alter the trajectory of the latent process, it might be reasonable to assume that the occurrence of an event modifies the latent process. In this case, one could use the same approach that we describe for modeling treatment's impact on the latent process, but substitute treatment times with event times. The impact of treatments and events on the latent process could also be modeled simultaneously. We do not consider this variation of the longitudinal submodel in our simulations or case study, but mention it now as it may be of interest in other settings.

A limitation of our joint model is its computational cost. For the case study, fitting the models required between 38 hours (when the model for treatment effect on the latent process was additive and the hazard had two treatment-related coefficients) and 146 hours (when the model for treatment effect on the latent process assumed the drift form and the hazard had two treatment-related coefficients). Computing the marginal log-likelihood for DIC/WAIC requires a similar amount of time, although parallelization can reduce DIC/WAIC calculations to under an hour from the analyst’s perspective. While this total computation time is small relative to the time needed to design and conduct an MRT, improving the speed of inference is important future work, as it would facilitate sensitivity analyses.

In the case study, we use DIC and WAIC to compare models that make different assumptions about how the longitudinal, recurrent event, and treatment-related components of the joint model are associated with each other. We use WAIC and DIC to compare joint models as a whole; \cite{zhang_2017} propose a version of DIC for joint longitudinal-survival models where DIC is decomposed into contributions from the longitudinal and survival submodels. Adapting this version of DIC to our setting could be an potentially interesting extension.

When modeling treatment in the longitudinal and recurrent event submodels via the functions $\bm{\mu}_i(t)$ and $\Tilde{\mu}_i(t)$, we make assumptions about both the specific shape of the treatment effect over time and the duration of active treatment, $\delta_a$ and $\delta_b$. In both our simulation study and the case study, we assume that the effect of treatment is at its maximum at the time of treatment and that the treatment ``dosage" decreases with the time since delivery. We focus on estimating the maximum effect of treatment ($\bm{\tau}$ and $\Tilde{\tau}$) at the time of delivery but assume the form of its decay is known. We do so largely because the MRT motivating this work does not have a particularly large sample size ($N = 64$). Here, DIC and WAIC can be used to compare joint models with different pre-specified treatment decay functions, but in other settings where more data are available, we could try to estimate additional parameters that characterize the treatment function (e.g., $\delta_a$ or $\delta_b$).

So far, we have assumed that treatment is binary; that is, either a prompt is sent or not sent. In reality, treatment in the motivating MRT can take three different levels: no prompt, a prompt encouraging engagement in a behavioral strategy requiring low effort, or a prompt to engage in a behavioral strategy requiring more effort. Extending the models presented in this paper to account for multiple levels of treatment would be straightforward, but would require more parameters. Additionally, our current definition of treatment depends only on the outcome of randomization. This definition corresponds to an intent-to-treat analysis. Information on subjects' engagement with the intervention (i.e., whether they carried out the suggested behavioral strategy) is self-reported in the EMA delivered one hour after randomization. Engagement with the intervention, called treatment compliance in other settings, could also be incorporated into the definition of treatment to enable an as-treated analysis.

Finally, in practice, many EMA questionnaires are designed such that event outcomes are measured in an interval censored way; that is, questions are often phrased as, ``Since the last EMA, have you [experienced some sort of event or engaged in some sort of behavior]''. Sometimes, as in our case study, this question is followed up with questions that prompt an individual to provide approximate times of the events or behaviors. Quite often, however, this subsequent set of questions is not asked and so the data only contain information on whether an event occurred within the time interval between consecutive EMAs. Thus, extending our method to handle this type of interval censored data could be practically useful.

\begin{acks}[Acknowledgments]

\end{acks}
\begin{funding}
MRA was supported in part by NIH grant numbers F31DA057048 and T32CA083654. JMGT was supported in part by NIH grant number R01CA129102. INS and WHD were supported in part by NIH grant number P50DA054039. INS was supported in part by NIH grant number R01DA039901.  WHD was supported in part by NIH grant number R01GM152549.  LNP was supported in part by R00CA252604.  LNP, DWW, and CYL were supported in part by R01CA224537, P30CA042014, U01CA229437, UL1TR002538, U54CA280812, and the Huntsman Cancer Foundation. The content is solely the responsibility of the authors and does not necessarily represent the official views of the National Institutes of Health.
\end{funding}

\begin{supplement}
\noindent Data from the smoking cessation trial are not available due to privacy concerns. Simulated example data and code are available on Github (\verb|www.github.com/madelineabbott/| \verb|OUF_JM_TX|). Additional supporting materials are available online.
\end{supplement}


\bibliographystyle{imsart-nameyear} 
\bibliography{references}       

\begin{thebibliography}{24}

\bibitem[\protect\citeauthoryear{Abbott et~al.}{2025a}]{abbott_2024}
\begin{barticle}[author]
\bauthor{\bsnm{Abbott},~\bfnm{M~R}\binits{M.~R.}}, \bauthor{\bsnm{Dempsey},~\bfnm{W~H}\binits{W.~H.}}, \bauthor{\bsnm{Nahum-Shani},~\bfnm{I}\binits{I.}}, \bauthor{\bsnm{Potter},~\bfnm{L~N}\binits{L.~N.}}, \bauthor{\bsnm{Wetter},~\bfnm{D~W}\binits{D.~W.}}, \bauthor{\bsnm{Lam},~\bfnm{C~Y}\binits{C.~Y.}} \AND \bauthor{\bsnm{Taylor},~\bfnm{J~M~G}\binits{J.~M.~G.}}
(\byear{2025}a).
\btitle{A {B}ayesian joint longitudinal-survival model with a latent stochastic process for intensive longitudinal data}.
\bjournal{Biometrics}
\bvolume{81}.
\end{barticle}
\endbibitem

\bibitem[\protect\citeauthoryear{Abbott et~al.}{2025b}]{abbott_2023}
\begin{barticle}[author]
\bauthor{\bsnm{Abbott},~\bfnm{M~R}\binits{M.~R.}}, \bauthor{\bsnm{Dempsey},~\bfnm{W~H}\binits{W.~H.}}, \bauthor{\bsnm{Nahum-Shani},~\bfnm{I}\binits{I.}}, \bauthor{\bsnm{Lam},~\bfnm{C~Y}\binits{C.~Y.}}, \bauthor{\bsnm{Wetter},~\bfnm{D~W}\binits{D.~W.}} \AND \bauthor{\bsnm{Taylor},~\bfnm{J~M~G}\binits{J.~M.~G.}}
(\byear{2025}b).
\btitle{A Continuous-Time Dynamic Factor Model for Intensive Longitudinal Data Arising from Mobile Health Studies}.
\bjournal{Psychometrika}
\bvolume{90}
\bpages{1-22}.
\end{barticle}
\endbibitem

\bibitem[\protect\citeauthoryear{Boruvka et~al.}{2018}]{boruvka_2018}
\begin{barticle}[author]
\bauthor{\bsnm{Boruvka},~\bfnm{A}\binits{A.}}, \bauthor{\bsnm{Almirall},~\bfnm{D}\binits{D.}}, \bauthor{\bsnm{Witkiewitz},~\bfnm{K}\binits{K.}} \AND \bauthor{\bsnm{Murphy},~\bfnm{S~A}\binits{S.~A.}}
(\byear{2018}).
\btitle{Assessing Time-Varying Causal Effect Moderation in Mobile Health}.
\bjournal{Journal of the American Statistical Association}
\bvolume{113}
\bpages{1112-1121}.
\end{barticle}
\endbibitem

\bibitem[\protect\citeauthoryear{Brigo and Mercurio}{2007}]{brigo_2007}
\begin{binbook}[author]
\bauthor{\bsnm{Brigo},~\bfnm{D}\binits{D.}} \AND \bauthor{\bsnm{Mercurio},~\bfnm{F}\binits{F.}}
(\byear{2007}).
\btitle{Interest Rate Models-Theory and Practice with Smile, Inflation and Credit},
\bedition{second} ed.
\bchapter{3}.
\bpublisher{Springer Finance}.
\end{binbook}
\endbibitem

\bibitem[\protect\citeauthoryear{Carpenter et~al.}{2017}]{carpenter_2017}
\begin{barticle}[author]
\bauthor{\bsnm{Carpenter},~\bfnm{B}\binits{B.}}, \bauthor{\bsnm{Gelman},~\bfnm{A}\binits{A.}}, \bauthor{\bsnm{Hoffman},~\bfnm{M~D}\binits{M.~D.}}, \bauthor{\bsnm{Lee},~\bfnm{D}\binits{D.}}, \bauthor{\bsnm{Goodrich},~\bfnm{B}\binits{B.}}, \bauthor{\bsnm{Betancourt},~\bfnm{M}\binits{M.}}, \bauthor{\bsnm{Brubaker},~\bfnm{M}\binits{M.}}, \bauthor{\bsnm{Guo},~\bfnm{J}\binits{J.}}, \bauthor{\bsnm{Li},~\bfnm{P}\binits{P.}} \AND \bauthor{\bsnm{Riddell},~\bfnm{A}\binits{A.}}
(\byear{2017}).
\btitle{Stan: A Probabilistic Programming Language}.
\bjournal{Journal of Statistical Software}
\bvolume{76}
\bpages{1--32}.
\end{barticle}
\endbibitem

\bibitem[\protect\citeauthoryear{Gelman, Carlin and Stern}{2014}]{gelman_2014}
\begin{binbook}[author]
\bauthor{\bsnm{Gelman},~\bfnm{A}\binits{A.}}, \bauthor{\bsnm{Carlin},~\bfnm{J~B}\binits{J.~B.}} \AND \bauthor{\bsnm{Stern},~\bfnm{H~S}\binits{H.~S.}}
(\byear{2014}).
\btitle{Bayesian Data Analysis},
\bedition{third} ed.
\bchapter{7}.
\bpublisher{Chapman \& Hall/CRC texts in statistical science}.
\end{binbook}
\endbibitem

\bibitem[\protect\citeauthoryear{Gelman, Hwang and Vehtari}{2014}]{gelman_2014b}
\begin{barticle}[author]
\bauthor{\bsnm{Gelman},~\bfnm{A}\binits{A.}}, \bauthor{\bsnm{Hwang},~\bfnm{J}\binits{J.}} \AND \bauthor{\bsnm{Vehtari},~\bfnm{A}\binits{A.}}
(\byear{2014}).
\btitle{Understanding predictive information criteria for {B}ayesian models}.
\bjournal{Statistics and Computing}
\bvolume{24}
\bpages{997–1016}.
\end{barticle}
\endbibitem

\bibitem[\protect\citeauthoryear{Ibrahim, Chu and Chen}{2010}]{ibrahim_2010}
\begin{barticle}[author]
\bauthor{\bsnm{Ibrahim},~\bfnm{J~G}\binits{J.~G.}}, \bauthor{\bsnm{Chu},~\bfnm{H}\binits{H.}} \AND \bauthor{\bsnm{Chen},~\bfnm{L~M}\binits{L.~M.}}
(\byear{2010}).
\btitle{Basic concepts and methods for joint models of longitudinal and survival data}.
\bjournal{Journal of Clinical Oncology}
\bvolume{26}
\bpages{2796–2801}.
\end{barticle}
\endbibitem

\bibitem[\protect\citeauthoryear{Klasnja et~al.}{2015}]{klasnja_2015}
\begin{barticle}[author]
\bauthor{\bsnm{Klasnja},~\bfnm{P}\binits{P.}}, \bauthor{\bsnm{Hekler},~\bfnm{E~B}\binits{E.~B.}}, \bauthor{\bsnm{Shiffman},~\bfnm{S}\binits{S.}}, \bauthor{\bsnm{Boruvka},~\bfnm{A}\binits{A.}}, \bauthor{\bsnm{Almirall},~\bfnm{D}\binits{D.}}, \bauthor{\bsnm{Tewari},~\bfnm{A}\binits{A.}} \AND \bauthor{\bsnm{Murphy},~\bfnm{S~A}\binits{S.~A.}}
(\byear{2015}).
\btitle{Microrandomized trials: An experimental design for developing just-in-time adaptive interventions}.
\bjournal{Health Psychology}
\bvolume{0}
\bpages{1220-8}.
\bdoi{10.1037/hea0000305}
\end{barticle}
\endbibitem

\bibitem[\protect\citeauthoryear{Klasnja et~al.}{2019}]{klasnja_2019}
\begin{barticle}[author]
\bauthor{\bsnm{Klasnja},~\bfnm{P}\binits{P.}}, \bauthor{\bsnm{Smith},~\bfnm{S}\binits{S.}}, \bauthor{\bsnm{Seewald},~\bfnm{N~J}\binits{N.~J.}}, \bauthor{\bsnm{Lee},~\bfnm{A}\binits{A.}}, \bauthor{\bsnm{Hall},~\bfnm{K}\binits{K.}}, \bauthor{\bsnm{Luers},~\bfnm{B}\binits{B.}}, \bauthor{\bsnm{Hekler},~\bfnm{E~B}\binits{E.~B.}} \AND \bauthor{\bsnm{Murphy},~\bfnm{S~A}\binits{S.~A.}}
(\byear{2019}).
\btitle{Efficacy of Contextually Tailored Suggestions for Physical Activity: A Micro-randomized Optimization Trial of {HeartSteps}}.
\bjournal{Annals of Behavioral Medicine}
\bvolume{53}
\bpages{573-582}.
\end{barticle}
\endbibitem

\bibitem[\protect\citeauthoryear{Liao et~al.}{2016}]{liao_2015}
\begin{barticle}[author]
\bauthor{\bsnm{Liao},~\bfnm{P}\binits{P.}}, \bauthor{\bsnm{Klasnja},~\bfnm{P}\binits{P.}}, \bauthor{\bsnm{Tewari},~\bfnm{A}\binits{A.}} \AND \bauthor{\bsnm{Murphy},~\bfnm{S~A}\binits{S.~A.}}
(\byear{2016}).
\btitle{Sample size calculations for micro-randomized trials in m{H}ealth}.
\bjournal{Statistics in Medicine}
\bvolume{35}
\bpages{1944-1971}.
\end{barticle}
\endbibitem

\bibitem[\protect\citeauthoryear{Merkle, Furr and Rabe-Hesketh}{2019}]{merkle_2019}
\begin{barticle}[author]
\bauthor{\bsnm{Merkle},~\bfnm{E~C}\binits{E.~C.}}, \bauthor{\bsnm{Furr},~\bfnm{D}\binits{D.}} \AND \bauthor{\bsnm{Rabe-Hesketh},~\bfnm{S}\binits{S.}}
(\byear{2019}).
\btitle{Bayesian Comparison of Latent Variable Models: Conditional Versus Marginal Likelihoods}.
\bjournal{Psychometrika}
\bvolume{84}
\bpages{802–829}.
\end{barticle}
\endbibitem

\bibitem[\protect\citeauthoryear{Nahum-Shani et~al.}{2018}]{nahum-shani_2018}
\begin{barticle}[author]
\bauthor{\bsnm{Nahum-Shani},~\bfnm{I}\binits{I.}}, \bauthor{\bsnm{Smith},~\bfnm{S~N}\binits{S.~N.}}, \bauthor{\bsnm{Spring},~\bfnm{B~J}\binits{B.~J.}}, \bauthor{\bsnm{Collins},~\bfnm{L~M}\binits{L.~M.}}, \bauthor{\bsnm{Witkiewitz},~\bfnm{K}\binits{K.}}, \bauthor{\bsnm{Tewari},~\bfnm{A}\binits{A.}} \AND \bauthor{\bsnm{Murphy},~\bfnm{S~A}\binits{S.~A.}}
(\byear{2018}).
\btitle{Just-in-Time Adaptive Interventions ({JITAI}s) in Mobile Health: Key Components and Design Principles for Ongoing Health Behavior Support}.
\bjournal{Annals of Behavioral Medicine}
\bvolume{52}
\bpages{446–462}.
\end{barticle}
\endbibitem

\bibitem[\protect\citeauthoryear{Nahum-Shani et~al.}{2021}]{nahum-shani_2021}
\begin{barticle}[author]
\bauthor{\bsnm{Nahum-Shani},~\bfnm{I}\binits{I.}}, \bauthor{\bsnm{Potter},~\bfnm{L~N}\binits{L.~N.}}, \bauthor{\bsnm{Lam},~\bfnm{C~Y}\binits{C.~Y.}}, \bauthor{\bsnm{Yap},~\bfnm{J}\binits{J.}}, \bauthor{\bsnm{Moreno},~\bfnm{A}\binits{A.}}, \bauthor{\bsnm{Stoffel},~\bfnm{R}\binits{R.}}, \bauthor{\bsnm{Wu},~\bfnm{Z}\binits{Z.}}, \bauthor{\bsnm{Wan},~\bfnm{N}\binits{N.}}, \bauthor{\bsnm{Dempsey},~\bfnm{W}\binits{W.}}, \bauthor{\bsnm{Kumar},~\bfnm{S}\binits{S.}}, \bauthor{\bsnm{Ertin},~\bfnm{E}\binits{E.}}, \bauthor{\bsnm{Murphy},~\bfnm{S~A}\binits{S.~A.}}, \bauthor{\bsnm{Rehg},~\bfnm{J~M}\binits{J.~M.}} \AND \bauthor{\bsnm{Wetter},~\bfnm{D~W}\binits{D.~W.}}
(\byear{2021}).
\btitle{The mobile assistance for regulating smoking ({MARS}) micro-randomized trial design protocol}.
\bjournal{Contemporary Clinical Trials}.
\bdoi{10.1016/j.cct.2021.106513}
\end{barticle}
\endbibitem

\bibitem[\protect\citeauthoryear{Qian et~al.}{2021}]{qian_2020}
\begin{barticle}[author]
\bauthor{\bsnm{Qian},~\bfnm{T}\binits{T.}}, \bauthor{\bsnm{Yoo},~\bfnm{H}\binits{H.}}, \bauthor{\bsnm{Klasnja},~\bfnm{P}\binits{P.}}, \bauthor{\bsnm{Almirall},~\bfnm{D}\binits{D.}} \AND \bauthor{\bsnm{Murphy},~\bfnm{S~A}\binits{S.~A.}}
(\byear{2021}).
\btitle{Estimating time-varying causal excursion effect in mobile health with binary outcomes}.
\bjournal{Biometrika}
\bvolume{108}
\bpages{507-527}.
\end{barticle}
\endbibitem

\bibitem[\protect\citeauthoryear{Rabbi et~al.}{2018}]{rabbi_2018}
\begin{barticle}[author]
\bauthor{\bsnm{Rabbi},~\bfnm{M}\binits{M.}}, \bauthor{\bsnm{{Philyaw Kotov}},~\bfnm{M}\binits{M.}}, \bauthor{\bsnm{Cunningham},~\bfnm{R}\binits{R.}}, \bauthor{\bsnm{Bonar},~\bfnm{E~E}\binits{E.~E.}}, \bauthor{\bsnm{Nahum-Shani},~\bfnm{I}\binits{I.}}, \bauthor{\bsnm{Klasnja},~\bfnm{P}\binits{P.}}, \bauthor{\bsnm{Walton},~\bfnm{M}\binits{M.}} \AND \bauthor{\bsnm{Murphy},~\bfnm{S}\binits{S.}}
(\byear{2018}).
\btitle{Toward Increasing Engagement in Substance Use Data Collection: Development of the Substance Abuse Research Assistant App and Protocol for a Microrandomized Trial Using Adolescents and Emerging Adults}.
\bjournal{JMIR Research Protocols}
\bvolume{7}.
\end{barticle}
\endbibitem

\bibitem[\protect\citeauthoryear{Rizopoulos}{2023}]{rizopoulos_2023}
\begin{bmisc}[author]
\bauthor{\bsnm{Rizopoulos},~\bfnm{D}\binits{D.}}
(\byear{2023}).
\btitle{Joint Modeling of Longitudinal and Time-to-Event Data with Applications in {R}}.
\bhowpublished{\url{https://www.drizopoulos.com/courses/EMC/ESP72.pdf}}.
\bnote{Accessed: 2024-06-17}.
\end{bmisc}
\endbibitem

\bibitem[\protect\citeauthoryear{Rizopoulos et~al.}{2024}]{rizopoulos_2024}
\begin{barticle}[author]
\bauthor{\bsnm{Rizopoulos},~\bfnm{D}\binits{D.}}, \bauthor{\bsnm{Taylor},~\bfnm{J~M~G}\binits{J.~M.~G.}}, \bauthor{\bsnm{Papageorgiou},~\bfnm{G}\binits{G.}} \AND \bauthor{\bsnm{Morgan},~\bfnm{T~M}\binits{T.~M.}}
(\byear{2024}).
\btitle{Using joint models for longitudinal and time-to-event data to investigate the causal effect of salvage therapy after prostatectomy}.
\bjournal{Statistical Methods in Medical Research}
\bvolume{33}
\bpages{894-908}.
\end{barticle}
\endbibitem

\bibitem[\protect\citeauthoryear{Shi, Wu and Dempsey}{2023}]{shi_2023}
\begin{barticle}[author]
\bauthor{\bsnm{Shi},~\bfnm{J}\binits{J.}}, \bauthor{\bsnm{Wu},~\bfnm{Z}\binits{Z.}} \AND \bauthor{\bsnm{Dempsey},~\bfnm{W}\binits{W.}}
(\byear{2023}).
\btitle{Assessing time-varying causal effect moderation in the presence of cluster-level treatment effect heterogeneity and interference}.
\bjournal{Biometrika}
\bvolume{110}
\bpages{645–662}.
\end{barticle}
\endbibitem

\bibitem[\protect\citeauthoryear{Taylor et~al.}{2014}]{taylor_2014}
\begin{barticle}[author]
\bauthor{\bsnm{Taylor},~\bfnm{J~M~G}\binits{J.~M.~G.}}, \bauthor{\bsnm{Shen},~\bfnm{J}\binits{J.}}, \bauthor{\bsnm{Kennedy},~\bfnm{E~H}\binits{E.~H.}}, \bauthor{\bsnm{Wang},~\bfnm{L}\binits{L.}} \AND \bauthor{\bsnm{Schaubel},~\bfnm{D~E}\binits{D.~E.}}
(\byear{2014}).
\btitle{Comparison of methods for estimating the effect of salvage therapy in prostate cancer when treatment is given by indication}.
\bjournal{Statistics in Medicine}
\bvolume{2}
\bpages{257--74}.
\end{barticle}
\endbibitem

\bibitem[\protect\citeauthoryear{Tran et~al.}{2021}]{tran_2021b}
\begin{barticle}[author]
\bauthor{\bsnm{Tran},~\bfnm{T~D}\binits{T.~D.}}, \bauthor{\bsnm{Lesaffre},~\bfnm{E}\binits{E.}}, \bauthor{\bsnm{Verbeke},~\bfnm{G}\binits{G.}} \AND \bauthor{\bsnm{Duyck},~\bfnm{J}\binits{J.}}
(\byear{2021}).
\btitle{Latent {O}rnstein-{U}hlenbeck models for {B}ayesian analysis of multivariate longitudinal categorical responses}.
\bjournal{Biometrics}
\bvolume{77}
\bpages{689--701}.
\bdoi{10.1111/biom.13292}
\end{barticle}
\endbibitem

\bibitem[\protect\citeauthoryear{Watanabe}{2010}]{watanabe_2010}
\begin{barticle}[author]
\bauthor{\bsnm{Watanabe},~\bfnm{S}\binits{S.}}
(\byear{2010}).
\btitle{Asymptotic equivalence of {B}ayes cross validation and widely applicable information criterion in singular learning theory}.
\bjournal{Journal of Machine Learning Research}
\bvolume{11}
\bpages{3571–3594}.
\end{barticle}
\endbibitem

\bibitem[\protect\citeauthoryear{Yu et~al.}{2004}]{yu_2004}
\begin{barticle}[author]
\bauthor{\bsnm{Yu},~\bfnm{M}\binits{M.}}, \bauthor{\bsnm{Law},~\bfnm{N~J}\binits{N.~J.}}, \bauthor{\bsnm{Taylor},~\bfnm{J~M~G}\binits{J.~M.~G.}} \AND \bauthor{\bsnm{Sandler},~\bfnm{H~M}\binits{H.~M.}}
(\byear{2004}).
\btitle{Joint longitudinal-survival-cure models and their application to prostate cancer}.
\bjournal{Statistica Sinica}
\bvolume{14}
\bpages{835-862}.
\end{barticle}
\endbibitem

\bibitem[\protect\citeauthoryear{Zhang et~al.}{2017}]{zhang_2017}
\begin{barticle}[author]
\bauthor{\bsnm{Zhang},~\bfnm{D}\binits{D.}}, \bauthor{\bsnm{Chen},~\bfnm{M~H}\binits{M.~H.}}, \bauthor{\bsnm{Ibrahim},~\bfnm{J~G}\binits{J.~G.}}, \bauthor{\bsnm{Boye},~\bfnm{M~E}\binits{M.~E.}} \AND \bauthor{\bsnm{Shen},~\bfnm{W}\binits{W.}}
(\byear{2017}).
\btitle{Bayesian Model Assessment in Joint Modeling of Longitudinal and Survival Data with Applications to Cancer Clinical Trials}.
\bjournal{Journal of Computational and Graphical Statistics}
\bvolume{26}
\bpages{121–133}.
\end{barticle}
\endbibitem

\end{thebibliography}


\begin{thebibliography}{8}

\bibitem[\protect\citeauthoryear{Abbott et~al.}{2025a}]{abbott_2023}
\begin{barticle}[author]
\bauthor{\bsnm{Abbott},~\bfnm{M~R}\binits{M.~R.}}, \bauthor{\bsnm{Dempsey},~\bfnm{W~H}\binits{W.~H.}}, \bauthor{\bsnm{Nahum-Shani},~\bfnm{I}\binits{I.}}, \bauthor{\bsnm{Lam},~\bfnm{C~Y}\binits{C.~Y.}}, \bauthor{\bsnm{Wetter},~\bfnm{D~W}\binits{D.~W.}} \AND \bauthor{\bsnm{Taylor},~\bfnm{J~M~G}\binits{J.~M.~G.}}
(\byear{2025}a).
\btitle{A Continuous-Time Dynamic Factor Model for Intensive Longitudinal Data Arising from Mobile Health Studies}.
\bjournal{Psychometrika}
\bvolume{90}
\bpages{1-22}.
\end{barticle}
\endbibitem

\bibitem[\protect\citeauthoryear{Abbott et~al.}{2025b}]{abbott_2024}
\begin{barticle}[author]
\bauthor{\bsnm{Abbott},~\bfnm{M~R}\binits{M.~R.}}, \bauthor{\bsnm{Dempsey},~\bfnm{W~H}\binits{W.~H.}}, \bauthor{\bsnm{Nahum-Shani},~\bfnm{I}\binits{I.}}, \bauthor{\bsnm{Potter},~\bfnm{L~N}\binits{L.~N.}}, \bauthor{\bsnm{Wetter},~\bfnm{D~W}\binits{D.~W.}}, \bauthor{\bsnm{Lam},~\bfnm{C~Y}\binits{C.~Y.}} \AND \bauthor{\bsnm{Taylor},~\bfnm{J~M~G}\binits{J.~M.~G.}}
(\byear{2025}b).
\btitle{A {B}ayesian joint longitudinal-survival model with a latent stochastic process for intensive longitudinal data}.
\bjournal{Biometrics}
\bvolume{81}.
\end{barticle}
\endbibitem

\bibitem[\protect\citeauthoryear{Gelman, Carlin and Stern}{2014}]{gelman_2014}
\begin{binbook}[author]
\bauthor{\bsnm{Gelman},~\bfnm{A}\binits{A.}}, \bauthor{\bsnm{Carlin},~\bfnm{J~B}\binits{J.~B.}} \AND \bauthor{\bsnm{Stern},~\bfnm{H~S}\binits{H.~S.}}
(\byear{2014}).
\btitle{Bayesian Data Analysis},
\bedition{third} ed.
\bchapter{7}.
\bpublisher{Chapman \& Hall/CRC texts in statistical science}.
\end{binbook}
\endbibitem

\bibitem[\protect\citeauthoryear{Merkle, Furr and Rabe-Hesketh}{2019}]{merkle_2019}
\begin{barticle}[author]
\bauthor{\bsnm{Merkle},~\bfnm{E~C}\binits{E.~C.}}, \bauthor{\bsnm{Furr},~\bfnm{D}\binits{D.}} \AND \bauthor{\bsnm{Rabe-Hesketh},~\bfnm{S}\binits{S.}}
(\byear{2019}).
\btitle{Bayesian Comparison of Latent Variable Models: Conditional Versus Marginal Likelihoods}.
\bjournal{Psychometrika}
\bvolume{84}
\bpages{802–829}.
\end{barticle}
\endbibitem

\bibitem[\protect\citeauthoryear{Prochaska, Vogel and Benowitz}{2022}]{prochaska_2022}
\begin{barticle}[author]
\bauthor{\bsnm{Prochaska},~\bfnm{J~J}\binits{J.~J.}}, \bauthor{\bsnm{Vogel},~\bfnm{E~A}\binits{E.~A.}} \AND \bauthor{\bsnm{Benowitz},~\bfnm{N}\binits{N.}}
(\byear{2022}).
\btitle{Nicotine delivery and cigarette equivalents from vaping a JUULpod}.
\bjournal{Tobacco Control}
\bvolume{31}
\bpages{e88-e93}.
\end{barticle}
\endbibitem

\bibitem[\protect\citeauthoryear{Rizopoulos}{2023}]{rizopoulos_2023}
\begin{bmisc}[author]
\bauthor{\bsnm{Rizopoulos},~\bfnm{D}\binits{D.}}
(\byear{2023}).
\btitle{Joint Modeling of Longitudinal and Time-to-Event Data with Applications in {R}}.
\bhowpublished{\url{https://www.drizopoulos.com/courses/EMC/ESP72.pdf}}.
\bnote{Accessed: 2024-06-17}.
\end{bmisc}
\endbibitem

\bibitem[\protect\citeauthoryear{Tran et~al.}{2021}]{tran_2021b}
\begin{barticle}[author]
\bauthor{\bsnm{Tran},~\bfnm{T~D}\binits{T.~D.}}, \bauthor{\bsnm{Lesaffre},~\bfnm{E}\binits{E.}}, \bauthor{\bsnm{Verbeke},~\bfnm{G}\binits{G.}} \AND \bauthor{\bsnm{Duyck},~\bfnm{J}\binits{J.}}
(\byear{2021}).
\btitle{Latent {O}rnstein-{U}hlenbeck models for {B}ayesian analysis of multivariate longitudinal categorical responses}.
\bjournal{Biometrics}
\bvolume{77}
\bpages{689--701}.
\bdoi{10.1111/biom.13292}
\end{barticle}
\endbibitem

\bibitem[\protect\citeauthoryear{Vehtari, Gelman and Gabry}{2017}]{vehtari_2017}
\begin{barticle}[author]
\bauthor{\bsnm{Vehtari},~\bfnm{A}\binits{A.}}, \bauthor{\bsnm{Gelman},~\bfnm{A}\binits{A.}} \AND \bauthor{\bsnm{Gabry},~\bfnm{J}\binits{J.}}
(\byear{2017}).
\btitle{Practical {B}ayesian model evaluation using leave-one-out cross-validation and {WAIC}}.
\bjournal{Statistics and Computing}
\bvolume{27}
\bpages{1413-1432}.
\end{barticle}
\endbibitem

\end{thebibliography}


\end{document}


\begin{frontmatter}
\title{Supplementary Material for: \\ Estimation of Time-Varying Treatment Effects in a Joint Model for Longitudinal and Recurrent Event Outcomes in Mobile Health Data}
\runtitle{Time-Varying Treatment Effects in a Joint Model}

\begin{aug}
\author[A]{\fnms{Madeline R}~\snm{Abbott}\ead[label=e1]{mrabbott@umich.edu}\orcid{0000-0002-5344-3732}},
\author[A]{\fnms{Jeremy M G}~\snm{Taylor}\ead[label=e2]{jmgt@umich.edu}}
\author[B]{\fnms{Inbal}~\snm{Nahum-Shani}\ead[label=e3]{inbal@umich.edu}}
\author[C]{\fnms{Lindsey N}~\snm{Potter}\ead[label=e4]{lindsey.potter@hci.utah.edu}}
\author[C]{\fnms{David W}~\snm{Wetter}\ead[label=e5]{david.wetter@hci.utah.edu}}
\author[C]{\fnms{Cho Y}~\snm{Lam}\ead[label=e6]{cho.lam@hci.utah.edu}}
\and
\author[A]{\fnms{Walter}~\snm{Dempsey}\ead[label=e7]{wdem@umich.edu}}

\address[A]{Department of Biostatistics, University of Michigan\printead[presep={,\ }]{e1,e2,e7}}

\address[B]{Institute for Social Research, University of Michigan\printead[presep={,\ }]{e3}}

\address[C]{Department of Population Health Sciences and Huntsman Cancer Institute, University of Utah\printead[presep={,\ }]{e4,e5,e6}}
\end{aug}

\end{frontmatter}


\doublespacing

\renewcommand{\figurename}{Web Figure}
\renewcommand{\tablename}{Web Table}
\renewcommand{\thesection}{Web Appendix \Alph{section}}

\section{Case Study}

\subsection{Defining the Recurrent Event Outcomes}\label{a:ouf_jm_tx:data_wrangling}

Poly-substance use is defined as engaging in any of the following activities: using marijuana, vaping, or smoking cigarettes.  At each ecological momentary assessment (EMA), participants are asked to respond to a set of questions in which they report if they used any of these substances since the last EMA, and if so, the approximate time of use.  More details on the questions used to assess each type of substance use, and our approach to converting these responses into event times suitable for modeling, are given below.  Note that these rules are defined specifically for the purpose of curating an event-time outcome appropriate for modeling, and are not intended to be used to draw general conclusions about comparable instances of substance use (for example, cigarette vs. vaping equivalence).

\vspace{0.5cm}

\paragraph*{Marijuana Use}

If participants report any marijuana use since the prior EMA, they are also asked to provide a single time corresponding to time of use.  Each reported time of marijuana use is considered a recurrent event.

\vspace{0.5cm}

\paragraph*{Cigarette Smoking}

If participants report smoking any cigarettes since the prior EMA, then they are prompted to respond to additional questions about how many cigarettes they smoked and approximately when they smoked the cigarettes.

If a participant reports smoking a partial or a single cigarette since the last EMA, then they are also asked to provide a single time corresponding to when they smoked.  We consider this time as a recurrent event time.

If a participant reports smoking more than one cigarette since the last EMA, then they are asked to respond to two additional questions.  These questions ask (i) when they smoked the first cigarette after the last EMA and (ii) when they smoked their most recent cigarette.  These questions result in an interval of time over which a participant has smoked a known number of cigarettes.  To convert this interval into a recurrent event time appropriate for modeling, we evenly distribute the reported number of cigarettes across the interval and then consider each time at which a cigarette is assumed to be smoked as a recurrent event time.  For example, if two cigarettes were smoked over an interval from A to B, then we would place one event at time A and one event at time B.  If three cigarettes were smoked over an interval from A to B, then we would place one event at time A, one event at time B, and one event halfway between times A and B.  If more than 10 cigarettes were reported smoked, then participants do not report the exact number but instead select the option of ``more than 10 cigarettes'' when filling out the EMA.  For cases when ``more than 10 cigarettes'' was selected, we assume that the cigarettes were smoked (and recurrent events occurred) at a rate of approximately 1 event per hour across the reported interval of time.

In some cases, an individual might smoke multiple cigarettes in a row, which could be viewed as a single episode of smoking, rather than multiple separate episodes.  To account for cases like these, we define an additional rule: if multiple cigarettes were reported smoked over an interval of less than an hour, then we consolidate these events into a single event and use the midpoint of the interval of time as the corresponding recurrent event time.

\vspace{0.5cm}

\paragraph*{Vaping}

Vaping is assessed in a similar way to cigarette smoking; however, vaping is reported in units of ``puffs''.  Before converting the responses to the vaping-related questions into recurrent event times, we first consider how many puffs constitute a single event.  One puff delivers much less nicotine than a single cigarette, and so a single puff is not equivalent to a single cigarette.  The nicotine contained in one e-cigarette pod is approximately equivalent to the nicotine contained in a pack of cigarettes (20 cigarettes); it also takes approximately 200 puffs to use up a pod \citep{prochaska_2022}.  Using these two facts, we assume that on average 10 puffs are equivalent to smoking one cigarette, and we define a single event of vaping as 0-15 puffs. It follows that 16-25 puffs are 2 events, 26-35 puffs are 3 events, and so on.  The conversion of puffs to events is summarized in Table \ref{a:tab:puff2event}.  After converting puffs to events, we next apply some rules to convert the information reported in the EMAs into recurrent event times.  

\begin{table}
\centering
\includegraphics[width=6cm]{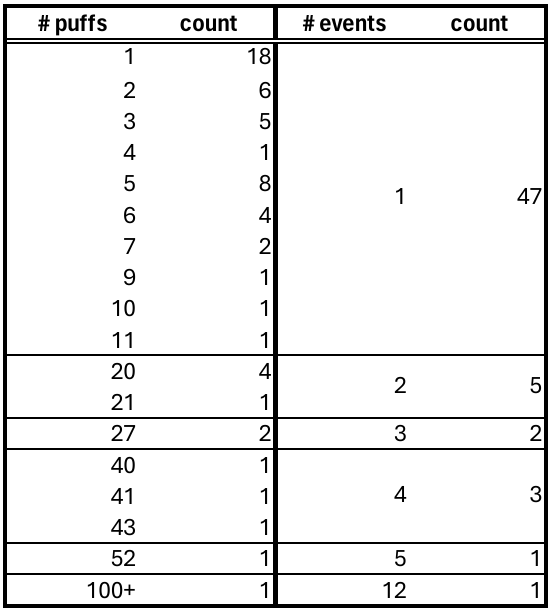}
\caption[Number of times that each number of puffs was observed and conversion of puffs to events.]{Number of times that each number of puffs was observed and conversion of puffs to events.}\label{a:tab:puff2event}
\end{table}

Vaping is assessed using questions that have the same structure as those used to assess cigarette smoking since the last EMA.  Participants are asked to report the total number of puffs since the prior EMA, the approximate time of the puff if only a single puff was taken, and the approximate time of the first and most recent puff if multiple puffs were reported.  After converting puffs to events, we apply rules similar to those used when converting cigarettes to event times.  When more than 1 but fewer than 16 puffs are reported, we convert these puffs into a single event; the approximate time of the event corresponds to the midpoint of the reported interval over which the puffs were taken.

\vspace{0.5cm}

\begin{figure}
\centering
\includegraphics[width=10cm]{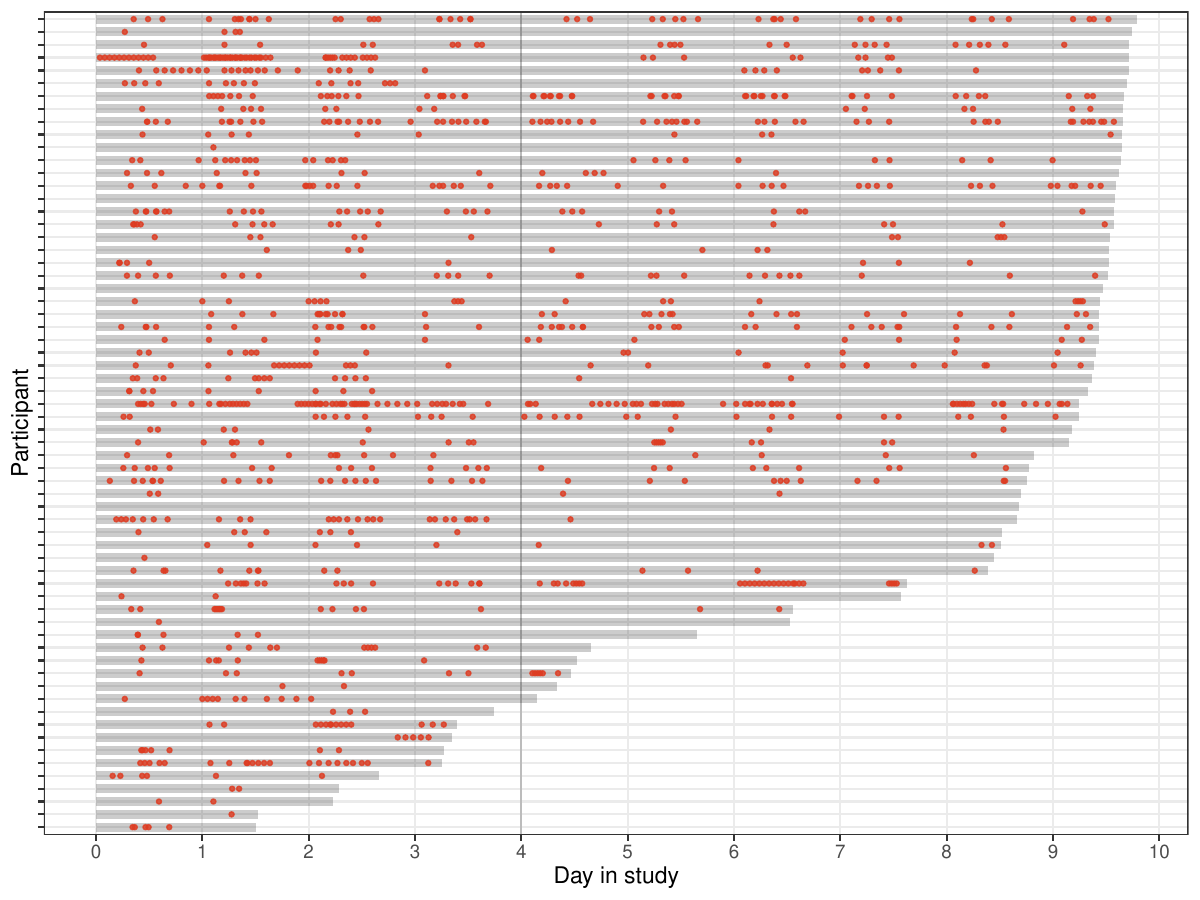}
\caption[Timing of recurrent poly-substance use events.]{Timing of recurrent poly-substance use events.  Grey bars indicate time periods when individuals are at risk of a recurrent event.  The vertical grey line at day 4 indicates the end of the quit day, which we use as the transition from the pre-quit to the post-quit period.}\label{a:fig:event_timeline}
\end{figure}

In the illustrative analysis in the main paper, we combine all instances of marijuana use, vaping, and cigarette smoking into a single recurrent event outcome called poly-substance use.  We do not distinguish between use of specific substances in our analysis.  In Web Figure \ref{a:fig:event_timeline}, we plot the timing of the recurrent poly-substance use events for each individual in the study.

\begin{figure}
\centering
\includegraphics[width=10cm]{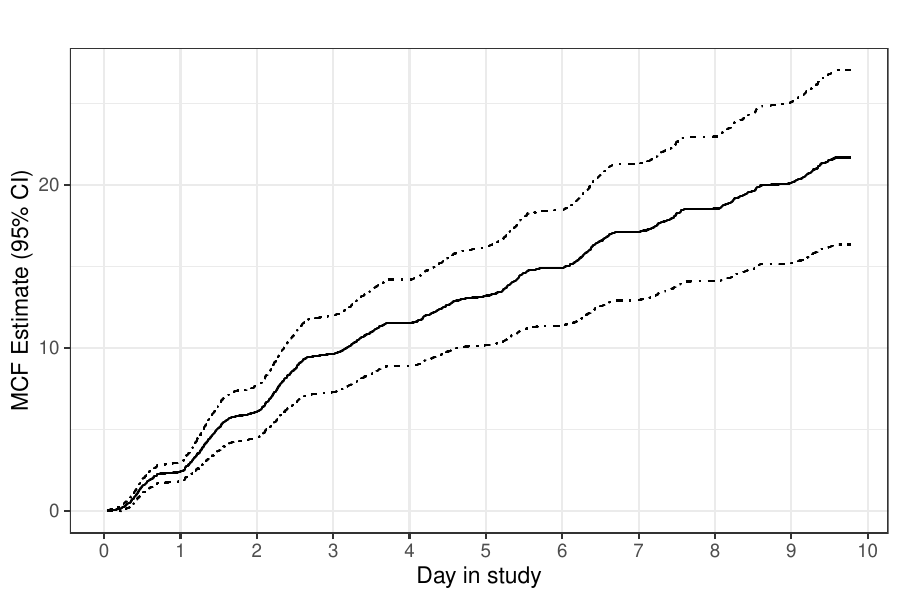}
\caption[Mean cumulative function for recurrent poly-substance use.]{Mean cumulative function for recurrent poly-substance use.}\label{a:fig:polysub_mcf}
\end{figure}

\begin{figure}
\centering
\includegraphics[width=10cm]{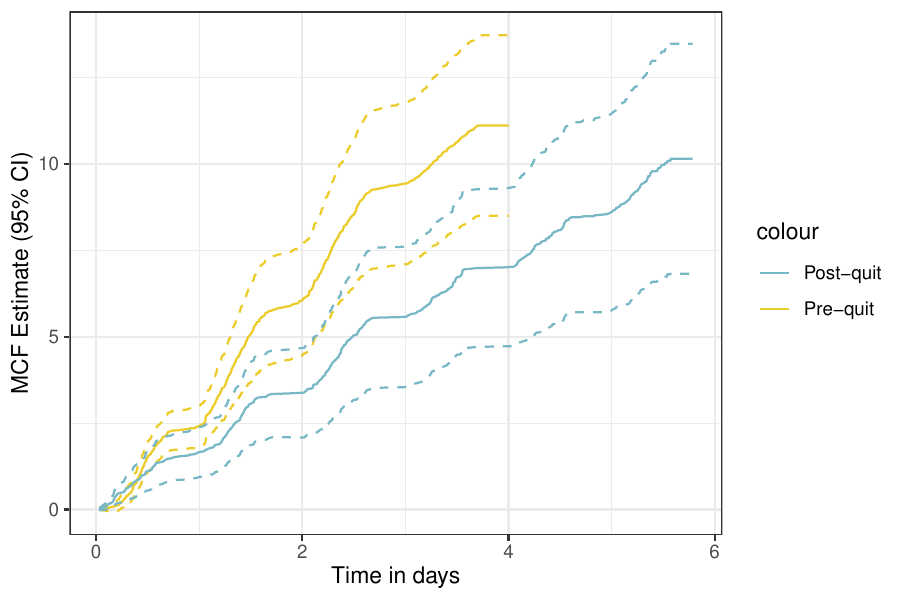}
\caption[Mean cumulative function for recurrent poly-substance use by pre- and post-quit periods.]{Mean cumulative function for recurrent poly-substance use by pre- and post-quit periods. For the pre-quit period, time 0 corresponds to time since the start of the study. For the post-quit period, time 0 corresponds to time since the end of the designated quit day (day 4).}\label{a:fig:polysub_mcf_preQ_vs_postQ}
\end{figure}

In Web Figure \ref{a:fig:polysub_mcf}, we plot the mean cumulative function (MCF) estimate across days in the study.  This plot tells use the expected cumulative number of events per person by each day in the study.  We can also look at the expected cumulative number of events per person  for the pre- and post-quit periods separately, as shown in Web Figure \ref{a:fig:polysub_mcf_preQ_vs_postQ}.

\subsection{Observed Data from Case Study}

\begin{figure}
\centering
\includegraphics[width=14cm]{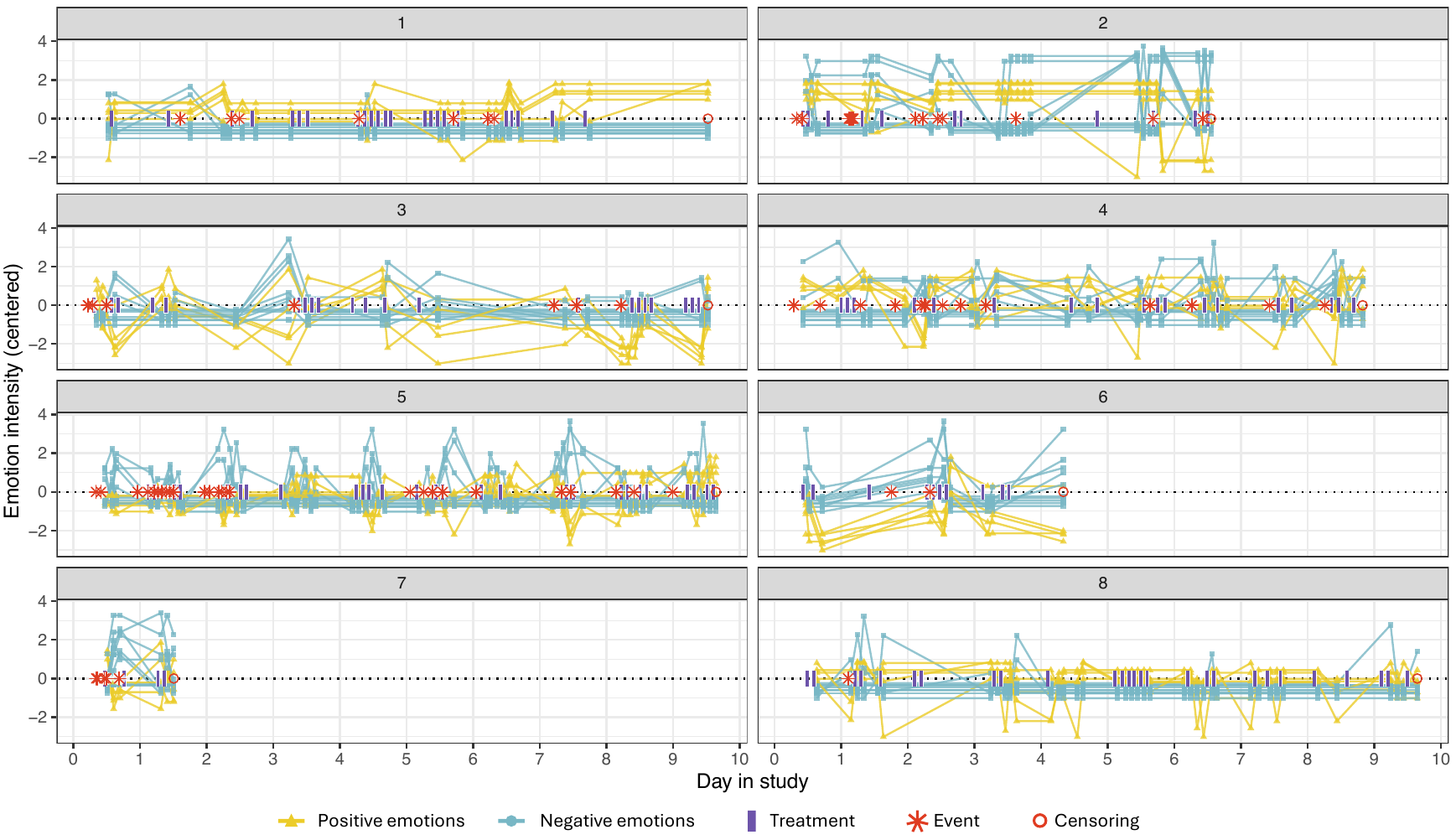}
\caption[Responses to emotion-related questions, timing of recurrent poly-substance use events, and timing of treatments for eight different participants in the motivating MRT.]{Responses to emotion-related questions, timing of recurrent poly-substance use events, and timing of treatments for eight different participants in the motivating MRT.}\label{fig:affsci_data}
\end{figure}

In Web Figure \ref{fig:affsci_data}, we plot data observed for eight different participants in the motivating micro-randomized trial (MRT).  These data consist of responses to the emotion-related questions (i.e., the measured longitudinal outcome), the timing of recurrent instances of poly-substance use (i.e., the recurrent event outcome), and the timing of app-based interventions (i.e., the repeated randomized treatment).

\subsection{Specifying the Hazard Model}\label{a:ouf_jm_tx:case_study_haz}

In the analysis of the motivating MRT data, our hazard model is

$$h_{ir}(t) = h_0(t) \exp\left[ \beta_1 \eta_{1i}(t) + \beta_2 \eta_{2i}(t) + \Tilde{\mu}_i(t) \right]$$

\noindent The baseline hazard is assumed to be log-normal, with separate parameters for the pre-quit vs. post-quit period.  Participants are instructed to quit on day 4 and so we use the end of day 4 as the transition between the pre-quit and post-quit periods.  To allow for period-specific parameters, we specify the baseline hazard as

\begin{align*}
    h_0(t) &= \text{I}(t < 4 \text{ days}) \times\frac{ \phi \left(log(t') | \mu_{0,pre}, \sigma_{0,pre} \right) }{t' \ \Phi \left(  \left(\mu_{0,pre} - log(t') \right) / \sigma_{0,pre} \right)} \\ &+ \text{I}(t \ge 4 \text{ days}) \times\frac{ \phi \left(log(t') | \mu_{0,post}, \sigma_{0,post} \right)  }{t' \ \Phi \left(  \left(\mu_{0,post} - log(t') \right) / \sigma_{0,post} \right)} 
\end{align*}

\noindent where $\phi$ is a normal PDF and $\Phi$ is the standard normal CDF.  Because we are using the clock-reset approach, time $t'$ corresponds to time since the most recent event.  $t$ is still time since the start of the study.

In one version of the hazard model, we assume that the treatment effect parameter in the treatment model for the hazard is constant throughout the study (i.e., we have one treatment parameter in the hazard model, $\Tilde{\tau}$.)  In another version of the hazard model, we allow the treatment model in the hazard to differ across the pre- and post-quit periods by re-defining the treatment function $\Tilde{\mu}_i(t)$ as
\begin{align*}
    \Tilde{\mu}_i(t) =& \sum_{t_{ia} \in \mathcal{A}_i(t)} \Tilde{\tau}_{\text{pre}} \left(1 - \frac{t - t_{ia}}{\delta_a} \right)_+ \times \text{I}(\text{study day } < 4 \text{ days}) \\& + \sum_{t_{ia} \in \mathcal{A}_i(t)} \Tilde{\tau}_{\text{post}} \left(1 - \frac{t - t_{ia}}{\delta_a} \right)_+ \times \text{I}(\text{study day } \ge 4 \text{ days})
\end{align*}

\noindent In $\Tilde{\mu}_i(t)$, time $t$ is the time since the start of the study.

\subsection{Additional Results from Fitting the Joint Longitudinal Recurrent Event Model}\label{a:ouf_jm_tx:more_case_study_results}

\begin{figure}
\centering
    \includegraphics[width=14cm]{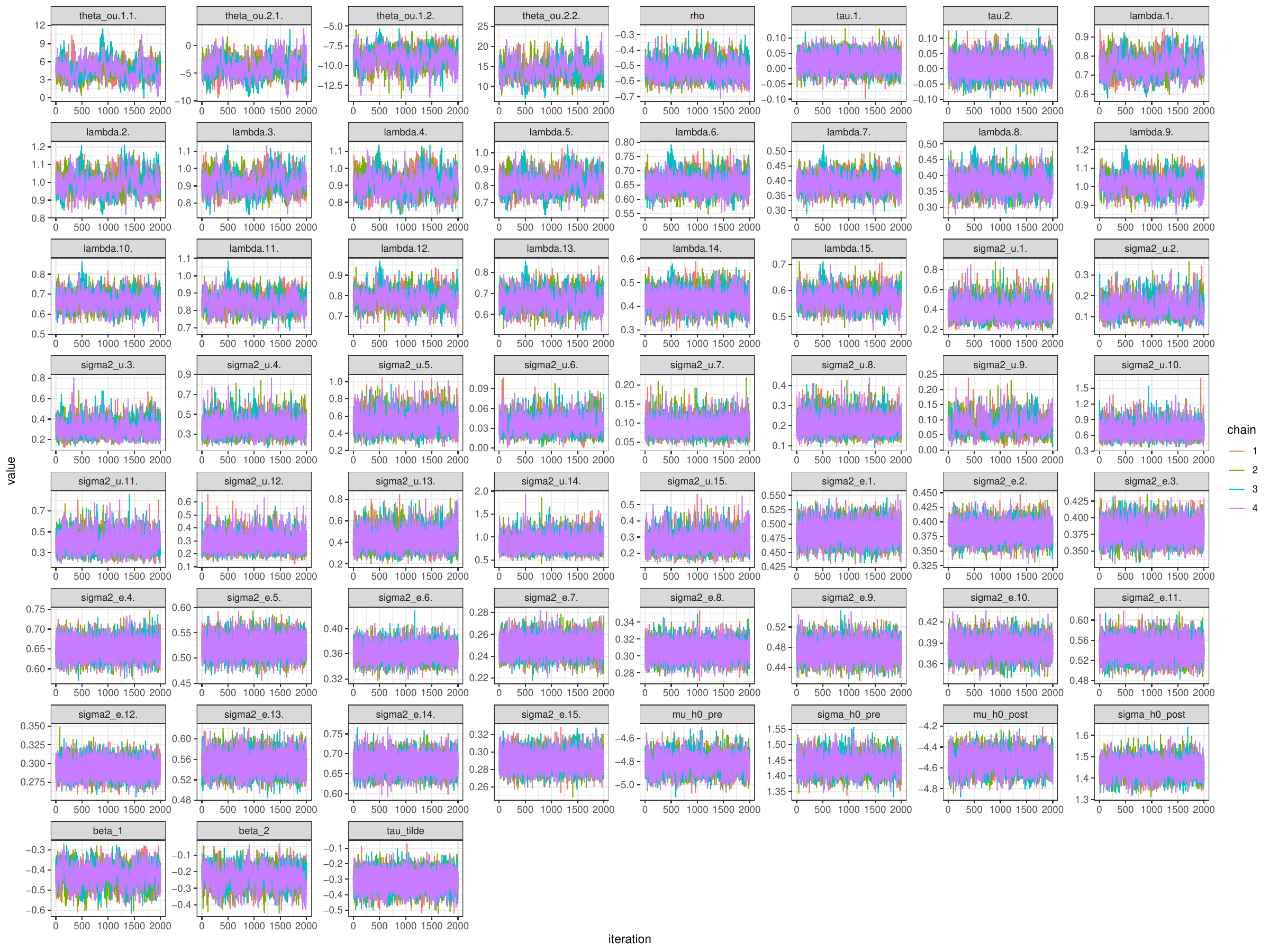}
    \caption[Trace plot for the joint model with an additive model for treatment effect in the longitudinal submodel and with a single treatment parameter in the hazard model.]{Trace plot for the joint model with an \textbf{additive} model for treatment effect in the longitudinal submodel and with a \textbf{single} treatment parameter in the hazard model.}\label{a:fig:trace_tx_add_hazB5}
\end{figure}

\begin{figure}
\centering
    \includegraphics[width=14cm]{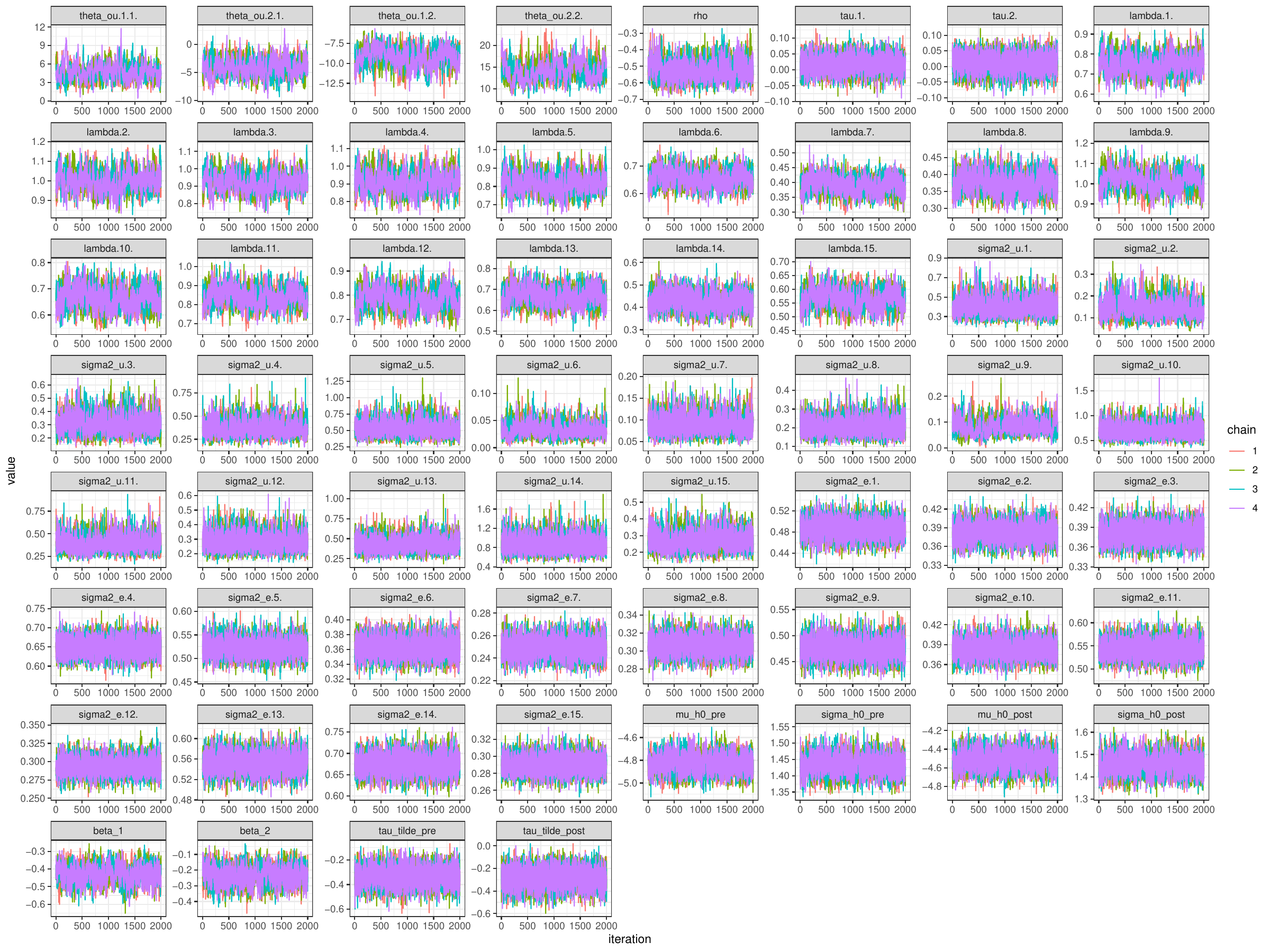}
    \caption[Trace plot for the joint model with an additive model for treatment effect in the longitudinal submodel and with separate pre-quit and post-quit treatment parameters in the hazard model.]{Trace plot for the joint model with an \textbf{additive} model for treatment effect in the longitudinal submodel and with \textbf{separate} pre-quit and post-quit treatment parameters in the hazard model.}\label{a:fig:trace_tx_add_haz42}
\end{figure}

\begin{figure}
\centering
    \includegraphics[width=14cm]{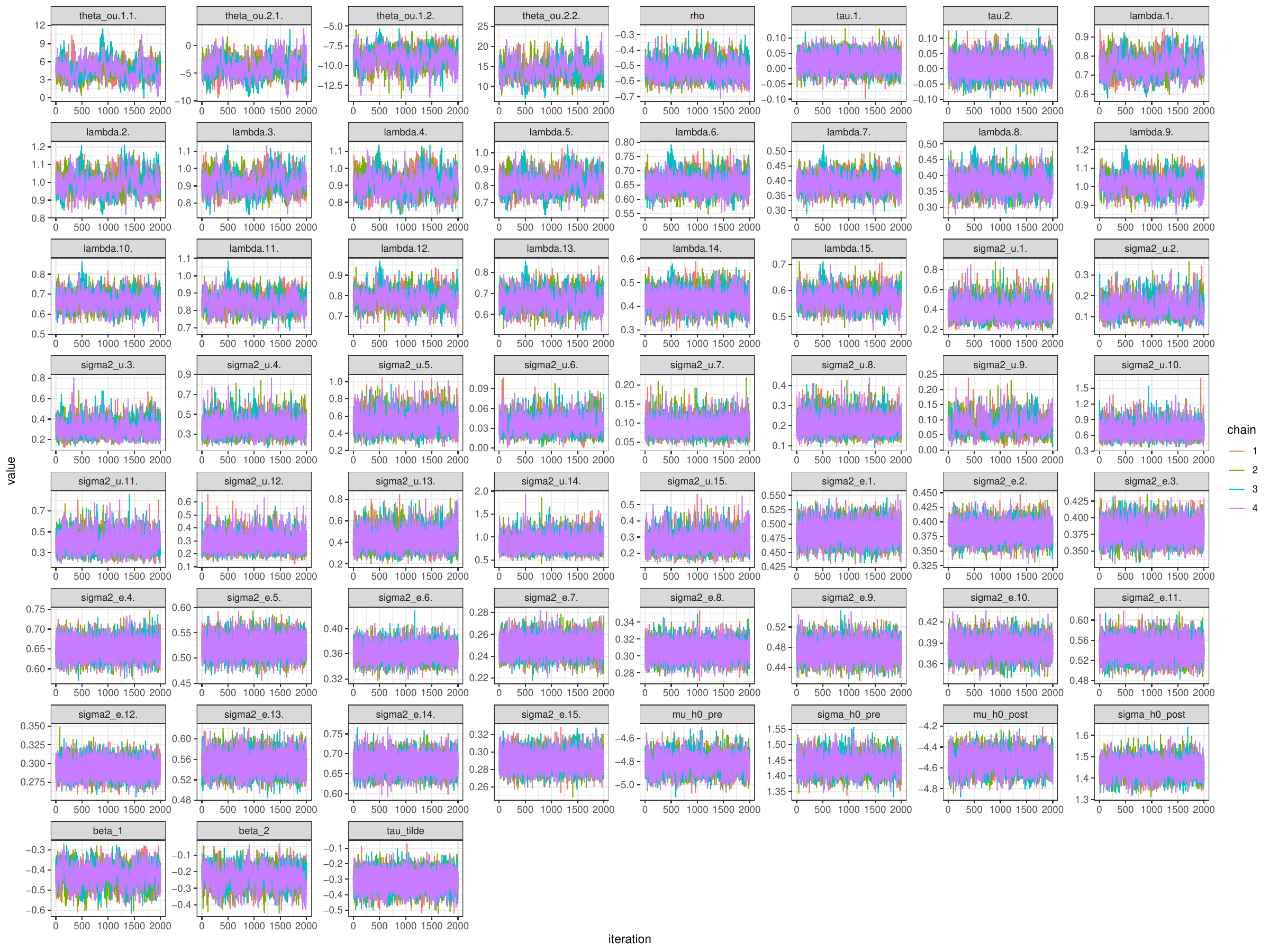}
    \caption[Trace plot for the joint model with a drift model for treatment effect in the longitudinal submodel and with a single treatment parameter in the hazard model.]{Trace plot for the joint model with a \textbf{drift} model for treatment effect in the longitudinal submodel and with a \textbf{single} treatment parameter in the hazard model.}\label{a:fig:trace_tx_drift_hazB5}
\end{figure}

\begin{figure}
\centering
    \includegraphics[width=14cm]{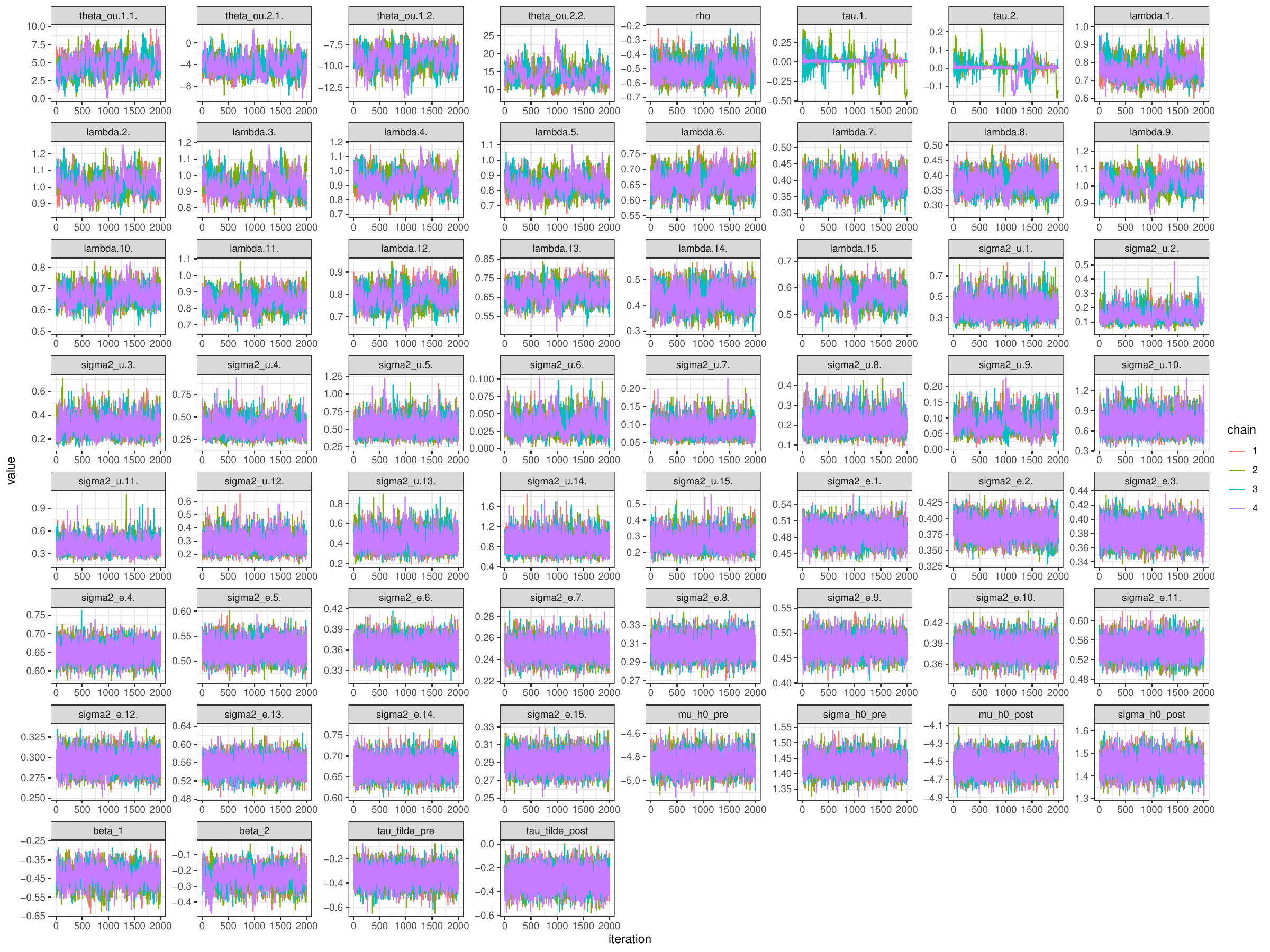}
    \caption[Trace plot for the joint model with a drift model for treatment effect in the longitudinal submodel and with separate pre-quit and post-quit treatment parameters in the hazard model.]{Trace plot for the joint model with a \textbf{drift} model for treatment effect in the longitudinal submodel and with \textbf{separate} pre-quit and post-quit treatment parameters in the hazard model.}\label{a:fig:trace_tx_drift_hazB4}
\end{figure}

We fit four different versions of the joint model, each of which corresponds to a different combination of treatment effect model in the longitudinal submodel (i.e., the additive or drift version of the treatment effect model) and hazard model (i.e., a hazard model that assumes the effect of treatment on the hazard is the same across all 10 days of the study or a hazard model that allows the effect of the treatment on the hazard to differ across the pre- vs. post-quit period.  Trace plots for each model are provided in Web Figures \ref{a:fig:trace_tx_add_hazB5}--\ref{a:fig:trace_tx_drift_hazB4}.

\subsubsection{Visualizing Decay in Correlation Over Time for the Latent Process}

\begin{figure}
 \begin{subfigure}[t]{7cm}
     \includegraphics[width=7cm]{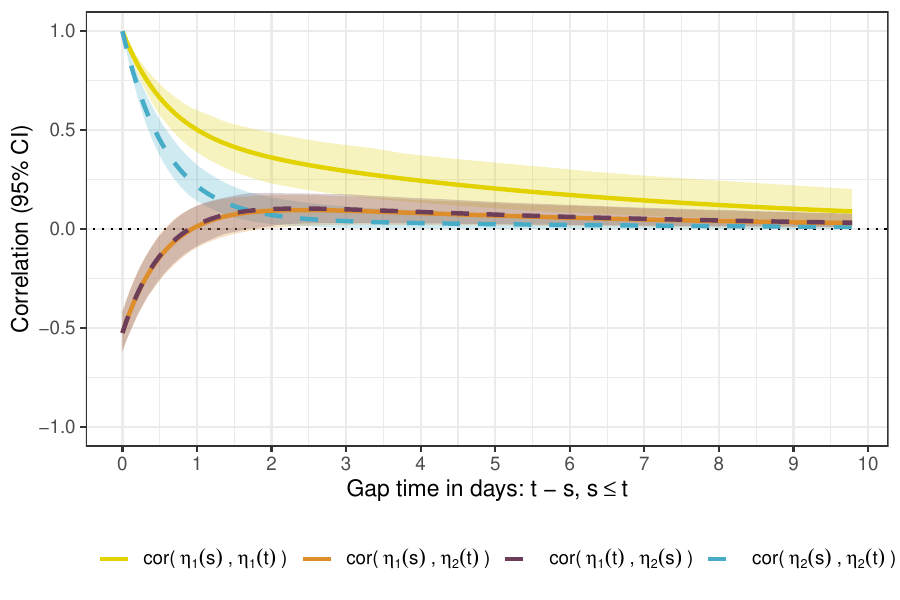}
     \caption{Model assumes an \textbf{additive} treatment effect on the latent process and a \textbf{single} treatment parameter in the hazard submodel.}
     \label{fig:a:xcorr_tx_add_hazB}
 \end{subfigure}
 \hfill
 \begin{subfigure}[t]{7cm}
     \includegraphics[width=7cm]{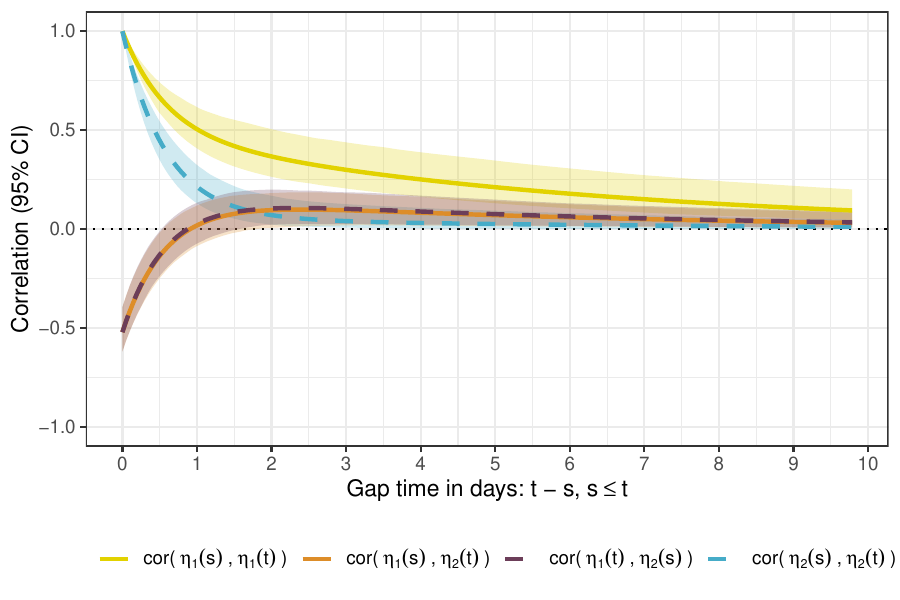}
     \caption{Model assumes an \textbf{additive} treatment effect on the latent process and \textbf{separate} pre- and post-quit treatment parameters in the hazard submodel.}
     \label{fig:a:xcorr_tx_add_hazB2}
 \end{subfigure}
 
 \medskip
 \begin{subfigure}[t]{7cm}
     \includegraphics[width=7cm]{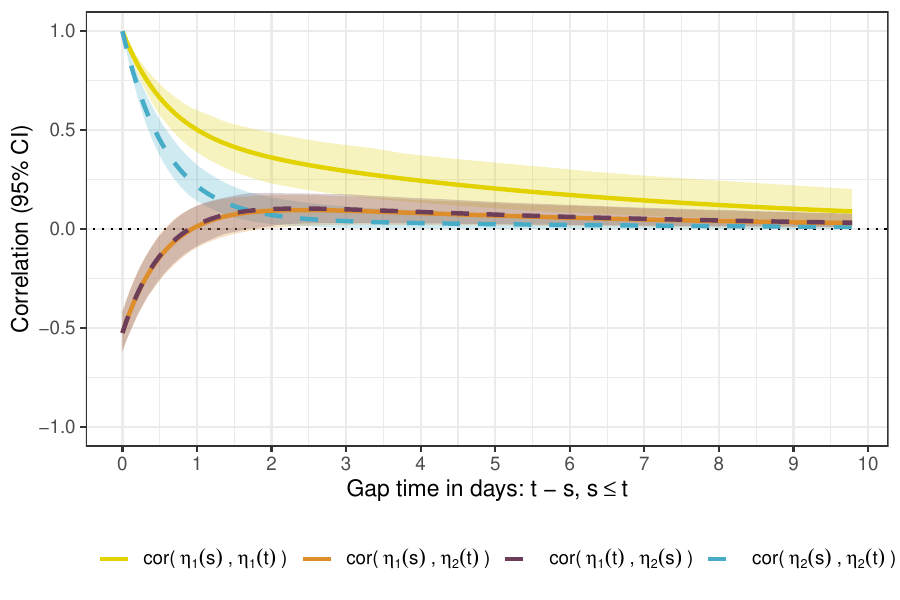}
     \caption{Model assumes a \textbf{drift} treatment effect on the latent process and a \textbf{single} treatment parameter in the hazard submodel.}
     \label{fig:a:xcorr_tx_drift_hazB}
 \end{subfigure}
 \hfill
 \begin{subfigure}[t]{7cm}
     \includegraphics[width=7cm]{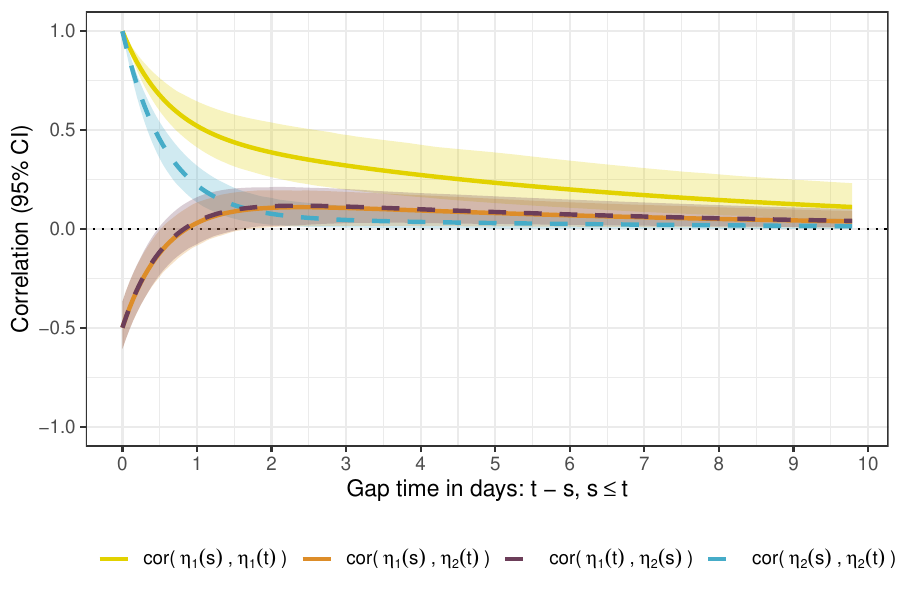}
     \caption{Model assumes a \textbf{drift} treatment effect on the latent process and \textbf{separate} pre- and post-quit treatment parameters in the hazard submodel.}
     \label{fig:a:xcorr_tx_drift_hazB2}
 \end{subfigure}
 \caption[Decay in cross- and auto-correlation in bivariate OU process from fitted joint models.]{Decay in cross- and auto-correlation in bivariate OU process from fitted joint models. For the different joint models, each plot shows the estimated decay in autocorrelation and cross-correlation between latent factors that represent positive affect ($\eta_1(t)$) and negative affect ($\eta_2(t)$) across increasing gap times, where time is on the scale of days.}\label{a:fig:xcorr}
\end{figure}

We can plot the decay in auto- and cross-correlation for the bivariate Ornstein-Uhlenbeck (OU) process estimated from the joint models.  These plots, shown in Web Figure \ref{a:fig:xcorr}, are created using the posterior means of $\bm{\theta}$ and $\rho$.

\subsubsection{Visualizing Estimates from the Treatment Effect Models}

The treatment-related parameters $\bm{\tau}$ in the longitudinal submodel have different interpretations when modeled as an additive effect vs. as drift and so to compare the estimates of $\bm{\tau}$ between models that assume different impacts of treatment, we can plot the terms that are added to the conditional mean of the OU process as a result of the treatment models.  That is, when treatment is modeled as an additive effect, the conditional expectation of the latent process is: $$\mathbb{E}\left[ \bm{\eta}_i(t)|\bm{\eta}_i(s) \right] = e^{-\bm{\theta} (t-s)}\bm{\eta}_i(s) + \bm{\mu}_i(t) - e^{-\bm{\theta} (t-s)} \bm{\mu}_i(s).$$  When treatment is modeled in the drift term, the conditional expectation of the latent process is:

\begin{align*}
   \mathbb{E}\left[ \bm{\eta}_i(t)|\bm{\eta}_i(s) \right] =& e^{-\bm{\theta} (t - s)} \bm{\eta}_i(s) + \\ & \sum_{t_{ia} \in \mathcal{A}_i(s - \delta_a, t)} \left[ \left( 1 - \frac{u - t_{ia}}{\delta_a} \right) e^{-\bm{\theta} (t - u)} \bm{\theta}^{-1} + \frac{1}{\delta_a} e^{-\bm{\theta} (t - u)} \right] \bm{\tau}\Bigg\rvert_{u = max(t_{ia}, s)}^{u = min(t, t_{ia} + \delta_a)}. 
\end{align*}

\begin{figure}
 \begin{subfigure}[t]{7cm}
     \includegraphics[width=7cm]{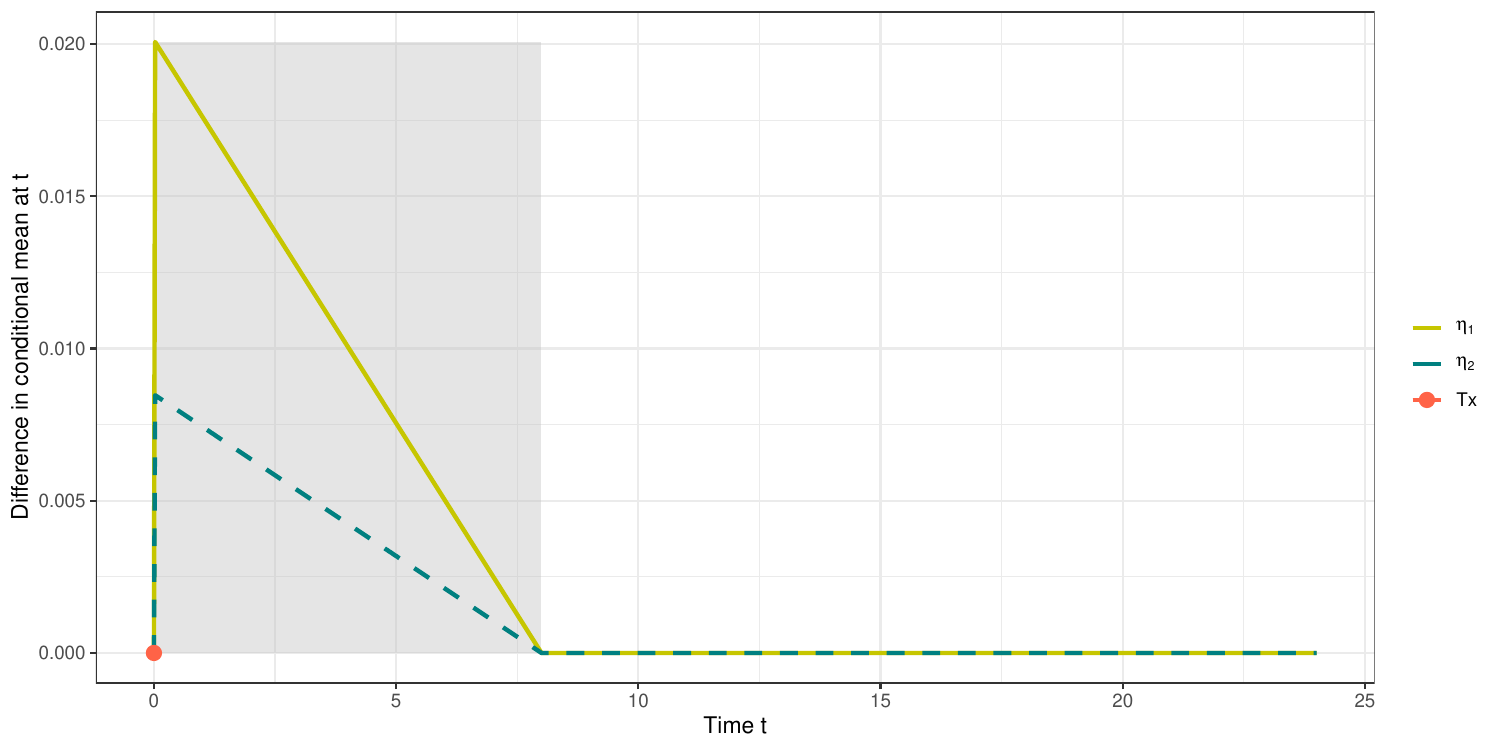}
     \caption{Model assumes an \textbf{additive} treatment effect on the latent process and a \textbf{single} treatment parameter in the hazard submodel.}
     \label{fig:a:tx_terms_tx_add_hazB}
 \end{subfigure}
 \hfill
 \begin{subfigure}[t]{7cm}
     \includegraphics[width=7cm]{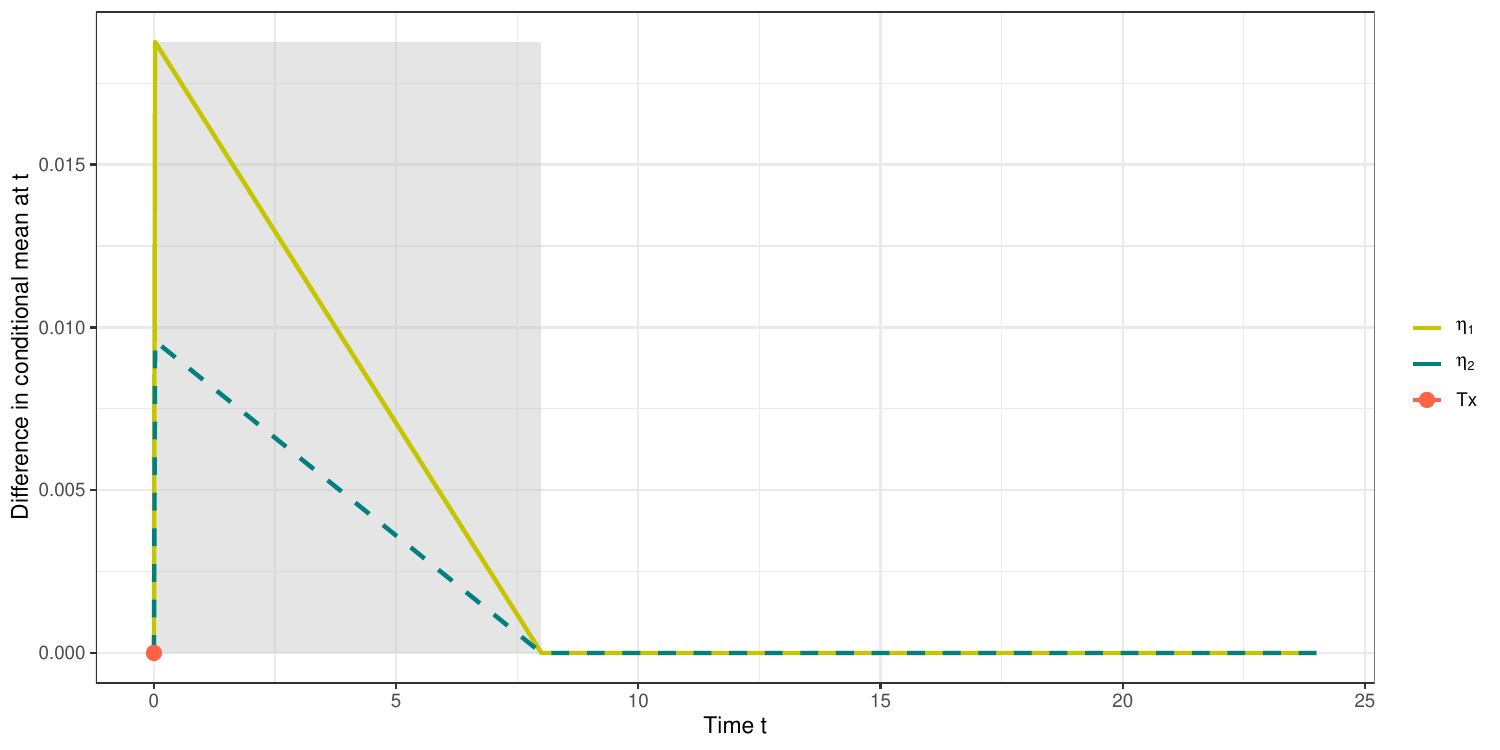}
     \caption{Model assumes an \textbf{additive} treatment effect on the latent process and \textbf{separate} pre- and post-quit treatment parameters in the hazard submodel.}
     \label{fig:a:tx_terms_tx_add_hazB2}
 \end{subfigure}
 
 \medskip
 \begin{subfigure}[t]{7cm}
     \includegraphics[width=7cm]{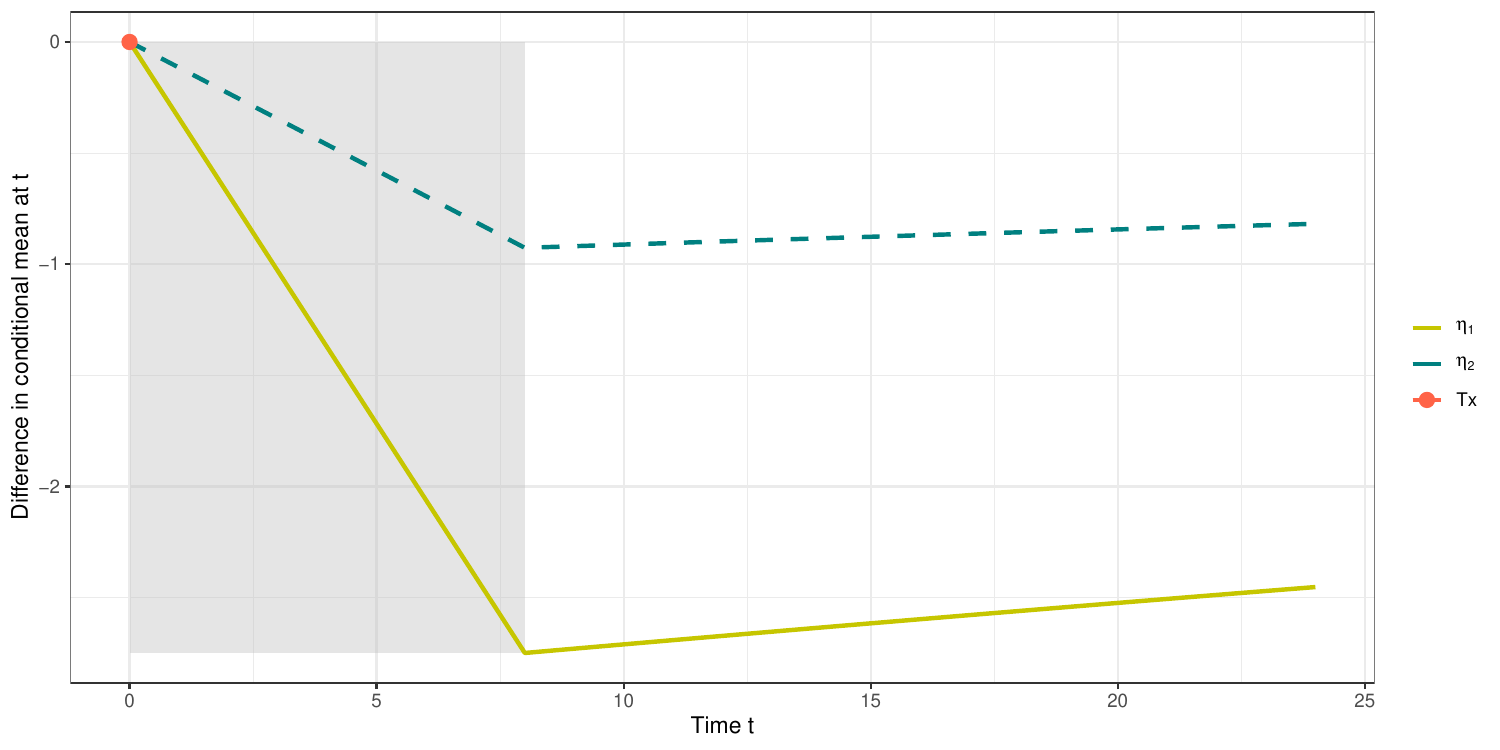}
     \caption{Model assumes a \textbf{drift} treatment effect on the latent process and a \textbf{single} treatment parameter in the hazard submodel.}
     \label{fig:a:tx_terms_xcorr_tx_drift_hazB}
 \end{subfigure}
 \hfill
 \begin{subfigure}[t]{7cm}
     \includegraphics[width=7cm]{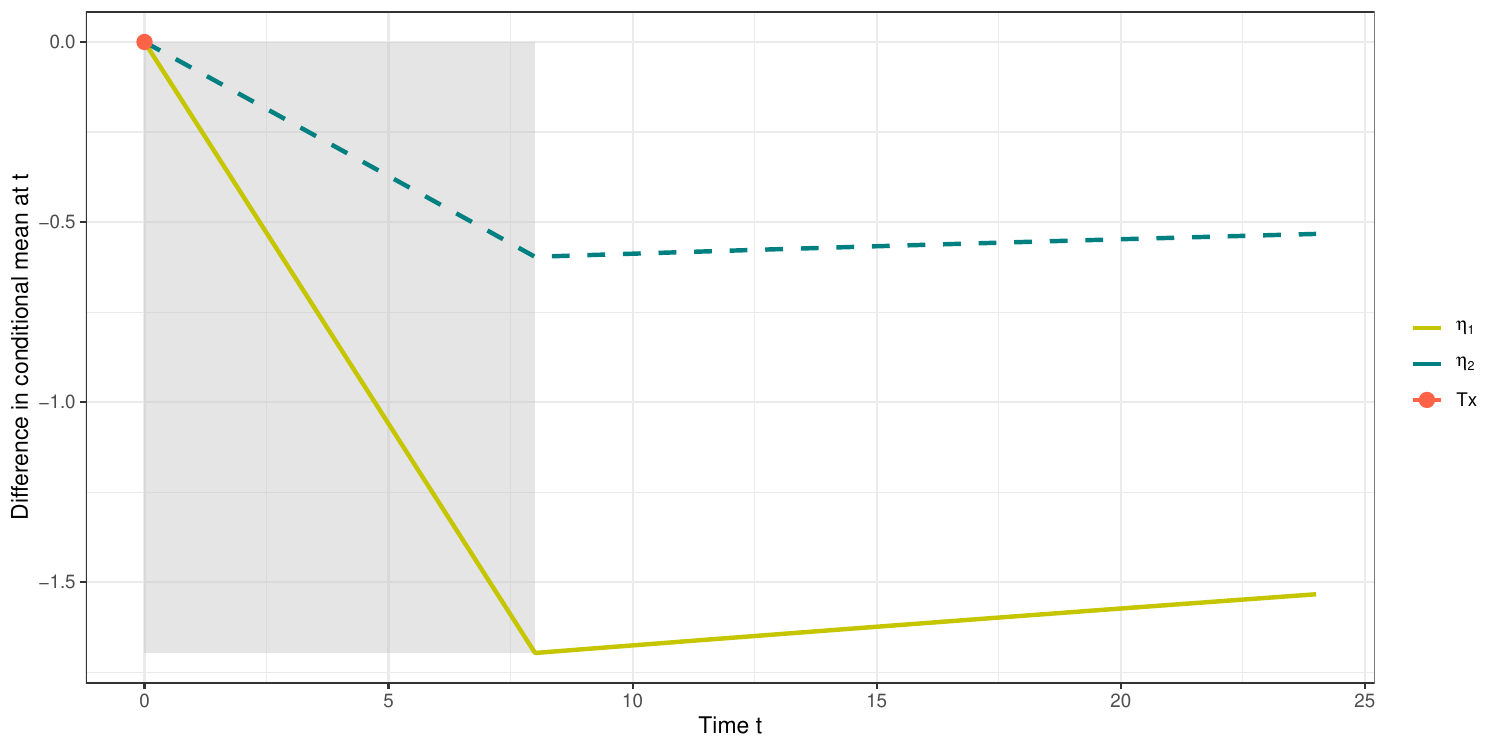}
     \caption{Model assumes a \textbf{drift} treatment effect on the latent process and \textbf{separate} pre- and post-quit treatment parameters in the hazard submodel.}
     \label{fig:a:tx_terms_tx_drift_hazB2}
 \end{subfigure}
 \caption[Estimated treatment-related terms from fitted joint models.]{Estimated treatment-related terms from fitted joint models.  For each fitted model, the plots show the estimated value of the difference between the fitted latent process and an OU process with a constant mean of 0, if a single treatment were to be sent at time 0.  If the treatment effect is modeled via a time-varying drift term, then this difference is $\int_0^t e^{-\bm{\theta}(t - u)} \bm{\mu}_i(u) du$.  If the treatment effect is modeled via an additive shift, then this term is $\bm{\mu}_i(t) - e^{-\bm{\theta} t} \bm{\mu}_i(0)$}\label{fig:a:tx_terms}
\end{figure}

To visualize the estimated effect of treatment on the latent process, we plot $$\bm{\mu}_i(t) - e^{-\bm{\theta} (t-s)} \bm{\mu}_i(s)$$ and $$\sum_{t_{ia} \in \mathcal{A}_i(s - \delta_a, t)} \left[ \left( 1 - \frac{u - t_{ia}}{\delta_a} \right) e^{-\bm{\theta} (t - u)} \bm{\theta}^{-1} + \frac{1}{\delta_a} e^{-\bm{\theta} (t - u)} \right] \bm{\tau}\Bigg\rvert_{u = max(t_{ia}, s)}^{u = min(t, t_{ia} + \delta_a)},$$ which correspond to the terms added to the conditional mean of the OU process as a result of our additive and drift treatment effects, respectively.  When plotting these terms, we assume that a single treatment was delivered at time $s = 0$.  Plots of these treatment-related terms are given in Web Figure \ref{fig:a:tx_terms}.

\subsubsection{Variability in the Longitudinal Outcomes}\label{a:ss:var_in_emotions}

The posterior estimates of parameters from the longitudinal submodel show a fair amount of variability in the random intercept variance and measurement error variance components.  We can compare these estimates to the variability that we see in the observed longitudinal outcomes.  Web Table \ref{a:tab:emotion_var} summarizes the average within-individual variance in the reported emotions and the variance of the within-individual means of reported emotions.  The rows are sorted by the rightmost column, corresponding to variance of the within-individual means.

\begin{table}
\centering
\begin{tabular}{r|l|r|r}
\hline
Model index & Emotion & Mean of ind.-specific variances & Var. of ind.-specific means\\
\hline
6 & angry & 0.537 & 0.258\\
\hline
7 & ashamed & 0.290 & 0.287\\
\hline
8 & guilty & 0.396 & 0.439\\
\hline
15 & hopeless & 0.436 & 0.542\\
\hline
9 & irritable & 0.843 & 0.718\\
\hline
2 & happy & 0.646 & 0.796\\
\hline
1 & grateful & 0.770 & 0.801\\
\hline
12 & sad & 0.509 & 0.853\\
\hline
13 & restless & 0.749 & 0.989\\
\hline
14 & bored & 0.798 & 1.002\\
\hline
11 & anxious & 0.769 & 1.090\\
\hline
3 & proud & 0.622 & 1.096\\
\hline
10 & lonely & 0.536 & 1.105\\
\hline
4 & relaxed & 0.859 & 1.151\\
\hline
5 & enthusiastic & 0.729 & 1.215\\
\hline
\end{tabular}
\caption[Empirical variability in measured longitudinal outcomes in the motivating MRT.]{Empirical variability in measured longitudinal outcomes in the motivating MRT.}\label{a:tab:emotion_var}
\end{table}

\section{Different Pathways for Treatment Effect}

\begin{figure}
\centering
\includegraphics[width=6cm]{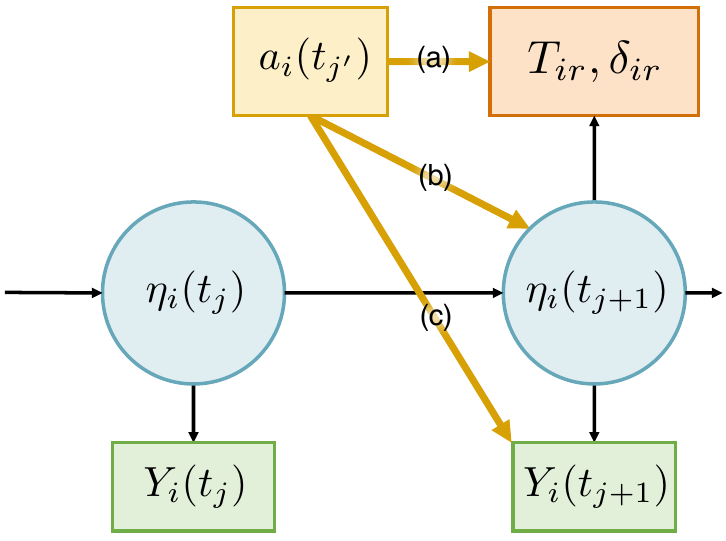}
\caption[Diagram illustrating potential mechanisms (a)--(c) for the effect of a single treatment at time $t_{j'}$, $a_i(t_{j'})$, on future values of the latent process $\eta_i(t_{j+1})$, the measured longitudinal outcomes $Y_i(t_{j+1})$, and the recurrent event outcome $(T_{ir}, \delta_{ir})$.]{Diagram illustrating potential mechanisms (a)--(c) for the effect of a single treatment at time $t_{j'}$, $a_i(t_{j'})$, on future values of the latent process $\eta_i(t_{j+1})$, the measured longitudinal outcomes $Y_i(t_{j+1})$, and the recurrent event outcome $(T_{ir}, \delta_{ir})$.  In the diagram, circles indicate that variables are latent and squares denote variables that are observed.  We omit the random intercept and measurement error terms from this diagram for simplicity; these terms are still included in the model definition.}
\label{a:fig:full_joint_model_dag_with_tx_over_time}
\end{figure}

The treatment effect could potentially be incorporated into our joint model in three different ways.  As described in the main paper, treatment could (a) directly impact the latent process, (b) alter the risk of recurrent events, or (c) change the measured longitudinal outcomes.  These potential pathways for treatment effect are summarized in Web Figure \ref{a:fig:full_joint_model_dag_with_tx_over_time}.  We only consider models for (a) and (b).

\section{Modeling the Impact of Treatment on the Latent Process}\label{a:ouf_jm_tx:hull_white_integral}

If we model the impact of treatment on the latent process by incorporating a time-varying drift term into the OU process (resulting in a process also known as a Hull-White process), then the solution to the stochastic differential equation (SDE) involves a term with an integral over the treatment effect function $\bm{\mu}_i(t)$.  As given in the main text, the conditional distribution of $\bm{\eta}_i(t)$ with time-dependent drift is

\begin{align}\label{a:eq:hw_dist}
    \bm{\eta}_i(t)|\bm{\eta}_i(s) \sim N_p\left( e^{-\bm{\theta} (t - s)} \bm{\eta}_i(s) + \int_s^t e^{-\bm{\theta} (t-u)} \bm{\mu}_i(u) du, \bm{V} - e^{- \bm{\theta} (t-s)} \bm{V} e^{- \bm{\theta}^{\top} (t-s)} \right)
\end{align}

\noindent We let $\mathcal{A}_i(s - \delta_a, t)$ denote the set of times at which treatments were sent to individual $i$ between time $s - \delta_a$ and time $t$; this set of treatment times corresponds to all treatments that are active at between times $s$ and $t$.  The analytic solution to this integral in Equation \ref{a:eq:hw_dist} is:

\begin{align*}
    \int_s^t e^{-\bm{\theta} (t - u)} \bm{\mu}_i(u) du &= \int_s^t e^{-\bm{\theta} (t - u)} \sum_{t_{ia} \in \mathcal{A}_i(s - \delta_a, t)} \bm{\tau} \left( 1 - \frac{u - t_{ia}}{\delta_a} \right)_+ du \\
    &= \sum_{t_{ia} \in \mathcal{A}_i(s - \delta_a, t)} \int_s^t e^{-\bm{\theta} (t - u)}  \bm{\tau} \left( 1 - \frac{u - t_{ia}}{\delta_a} \right)_+ du \\
    &= \sum_{t_{ia} \in \mathcal{A}_i(s - \delta_a, t)} \int_{max(t_{ia}, s)}^{min(t, t_{ia} + \delta_a)} e^{-\bm{\theta} (t - u)}  \bm{\tau} \left( 1 - \frac{u - t_{ia}}{\delta_a} \right) du \\
    &= \sum_{t_{ia} \in \mathcal{A}_i(s - \delta_a, t)} \left[ \left( 1 - \frac{u - t_{ia}}{\delta_a} \right) e^{-\bm{\theta} (t - u)} \bm{\theta}^{-1} + \frac{1}{\delta_a} e^{-\bm{\theta} (t - u)} \right] \bm{\tau}\Bigg\rvert_{max(t_{ia}, s)}^{min(t, t_{ia} + \delta_a)}
\end{align*}

\noindent Plugging this result into Equation \ref{a:eq:hw_dist}, we can re-write the distribution in an analytic form:

\begin{equation}
    \begin{aligned}
    \bm{\eta}_i(t)|\bm{\eta}_i(s) \sim N_p \Bigg( & e^{-\bm{\theta} (t - s)} \bm{\eta}_i(s) \\ & + \sum_{t_{ia} \in \mathcal{A}_i(s - \delta_a, t)} \left[ \left( 1 - \frac{u - t_{ia}}{\delta_a} \right) e^{-\bm{\theta} (t - u)} \theta^{-1} + \frac{1}{\delta_a} e^{-\bm{\theta} (t - u)} \right] \bm{\tau}\Bigg\rvert_{u = max(t_{ia}, s)}^{u = min(t, t_{ia} + \delta_a)}, \\ & \bm{V} - e^{- \bm{\theta} (t-s)} \bm{V} e^{- \bm{\theta}^{\top} (t-s)} \Bigg)
\end{aligned}
\end{equation}

\section{Prior Distributions}\label{a:ouf_jm_tx:priors}

When fitting our joint model, we base our prior distributions on those used in \cite{tran_2021b}.  The priors we use are:

\begin{align*}
    &\lambda_k \sim \text{half-}N(1, \sigma^2_\lambda); k = 1, ..., K \\
    &\sigma_\lambda \sim \text{half-Cauchy}(0, 5) \\
    &\theta_{OU_{11}}, \theta_{OU_{21}}, \theta_{OU_{12}}, \theta_{OU_{22}} \sim N(0, 10^2) \\
    &\rho \sim \text{Uniform}(-0.999999, 0.999999) \\
    &\sigma_{u_k} \sim \text{half-}Cauchy(0, 5); k = 1, ..., K \\
    &\sigma_{u_{\epsilon}} \sim \text{half-}Cauchy(0, 5); k = 1, ..., K \\
    &\beta_0, \beta_1, \beta_2 \sim N(0, 5^2) \\
    &\tau_1, \tau_2 \sim N(0, 5^2) \\
    &\Tilde{\tau} \sim N(0, 5^2)
\end{align*}

\section{Simulation Study}\label{a:ouf_jm_tx:sim_study}

True model parameters, which are informed by the longitudinal models fit to data from similar mobile health (mHealth) smoking cessation studies, are given below.  Setting 1's longitudinal submodel parameter values are roughly similar to those estimated in the case study in \cite{abbott_2023} and setting 2's longitudinal submodel parameter values are roughly similar to those estimated in the case study in \cite{abbott_2024}.

\begin{itemize}
    \item[] \textbf{Setting 1: measurement submodel} $$\bm{\Lambda} = \begin{bmatrix}
        0.9 & 0 \\ 0.5 & 0 \\ 0 & 1 \\ 0 & 0.8
    \end{bmatrix}, \bm{\Sigma}_u = \begin{bmatrix}
       0.16 & 0 & 0 & 0 \\ 0 & 0.25 & 0 & 0 \\ 0 & 0 & 0.64 & 0 \\ 0 & 0 & 0 & 1.00 
    \end{bmatrix}, \bm{\Sigma}_{\epsilon} = \begin{bmatrix} 0.04 & 0 & 0 & 0 \\ 0 & 0.36 & 0 & 0 \\ 0 & 0 & 0.09 & 0 \\ 0 & 0 & 0 & 0.49 
    \end{bmatrix}$$
    \item[] \textbf{Setting 1: structural submodel} $$\bm{\tau} = [2, -1]^{\top}, \bm{\theta} = \begin{bmatrix}
        2.4 & 1.2 \\ 2.9 & 3.6 \end{bmatrix}, \bm{\sigma} =  \begin{bmatrix}
        1.78 & 0 \\ 0 & 1.80 \end{bmatrix} \Longrightarrow \rho = -0.68 $$
    \item[] \textbf{Settings 1: event-time submodel}
    \begin{enumerate}
        \item $\beta_0 = -1.8, \beta_1 = -0.5, \beta_2 = 0.5, \Tilde{\tau} = -0.8$ 
        \item $\beta_0 = -1.5, \beta_1 = -0.5, \beta_2 = 0.5, \beta_3 = 0.4, \Tilde{\tau} = -0.8$ 
    \end{enumerate}
    \item[]  
    \item[] \textbf{Setting 2: measurement submodel} $$\bm{\Lambda} = \begin{bmatrix}
        0.4 & 0 \\ 0.25 & 0 \\ 0 & 0.5 \\ 0 & 0.6
    \end{bmatrix}, \bm{\Sigma}_u = \begin{bmatrix}
       0.16 & 0 & 0 & 0 \\ 0 & 0.16 & 0 & 0 \\ 0 & 0 & 0.25 & 0 \\ 0 & 0 & 0 & 0.16 
    \end{bmatrix}, \bm{\Sigma}_{\epsilon} = \begin{bmatrix} 0.04 & 0 & 0 & 0 \\ 0 & 0.01 & 0 & 0 \\ 0 & 0 & 0.09 & 0 \\ 0 & 0 & 0 & 0.04
    \end{bmatrix}$$ 
    \item[] \textbf{Setting 2: structural submodel} $$\bm{\tau} = [2, -1]^{\top}, \bm{\theta} = \begin{bmatrix} 
        10.2 & 5.1 \\ 4.9 & 10 \end{bmatrix}, \bm{\sigma} =  \begin{bmatrix}
        3.92 & 0 \\ 0 & 3.89 \end{bmatrix} \Longrightarrow \rho = -0.50 $$
    \item[] \textbf{Settings 2: event-time submodel}
    \begin{enumerate} 
        \item $\beta_0 = -1.8, \beta_1 = -0.5, \beta_2 = 0.5, \Tilde{\tau} = -0.8$ 
        \item $\beta_0 = -1.5, \beta_1 = -0.5, \beta_2 = 0.5, \beta_3 = 0.4, \Tilde{\tau} = -0.8$ 
    \end{enumerate}
\end{itemize}

The true hazard models used to generate the recurrent event outcomes are:
\begin{enumerate}
    \item $h_{ir}(t) = h_0 \exp\big\{\beta_1 \eta_{1i}(t) + \beta_2 \eta_{2i}(t) + \Tilde{\mu}_i(t) \big\}$
    \item $h_{ir}(t) = h_0 \exp\big\{\beta_1 \eta_{1i}(t) + \beta_2 \eta_{2i}(t) + \beta_3 g(t - t_{i,r-1}) + \Tilde{\mu}_i(t) \big\}$
\end{enumerate}

\begin{figure}
\centering
\includegraphics[width=12cm]{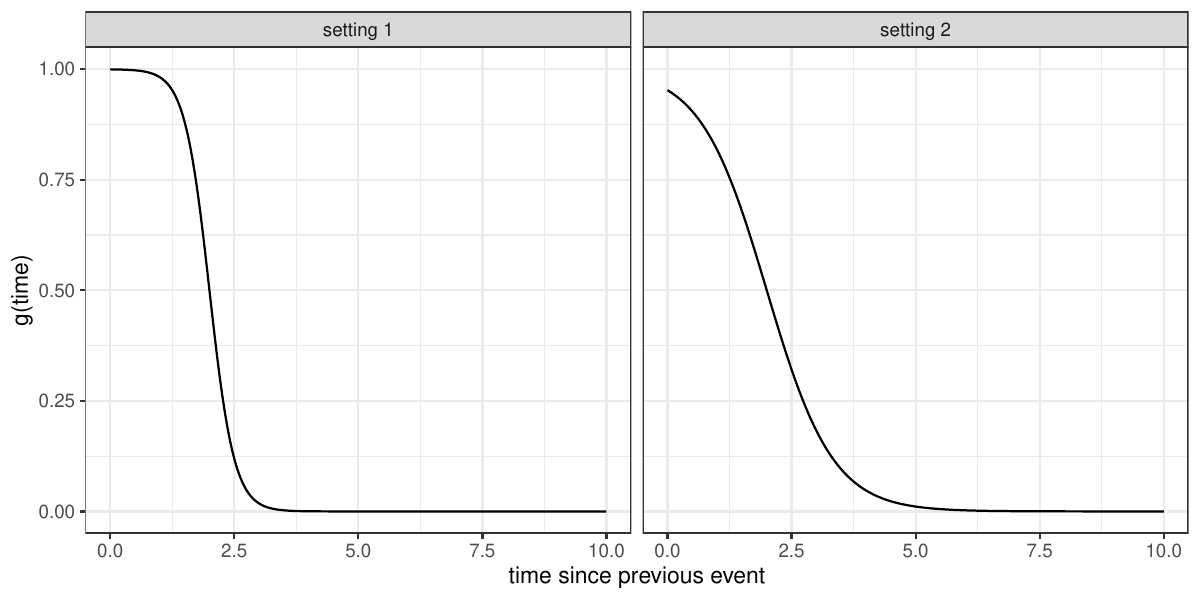}
\caption[Curves for the $g(\cdot)$ functions that capture the association between the hazard of an event as a function of time (in days) since the most recent prior event.]{Curves for the $g(\cdot)$ functions that capture the association between the hazard of an event as a function of time (in days) since the most recent prior event.}\label{a:fig:g_functions}
\end{figure}

\noindent We generate event outcomes from both of these hazard models using the true parameter values for setting 1 and setting 2.  For setting 1, $g(x) = \frac{1}{1 + \exp\{4(x - 2)\}}$ and for setting 2, $g(x) = \frac{1}{1 + \exp\{1.5(x - 2)\}}$. In both of these hazard models, we assume that the baseline hazard is constant, $h_0 = \exp(\beta_0)$.  In Web Figure \ref{a:fig:g_functions}, we plot the curves for the $g(\cdot)$ functions that capture the association between the hazard of an event as a function of time since the most recent prior event.

\begin{figure}
\centering
\includegraphics[width=12cm]{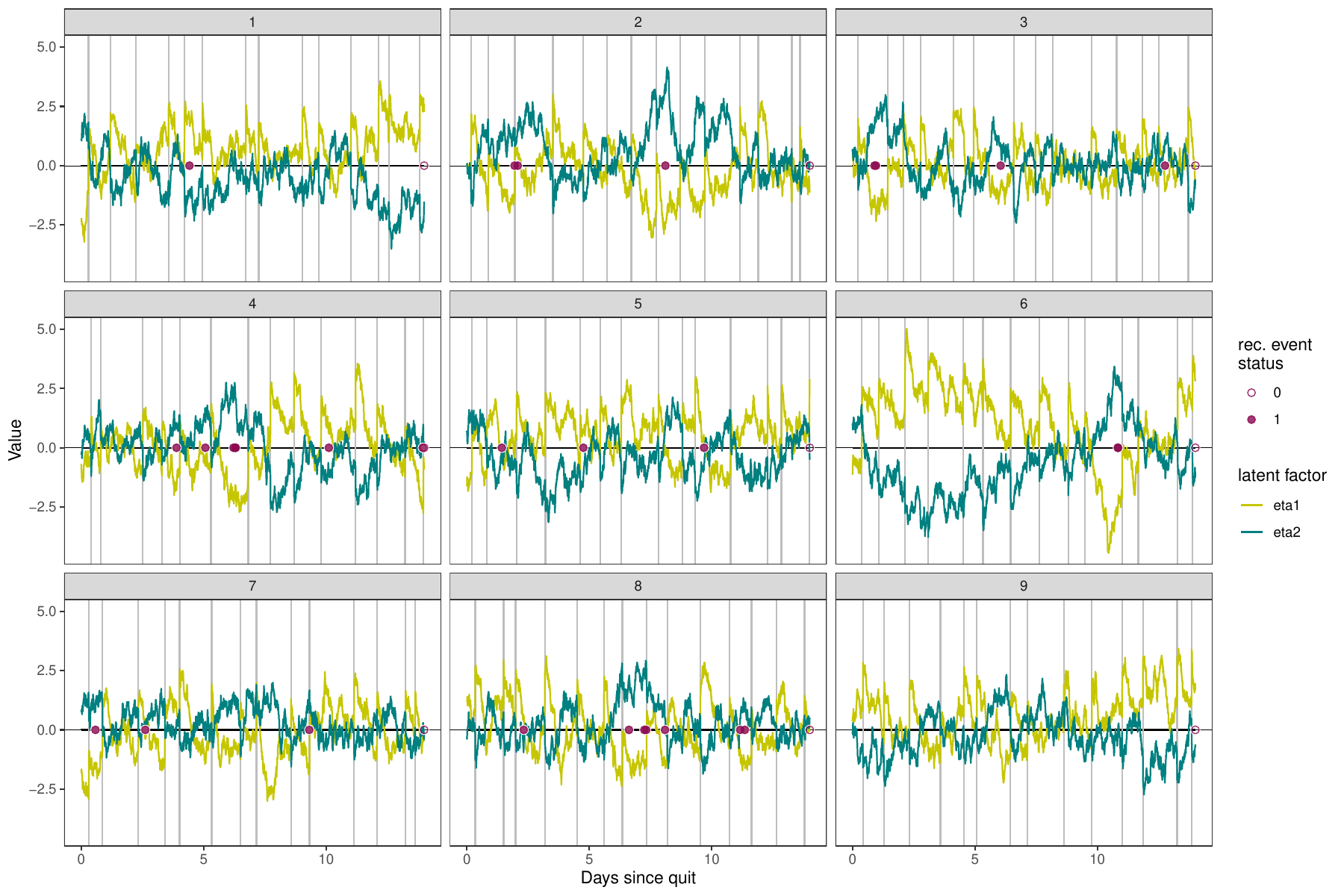}
\caption[Setting 1: Treatment effect is additive and the hazard takes the form of model 1.]{\textbf{Setting 1}: Treatment effect is \textbf{additive} and the hazard takes the form of \textbf{model 1}. Green lines show the trajectory of the bivariate latent process and vertical grey bars indicate the timing of the treatments, which are sent randomly once per day and are assumed to have an effect that lasts half a day ($\delta_a = 0.5$). Events are shown as solid pink dots and censoring times are indicated with open pink dots.}\label{a:fig:ou_traj_s1_add_haz_4}
\end{figure}

\begin{figure}
\centering
\includegraphics[width=12cm]{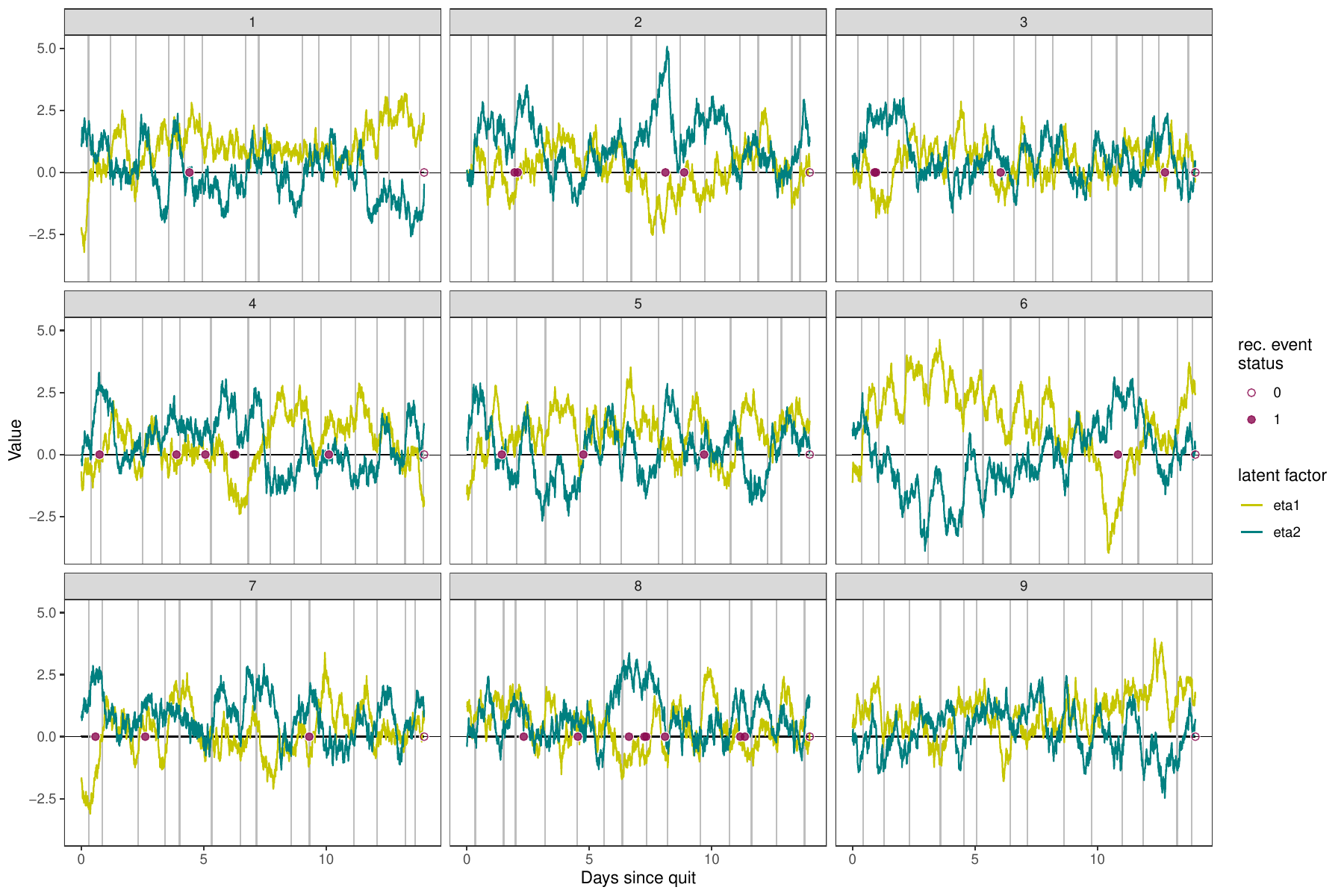}
\caption[Setting 1: Treatment effect is modeled as drift and the hazard takes the form of model 1.]{\textbf{Setting 1}: Treatment effect is modeled as \textbf{drift} and the hazard takes the form of \textbf{model 1}. Green lines show the trajectory of the bivariate latent process and vertical grey bars indicate the timing of the treatments, which are sent randomly once per day and are assumed to have an effect that lasts half a day ($\delta_a = 0.5$). Events are shown as solid pink dots and censoring times are indicated with open pink dots.}\label{a:fig:ou_traj_s1_drift_haz_4}
\end{figure}

\begin{figure}
\centering
\includegraphics[width=12cm]{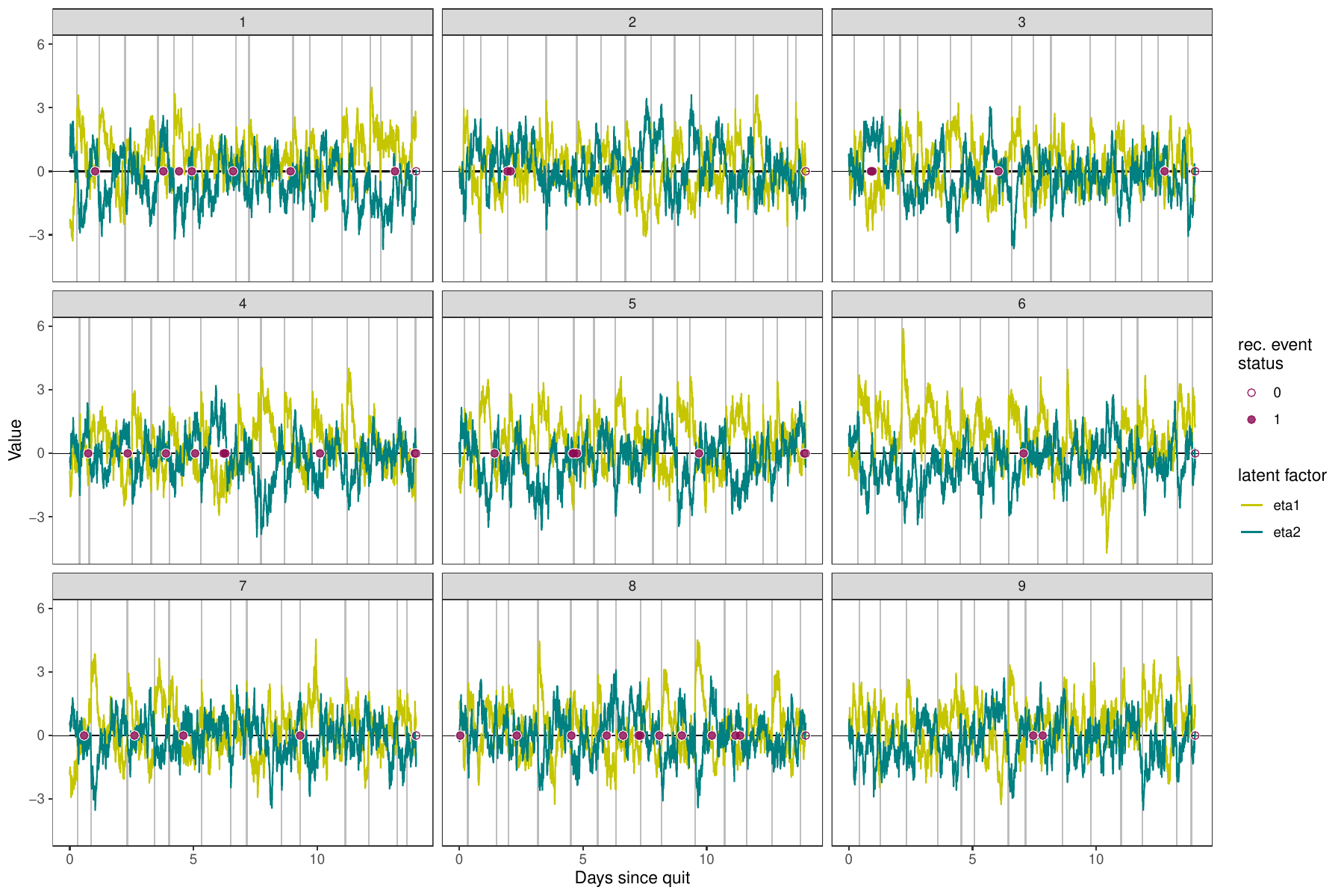}
\caption[Setting 2: Treatment effect is additive and the hazard takes the form of model 1.]{\textbf{Setting 2}: Treatment effect is \textbf{additive} and the hazard takes the form of \textbf{model 1}. Green lines show the trajectory of the bivariate latent process and vertical grey bars indicate the timing of the treatments, which are sent randomly once per day and are assumed to have an effect that lasts half a day ($\delta_a = 0.5$). Events are shown as solid pink dots and censoring times are indicated with open pink dots.}\label{a:fig:ou_traj_s2_add_haz_4}
\end{figure}

\begin{figure}
\centering
\includegraphics[width=12cm]{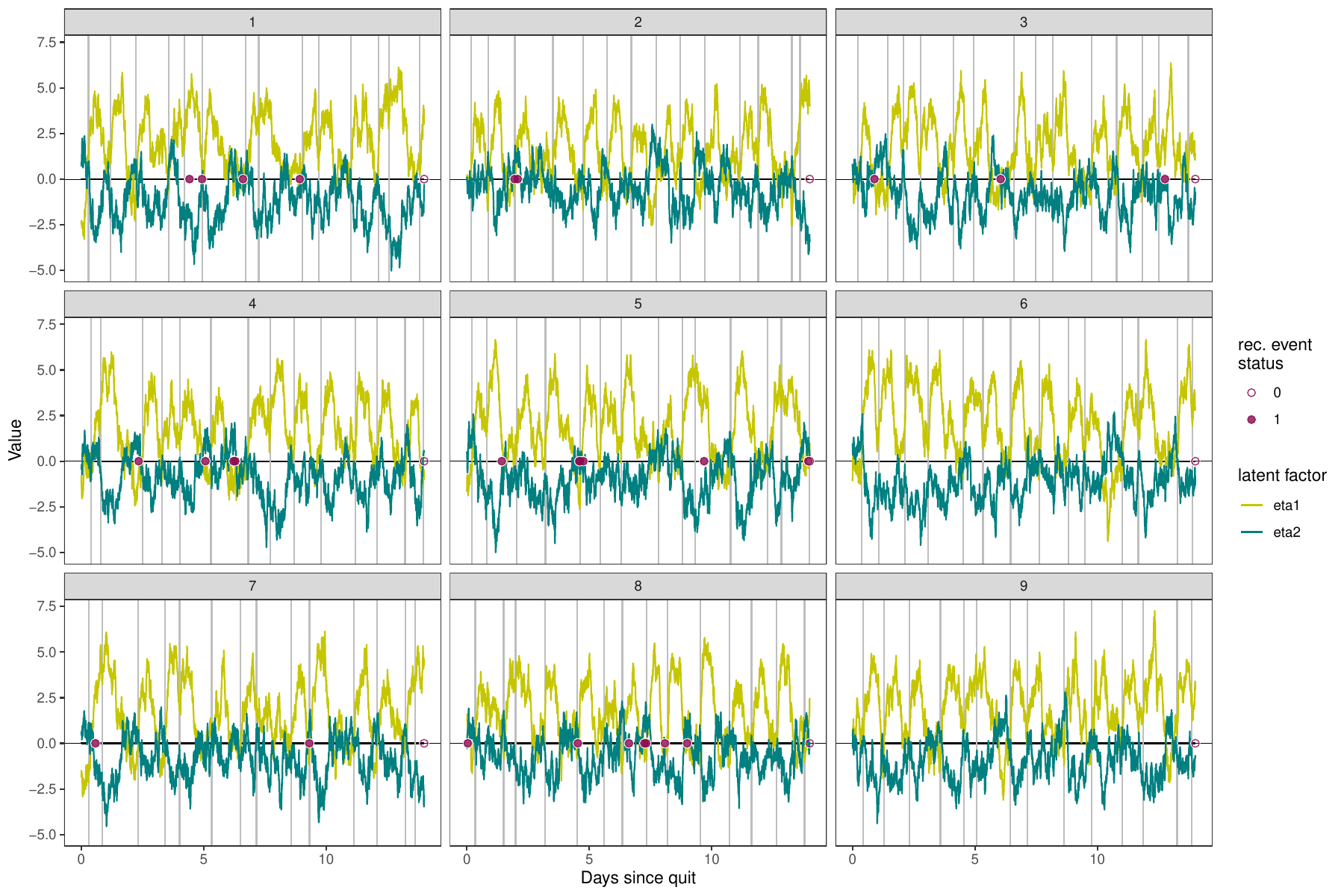}
\caption[Setting 2: Treatment effect is modeled as drift and the hazard takes the form of model 1.]{\textbf{Setting 2}: Treatment effect is modeled as \textbf{drift} and the hazard takes the form of \textbf{model 1}. Green lines show the trajectory of the bivariate latent process and vertical grey bars indicate the timing of the treatments, which are sent randomly once per day and are assumed to have an effect that lasts half a day ($\delta_a = 0.5$). Events are shown as solid pink dots and censoring times are indicated with open pink dots.}\label{a:fig:ou_traj_s2_drift_haz_4}
\end{figure}

We also plot the trajectories of the latent bivariate OU process for a subset of individuals in a single simulated dataset, illustrating both the case when the treatment effect is modeled as an additive term and as drift, in Web Figures \ref{a:fig:ou_traj_s1_add_haz_4}-\ref{a:fig:ou_traj_s2_drift_haz_4}.  We just provide these plots when using hazard model 1 to generate the recurrent event outcomes.

\subsection{Additional Simulations for $\beta_3$}\label{a:ouf_jm_tx:sim_study_beta3}

\begin{figure}
    \centering
    \captionsetup{width=15cm}
    \includegraphics[width=15cm]{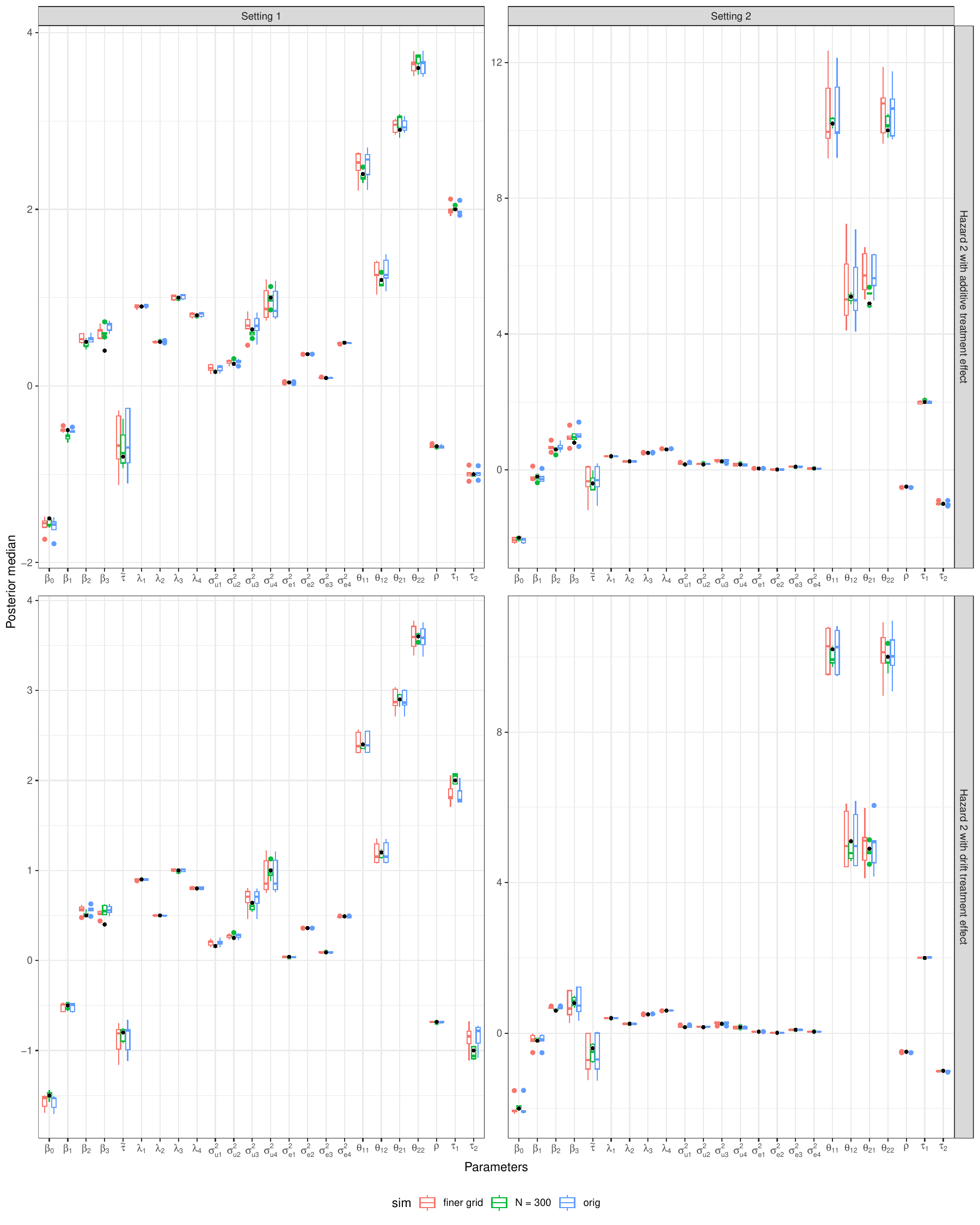}
    \caption[For data generated under settings 1 and 2 with hazard model 2 and different treatment effect models in the longitudinal submodel, we use box plots to summarize the distribution of the posterior medians for all parameters across the 5 simulated datasets.]{For data generated under settings 1 and 2 with hazard model 2 and different treatment effect models in the longitudinal submodel, we use box plots to summarize the distribution of the \textbf{posterior medians for all parameters} across the 5 simulated datasets. The original simulation design has a sample size of $N = 100$ and assumes a grid width of 12-hour intervals for approximating the cumulative hazard function via a midpoint rule (sim: orig).  We modify this original simulation design by either increasing the sample size to $N = 300$ (sim: N = 300) or decreasing the grid width to 4-hour intervals (sim: finer grid).  True parameter values are indicated with colored dots.} \label{a:fig:post_medians_bonus}
\end{figure}

\begin{figure}
    \centering
    \captionsetup{width=15cm}
    \includegraphics[width=15cm]{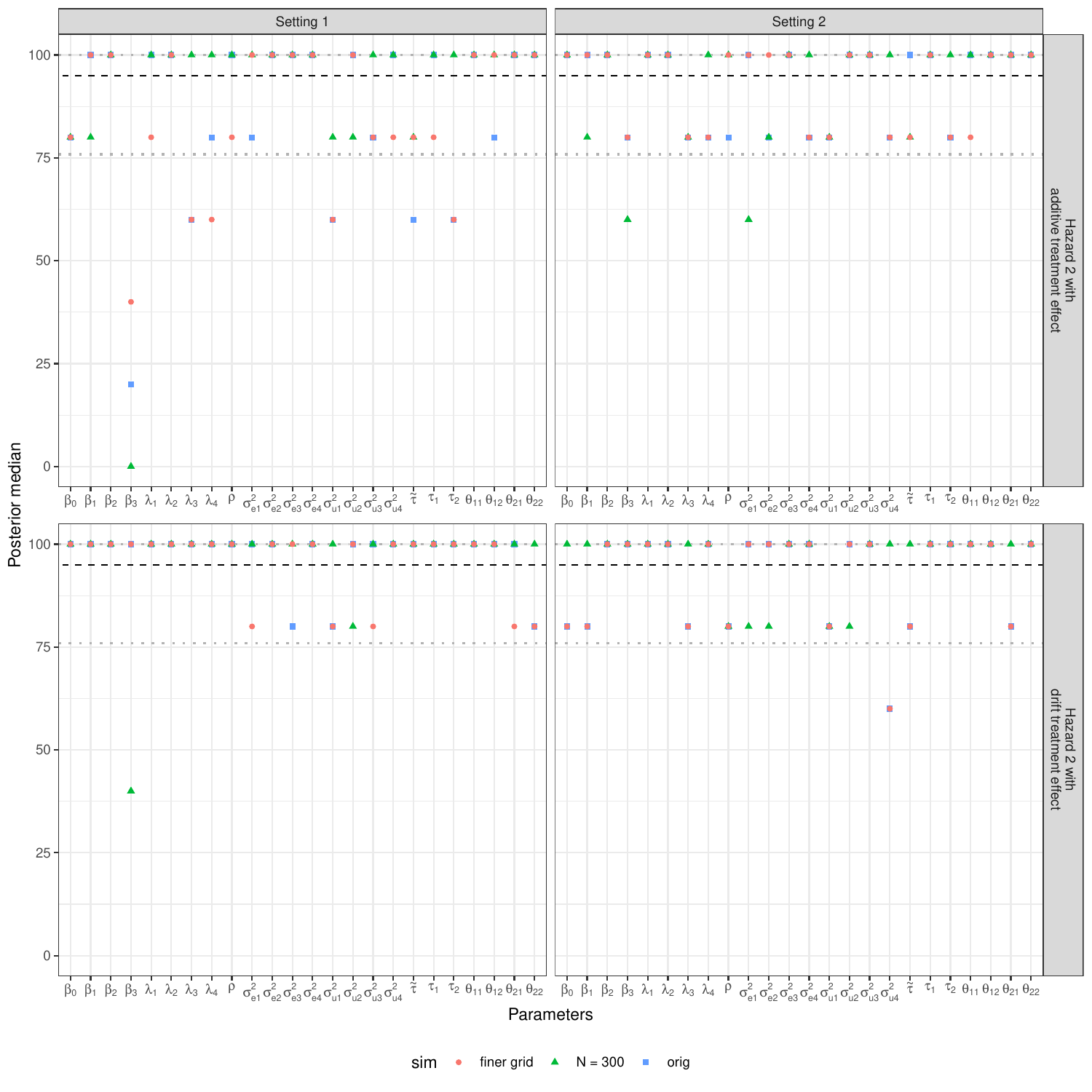}
    \caption[For data generated under settings 1 and 2 with hazard model 2 and different treatment effect models in the longitudinal submodel, we summarize the coverage rate of 95\% credible intervals across the 5 simulated datasets with the colored dots.]{For data generated under settings 1 and 2 with hazard model 2 and different treatment effect models in the longitudinal submodel, we summarize the \textbf{coverage rate of 95\% credible intervals} across the 5 simulated datasets with the colored dots.  The original simulation design has a sample size of $N = 100$ and assumes a grid width 12-hour intervals for approximating the cumulative hazard function via a midpoint rule (sim: orig).  We modify this original simulation design by either increasing the sample size to $N = 300$ (sim: N = 300) or decreasing the grid width to 4-hour intervals (sim: finer grid).  The black horizontal dashed lines indicate target coverage and the dotted grey lines corresponds to the upper and lower bounds of a 95\% binomial proportion confidence interval for a probability of 0.95.} \label{a:fig:post_cov_bonus}
\end{figure}

In the simulation results shown in the main paper, we see some bias in the posterior medians for parameter $\beta_3$ in the recurrent event submodel when assuming hazard model 2.  This bias results in a coverage rate that is lower than nominal.  The parameter $\beta_3$ is the coefficient that captures the association between the hazard of the $r^{th}$ recurrent event and the time since the $(r-1)^{th}$ event.  We investigate two factors that might be contributing to this bias: (a) the finite sample size of our simulated datasets and (b) the choice of grid width used when approximating the cumulative hazard function via a midpoint rule.  To investigate (a), we generate data with a sample size of $N = 300$ individuals, an increase from the $N = 100$ sample size used in the original simulation study.  To investigate (b), we increase the density of the grid used in the midpoint approximation of the cumulative hazard function from one point approximately each 12 hours to one point about each 4 hours.  We compare the posterior medians and coverage rates for these additional simulations to a subset of the original simulations in Web Figures \ref{a:fig:post_medians_bonus} and \ref{a:fig:post_cov_bonus}.  We consider 5 replicates.

The results of this small supplemental simulation study suggest that the bias in the posterior median for $\beta_3$ is likely related to fitting a complex model to a fairly small dataset.  The bias in the posterior median results in low coverage.  In setting 2, we see a decrease in the bias for this parameter estimate when we increase the sample size from 100 independent individuals to 300 independent individuals.  In setting 1, we see less of a decrease in bias when increasing the sample size.  This differential decrease in bias could be partially due to the differences in the amount of temporal correlation across these two settings: in setting 1, the longitudinal latent process has slower decay in correlation over time than in setting 2, and so accurately estimating $\beta_3$, the coefficient on another function of time, could be more challenging in setting 1 than in setting 2.  If we were to substantially increase the sample size, perhaps to $3000$, we would expect to see less bias in both settings.

We also hypothesized that the bias in $\beta_3$ could be related to the coarseness of the approximation used to evaluate the cumulative hazard function.  As mentioned in the main text, we approximate the cumulative hazard function by replacing the integral with a sum over a fine grid of points; we then evaluate this sum using a midpoint rule.  If the grid used in the approximation is too coarse, then it will be poor.  However, we found that decreasing the coarseness of the grid by a factor of 3 did not substantially impact the bias in our estimates.

\section{Model Comparisons}\label{a:ouf_jm_tx:lppd_calcs}

To compare the fit of different models, we consider two different measures of predictive accuracy: deviance information criterion (DIC) and widely applicable information criterion (WAIC).  DIC and WAIC are both based on estimates of the log pointwise predictive density (lppd).  For a general model with data $y_1, ..., y_N$ and parameters $\Theta$, \cite{vehtari_2017} define the lppd as
\begin{align*}
    \text{lppd} = \sum_{i = 1}^{N} log \left(  p(y_i | y) \right) = \sum_{i = 1}^{N} log \left(  \int p(y_i | \Theta) p (\Theta | y) d\Theta \right)
\end{align*}

\noindent The lppd can be computed as
\begin{align*}
    \widehat{\text{lppd}} = \sum_{i = 1}^{N} log \left( \frac{1}{S} \sum_{s = 1}^{S} p(y_i | \Theta^s) \right)
\end{align*}

\noindent where $\Theta^s, s = 1, ..., S$ are draws from the ``usual'' posterior distribution, $p(\Theta | y)$ is the posterior distribution, and  $p(y_i | y)$ is the posterior predictive distribution.  In our joint model, the data $y_i$ would consist of the longitudinal outcomes and the recurrent event outcomes.

Continuing with generic notation using $y$ as the data and $\Theta$ as the model parameters, \cite{gelman_2014} define DIC and WAIC as:

\begin{itemize}
    \item DIC: $$DIC = -2log p (y | \hat{\Theta}) + 2p_{DIC}$$ where $\hat{\Theta}$ is the posterior mean and $p_{\text{DIC}}$ is the effective number of parameters, with $$p_{\text{DIC}} = 2 \left( log p (y | \hat{\Theta}) - \frac{1}{S} \sum_{s = 1}^S log p (y | \Theta^s)  \right)$$ where $s$ indexes posterior samples.
    \item WAIC: $$WAIC = -2 \text{lppd} + 2p_{\text{WAIC}2}$$ where the computed log pointwise predictive density is $$\widehat{\text{lppd}} = \sum_{i = 1}^N log\left( \frac{1}{S} \sum_{s = 1}^S p(y_i | \Theta^s) \right)$$ and the effective number of parameters is $$p_{\text{WAIC}2} = \sum_{i = 1}^N V_{s = 1}^S \left( log p (y_i | \Theta^s) \right)$$ and $V_{s = 1}^S (a_s) = \frac{1}{S-1} \sum_{s = 1}^S (a_s - \Bar{a})^2$.
\end{itemize}

\noindent \cite{gelman_2014} also discuss alternative ways to calculate the effective sample size, but the definitions above use the suggested/standard approaches.

Because our setting involves latent variables, we must carefully consider the form of the likelihood that we would like to use in our calculations of the log pointwise predictive density.  That is, is it more appropriate to use a version of the likelihood that conditions on the latent factors and all other model parameters, or should we use a version of the likelihood in which we marginalize over the latent factors?  Working in the context of latent factor models commonly used in psychometric research, \cite{merkle_2019} discuss important differences between these conditional and marginal versions of the likelihood, and how these different definitions target different types of predictive accuracy.  They recommend working with the marginal version of the likelihood, which aligns with the recommendations for joint models given in \cite{rizopoulos_2023}.  Before we define the computed log pointwise predictive density in our setting, we define some general notation:

\begin{itemize}
    \item $y$ = data (in our setting, this includes both the longitudinal and recurrent event outcomes)
    \item $\Theta$ = all parameters in our model, excluding the latent process $\eta$ and its parameters $\theta_{OU}$ and $\sigma_{OU}$
    \item $\psi$ = OU process parameters $\theta_{OU}$ and $\sigma_{OU}$
    \item $f_c$ = joint model likelihood conditional on the latent factors
    \item $f_m$ = joint model likelihood marginalized over the latent factors
\end{itemize}

\noindent Then, the lppd can be computed as:

\begin{align}\label{eq:lppd1}
    \widehat{\text{lppd}} &= \sum_{i = 1}^{N} log \mathbb{E}_{\Theta , \psi | y} \left\{ f_m(y_i | \Theta, \psi) \right\} \\
    &= \sum_{i = 1}^{N} log \left[ \int f_m(y_i | \Theta, \psi) p(\Theta, \psi | y) d\Theta d\psi \right] \\
    &= \sum_{i = 1}^{N} log \left[ \int  \mathbb{E}_{\eta | \psi} \left\{ f_c(y_i | \eta_i, \Theta) \right\} p(\Theta, \psi | y) d\Theta d\psi \right]
\end{align}

\noindent Focus now on $\mathbb{E}_{\eta | \psi} \left\{ f_c(y_i | \eta_i, \Theta) \right\}$, which requires integrating over the latent process:

\begin{align*}
    \mathbb{E}_{\eta | \psi} \left\{ f_c(y_i | \eta_i, \Theta) \right\} = \int f_c(y_i | \eta_i, \Theta)  p(\eta_i | \psi) d\eta_i
\end{align*}

\noindent To directly approximate this integral with a sum over sampled values of $\eta$ would require having draws of $\eta$ from the distribution $p(\eta_i | \psi)$.  The posterior samples of $\eta$ that are generated during model fitting are conditional on the data, $p(\eta_i | y)$.  We can use the posterior samples and importance sampling to approximate the integral.

\begin{align}\label{eq:imp_samp}
    \mathbb{E}_{\eta | \psi} \left\{ f_c(y_i | \eta_i, \Theta) \right\} &= \int f_c(y_i | \eta_i, \Theta) p(\eta_i | \psi) d\eta_i \\
    &= \int f_c(y_i | \eta_i, \Theta) \frac{p(\eta_i | \psi)}{p(\eta_i | y)} p(\eta_i | y) d\eta_i \\
    &\approx \frac{1}{M} \sum_{m = 1}^{M} f_c(y_i | \eta^m_i, \Theta) \frac{p(\eta^m_i | \psi)}{p(\eta^m_i | y)}
\end{align}

\noindent where $\eta^m_i$ is sampled from the unconditional marginal distribution of $\eta_i | y$, as given in Appendix C of \cite{merkle_2019}.  This unconditional marginal distribution of $\eta_i | y$ is a normal distribution with a mean and variance that are based on the marginal mean and variance of the usual posterior samples of $\eta_i$.  If we follow this approach to generate samples of $\eta^m_i$, then we can plug Equation \ref{eq:imp_samp} in to Equation \ref{eq:lppd1}:

\begin{align}\label{eq:lppd2}
    \widehat{\text{lppd}} &= \sum_{i = 1}^{N} log \left[ \int  \mathbb{E} \left\{ f_c(y_i | \eta_i, \Theta) \right\} p(\Theta, \psi | y) d\Theta d\psi \right] \\
    & \approx \sum_{i = 1}^{N} log \Bigg[ \int \bigg[ \frac{1}{M} \sum_{m = 1}^{M} f_c(y_i | \eta^m_i, \Theta) \frac{p(\eta^m_i | \psi )}{p(\eta^m_i | y )} \bigg] p(\Theta, \psi | y) d\Theta d\psi  \Bigg] \\
    & \approx \sum_{i = 1}^{N} log \Bigg[ \frac{1}{S} \sum_{s = 1}^{S} \bigg[ \frac{1}{M} \sum_{m = 1}^{M} f_c(y_i | \eta^m_i, \Theta^s) \frac{p(\eta^m_i | \psi^s )}{p(\eta^m_i | y )} \bigg]  \Bigg]
\end{align}

\noindent where $\Theta^s, \psi^s$ are the usual posterior samples.  Then, to compute lppd for one individual, we would:
\begin{enumerate}
    \item Sample $M$ values of $\eta^m_i$ from the unconditional marginal distribution of $\eta_i | y$, which does not depend on $\Theta$ or $\psi$ if we use a version of the approach in \cite{merkle_2019}. Note that because we are calculating the empirical covariance matrix for the entire vector $\eta_i$, the covariance matrix is large and can be unstable.  To avoid non-positive definite covariance matrices, we can add a very small amount to the diagonal of the empirical covariance matrix.
    \item For each posterior sample $s = 1, ..., S$ of $\Theta^s$ and $\psi^s$, compute the density across the $M$ sampled values of $\eta^m_i$.
\end{enumerate}

\noindent We repeat steps 1 and 2 for each individual $i = 1, ..., N$ and sum up the values of the marginal log-likelihood to get $\widehat{\text{lppd}}$.  We can make this approach more explicit in our definition of DIC and WAIC:
\begin{align*}
    DIC =& -2log p (y | \hat{\Theta}) + 2p_{DIC} \\
    =& -2log p(y | \hat{\Theta}) + 4 \left[ log p(y | \hat{\Theta}) - \frac{1}{S} \sum_{s = 1}^S log p(y | \Theta^s) \right] \\
    =& -2 \sum_{i = 1}^{N} log\left[ \frac{1}{M} \sum_{m = 1}^M p(y_i | \hat{\Theta}, \eta_i^m) \frac{p(\eta_i^m | \hat{\psi})}{p(\eta_i^m | y)} \right] \\ & + 4 \Biggl[ \sum_{i = 1}^{N} log\Bigl[ \frac{1}{M} \sum_{m = 1}^M p(y_i | \hat{\Theta}, \eta_i^m) \frac{p(\eta_i^m | \hat{\psi})}{p(\eta_i^m | y)} \Bigr] \\ & \hspace{1cm}- \frac{1}{S} \sum_{s = 1}^{S} \Bigl[ \sum_{i = 1}^{N} log \bigl[ \frac{1}{M} \sum_{m = 1}^{M} p(y_i | \Theta^s, \eta_i^m) \frac{p(\eta_i^m | \psi^s)}{p(\eta_i^m | y)} \bigr] \Bigr] \Biggr] \\\\[10pt]
    WAIC =& -2 \widehat{\text{lppd}} + 2 p_{WAIC2} \\
    =& -2 \sum_{i = 1}^{N} log \Biggl[ \frac{1}{S} \sum_{i = 1}^{S} \Bigr[ \frac{1}{M} \sum_{m = 1}^{M} p(y_i | \Theta^s, \eta_i^m) \frac{p(\eta_i^m | \psi^s)}{p(\eta_i^m | y)} \Bigr] \Biggr] \\ & + 2 \sum_{i = 1}^{N} V_{s = 1}^{S} \Biggl( 
log \Bigl[ \frac{1}{M} \sum_{m = 1}^M p(y_i | \Theta^s, \eta_i^m) \frac{p(\eta_i^m | \psi^s)}{p(\eta_i^m | y)} \Bigr] \Biggr)
\end{align*}

\noindent In the definitions above, $s$ indexes posterior samples, $\hat{\Theta}$ and $\hat{\psi}$ are posterior means, $m$ indexes samples of the latent process drawn from the approximate unconditional marginal distribution of $\eta_i | y$, and $V_{s = 1}^S(\cdot)$ is the sample variance.  $y$ contains both our longitudinal and recurrent event data (written as $(Y, T, \delta)$ in the main paper).  In the main paper, we subsample every 5th iteration of the final 1,000 posterior samples across all chains, resulting in $S = 800$. We use $M = 25$ when calculating DIC and WAIC.

\vspace{0.5cm}

\begin{figure}
    \centering
    \captionsetup{width=15cm}
    \includegraphics[width=10cm]{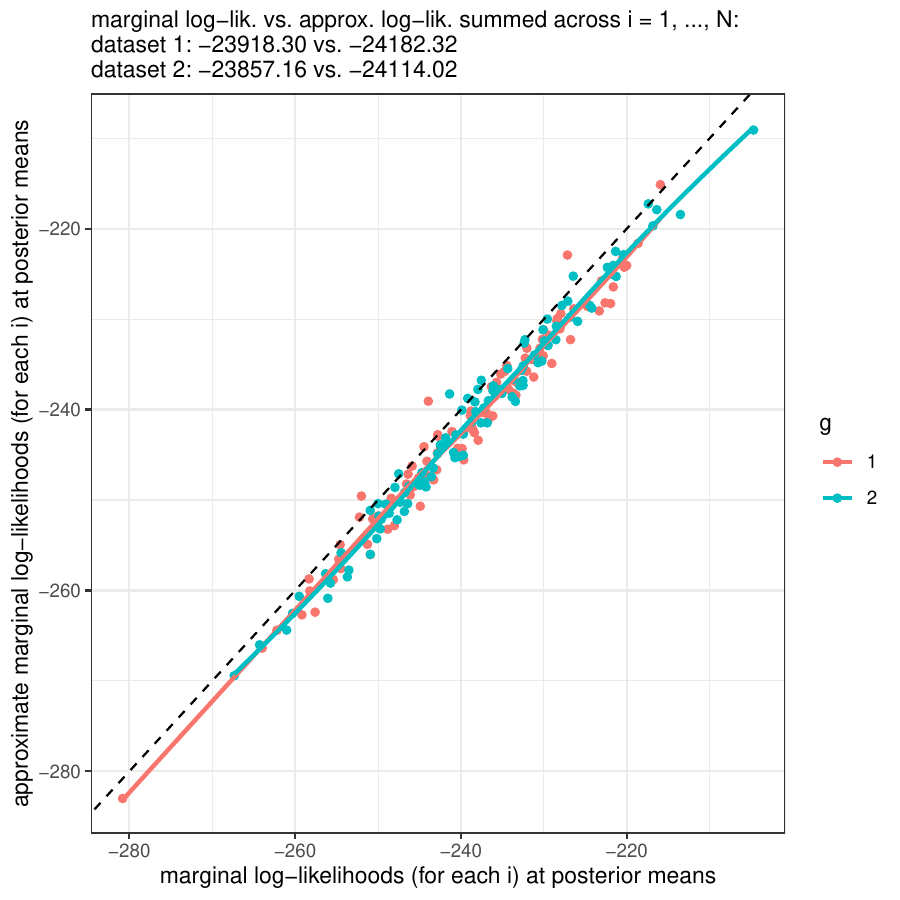}
    \caption[Comparison of approximate log-likelihood values for each individual evaluated at the posterior mean to the exact log-likelihood for each individual using the true marginal distribution evaluated at the posterior mean.]{Comparison of approximate log-likelihood values for each individual evaluated at the posterior mean to the exact log-likelihood for each individual using the true marginal distribution evaluated at the posterior mean. Results from two simulated datasets and two fitted models are shown using red and blue.}\label{a:fig:marg_llk_vs_approx_marg_llk_for_ouf}
\end{figure}

\begin{figure}
    \centering
    \captionsetup{width=15cm}
    \includegraphics[width=15cm]{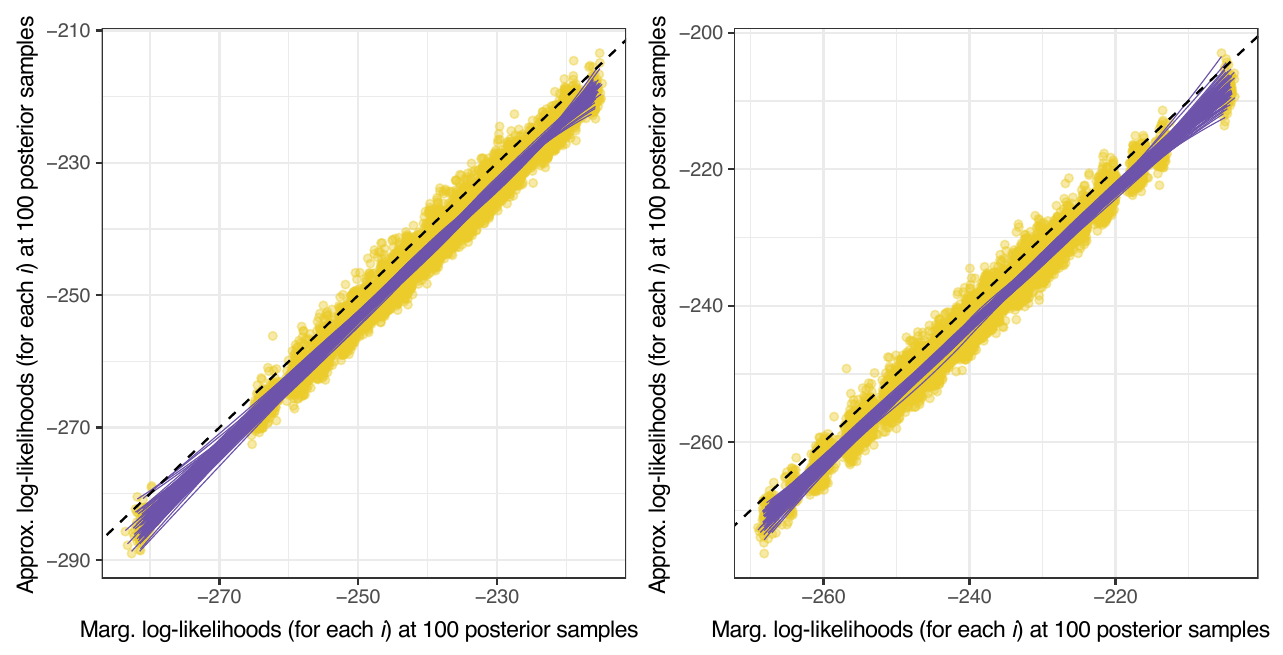}
    \caption[Comparison of approximate log-likelihood values for each individual evaluated at each posterior sample to the exact log-likelihood for each individual using the true marginal distribution evaluated at each posterior sample $s$.]{Comparison of approximate log-likelihood values for each individual evaluated at each posterior sample to the exact log-likelihood for each individual using the true marginal distribution evaluated at each posterior sample $s$.  Note that we subsample 100 of the total 1000 posterior samples. Results from one simulated dataset are on the left, results from the other dataset are on the right. Linear best-fit lines are plotted across individuals for each set of posterior samples $s$. }\label{a:fig:marg_llk_vs_approx_marg_llk_for_ouf_by_post_sample}
\end{figure}

To confirm that our approximate approach to calculate the marginal log-likelihood works reasonably well, we try calculating the marginal log-likelihood using the approximate approach described above and using the exact marginal distribution for just the longitudinal submodel with the additive treatment effect model.  For this longitudinal submodel, we can derive the marginal distribution algebraically.  We calculate the approximate and exact log-likelihoods at the posterior means for two simulated datasets and compare the results in Web Figure \ref{a:fig:marg_llk_vs_approx_marg_llk_for_ouf}.  If our approximation works well, then we expect the value of the marginal log-likelihood for each individual $i$ to be roughly the same for both the approximate marginal log-likelihood vs. the algebraic/exact marginal log-likelihod evaluated at the posterior mean.  Web Figure \ref{a:fig:marg_llk_vs_approx_marg_llk_for_ouf} shows this to be true, as indicated by the colored lines and points falling along the 0-1 axis.  We do see some bias, but because this bias is consistent, it should not be problematic for the purposes of model selection.  We can also repeat the same comparison but for the approximate and exact marginal log-likelihoods evaluated each posterior sample $\bm{\Theta}^s$, rather than the posterior mean $\hat{\bm{\Theta}}$ (see Web Figure \ref{a:fig:marg_llk_vs_approx_marg_llk_for_ouf_by_post_sample}).

\subsection{Case Study Results}

In Web Table \ref{a:tab:dic}, we provide DIC for each of the joint models fit to the case study data.  Of the 8000 posterior samples across 4 chains, we subsampled every 5th iteration of the final 1,000 posterior samples (resulting in $S = 800$) to use in our WAIC and DIC calculations.  We used $M = 25$ for generating marginal posterior samples of the latent process $\eta$.

\begin{table}
\resizebox{15cm}{!}{
\begin{tabular}{r|cc}
\textbf{DIC} & \multicolumn{2}{c}{\textbf{Impact of treatment on latent process}} \\ \hline
\textbf{Hazard model} & \multicolumn{1}{c}{Additive} & \multicolumn{1}{c}{Drift} \\ \hline
Single treatment parameter & 187,857.3 & 187,431.8 \\ \hline
Separate pre- and post-quit treatment parameters & 188,665.5 & \textbf{186,805.5}
\end{tabular}}
\caption[DIC for the joint models fit to the motivating MRT data.]{DIC for the joint models fit to the motivating MRT data. When approximating the marginal log-likelihood, we subsampled every 5th iteration of the final 1,000 posterior samples to use in our calculations. The lowest value of DIC is indicated in bold text.}\label{a:tab:dic}
\end{table}

\section{Posterior Predictive Check}\label{a:ouf_jm_tx:ppcs}

In addition to selecting among the fitted models using information criteria, we can examine the fit of the models using a posterior predictive check.  When designing our posterior predictive checks, we focus on the event-time submodel as we are most interested in assessing the form of the hazard model used to model the recurrent instances of poly-substance use as a function of a time-varying treatment effect and latent variables that capture affective states.

The goal of our posterior predictive check is to assess the fit of the model given the data at hand and so we use posterior samples from the full joint posterior distribution in our checks.  Note that these posterior predictive checks are distinct from forecasting, which would require a sequential approach, rather than using samples from the full joint posterior distribution.

Our posterior predictive checks involve generating a sequence of predicted events based on a set, $s$, of posterior samples of parameters and latent process trajectories.  The timing of the predicted events is compared to the timing of the observed events using a mean cumulative function (MCF).  We predict events between a specified window of time defined using $t_{start}$ and $t_{stop}$, $t_{start} < t_{stop}$.  The predicted events are conditional on the observed event history up until $t_{start}$; the observed treatment history through $t_start$, $\mathcal{R}_i(t_{start})$; and the usual set of posterior samples of the latent process $\eta$ until $t_{stop}$, $\hat{\mathcal{H}}^s_i(t_{stop}) = \{ \hat{\eta}^s_i(t) : 0 \leq t \leq t_{stop} \}$.  Note that if we were to have specified our structural submodel to depend on the event process, then we would need to re-generate posterior samples of the latent process $\eta$ between $t_{start}$ and $t_{stop}$ as part of our posterior predictive check.  But, because we defined our structural submodel such that it is not a function of the recurrent events, we can just use the usual posterior $\eta$ trajectories.

When predicting events, we also condition on the treatment history until time $t_{stop}$, $\mathcal{A}_i(t_{stop})$. Recall that when specifying the structural submodel, we have allowed the latent process trajectory to depend on the treatment history.  If we want to use the usual posterior samples of $\eta$ to predict events, we also need to condition on the observed treatment history until time $t_{stop}$.  If we want to re-define the timing of treatments between $t_{start}$ and $t_{stop}$ to, for example, demonstrate how the timing of delivering an intervention---or not delivering an intervention at all---would impact the predicted event time, then we would also need to regenerate our posterior samples of $\eta$ between $t_{start}$ and $t_{stop}$.  Because we are focused on checking the fit of the model to the data at hand, we simply assume that the complete treatment history is known.

To generate posterior predictions of events between $t_{start}$ and $t_{stop}$ for a single set of posterior samples indexed by $s$, we take the following steps:

\begin{enumerate}
    \item Condition on observed events $\mathcal{R}_i(t_{tstart})$, the posterior trajectory of the latent process $\mathcal{H}_i^s(t_{stop})$, the observed treatment sequence $\mathcal{A}_i(t_{stop})$, and posterior samples of parameters from the recurrent event submodel, $\bm{\Theta}_R^s$.
    \item Divide the window from $t_{start}$ to $t_{stop}$ into a finer grid of points (this is the same grid used in the midpoint approximation when fitting the model).  Call these points $t_{start} = t_0, t_1, t_2, ..., t_{u}, t_{u+1} = t_{stop}$.
    \item Start with $t_{start} = t_0$.  Then, generate a total number of new events between $t_0$ and $t_1$ by drawing from $$R^s_{i}(t_{0}, t_{1}) \sim Poisson\left(\hat{h}^s_i(t_{1} | \mathcal{R}_i(t_{start}), \mathcal{H}^s_i(t_1), \mathcal{A}_i(t_1), \bm{\Theta}^s_R) \times (t_{1} - t_0) \right).$$  Note that we have conditioned on event history $\mathcal{R}_i(t_{0})$.  If the generated value of $R^s_{i}(t_{0}, t_{1}) > 0$, then distribute the events across the interval between $t_{0}$ and $t_1$ by sampling the predicted event times from $Uniform(t_{0}, t_1)$.
    \item Then, repeat the previous step across the rest of the grid indexed by $u$, generating events by sampling from $$R^s_{i}(t_u, t_{u+1}) \sim Poisson\left(\hat{h}^s_i(t_{u} | \mathcal{R}_i(t_{start}), R^s_i(t_{start}, t_{u}), \mathcal{H}^s_i(t_u), \mathcal{A}_i(t_u), \bm{\Theta}^s_R) \times (t_{u+1} - t_u) \right),$$ which conditions on the observed events $\mathcal{R}_i(t_{start})$ and the predicted events $R^s_i(t_{start}, t_{u})$. 
\end{enumerate}

\begin{figure}
    \centering
    \includegraphics[width=14cm]{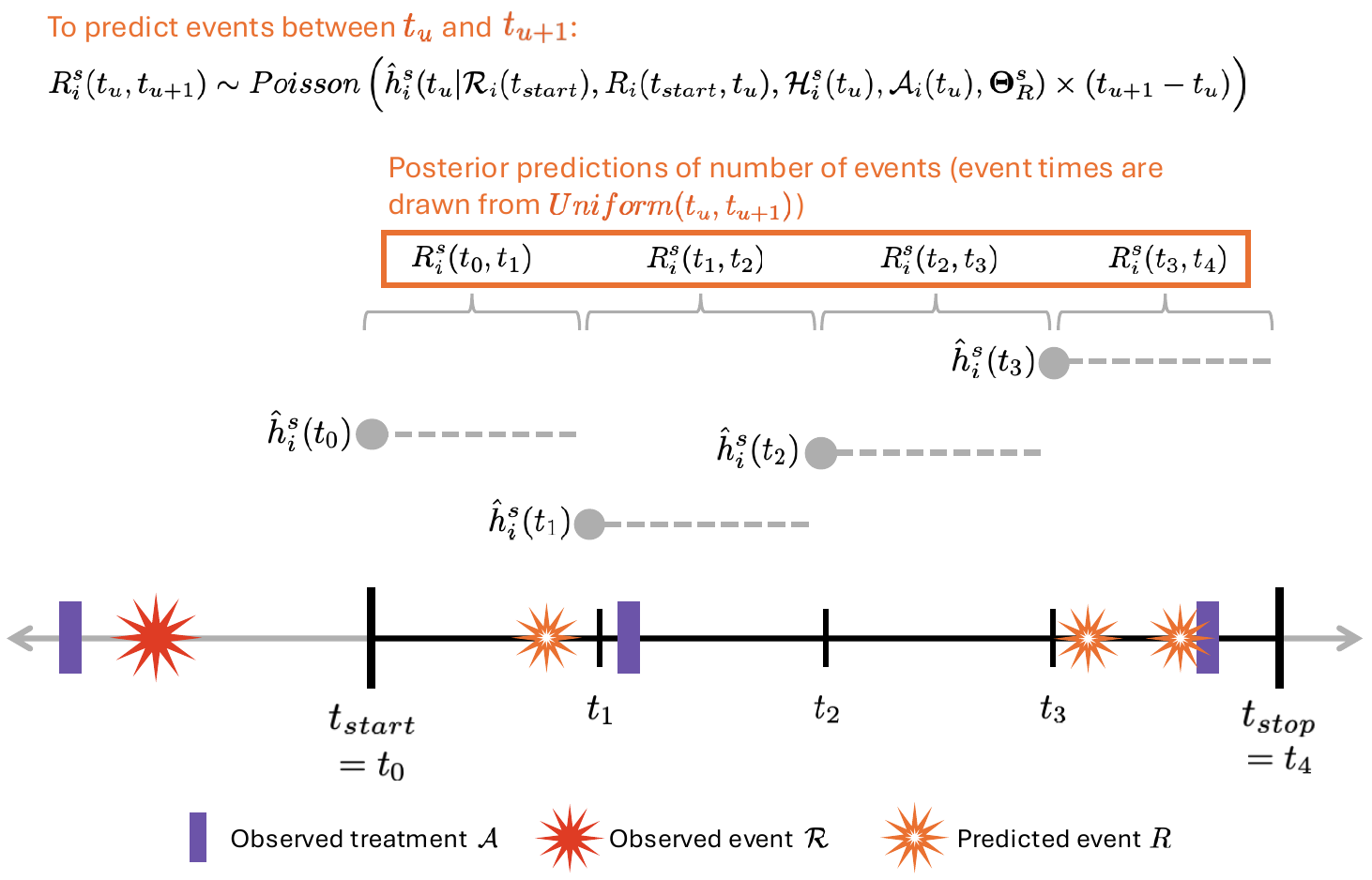}
    \caption{To generate posterior predictions of events over the time interval between $t_{start}$ and $t_{stop}$ for a single set of posterior samples $s$, we sequentially predict the number of observed events within each subinterval defined by a finer set of grid points, conditioning on the observed event history prior to $t_{start}$ and the posterior predicted event history between $t_{start}$ and prior grid points.}
    \label{fig:ppc_sketch}
\end{figure}

We repeat the steps outlined above (and described in Web Figure \ref{fig:ppc_sketch}) for all posterior samples $s = 1, ..., S$. For each $s$, we use the predicted events to generate a mean cumulative function (MCF).  To summarize the $S$ posterior MCFs, we compute the median and $5^{th}$ and $95^{th}$ percentiles at each time point.  The predicted event times are sparse so rather than computing time-specific percentiles using the actual times, we compute rolling percentiles across bins of width 0.2 (e.g., the median of all MCF values within +/- 0.1 of current time).  We then compare this posterior predicted median MCF to the MCF for the observed events.

\subsection{Case Study Results}

\begin{figure}[h!]
     \centering
     \begin{subfigure}[b]{14cm}
         \centering
         \includegraphics[width=14cm]{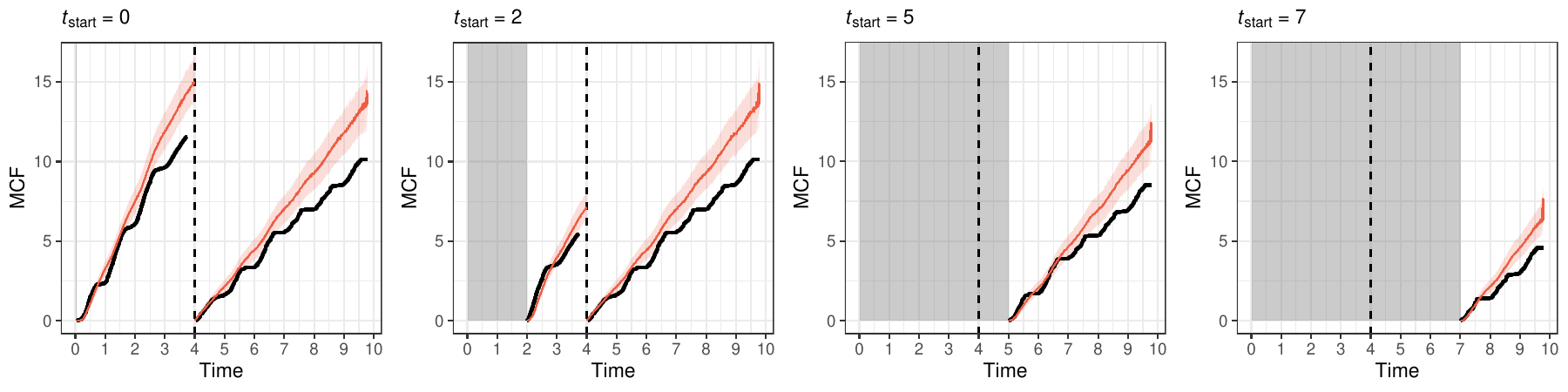}
         \caption{Model assumes an \textbf{additive} treatment effect on the latent process and a \textbf{single} treatment parameter in the hazard submodel.}
         \label{fig:1}
     \end{subfigure}
     \begin{subfigure}[b]{14cm}
         \centering
         \includegraphics[width=14cm]{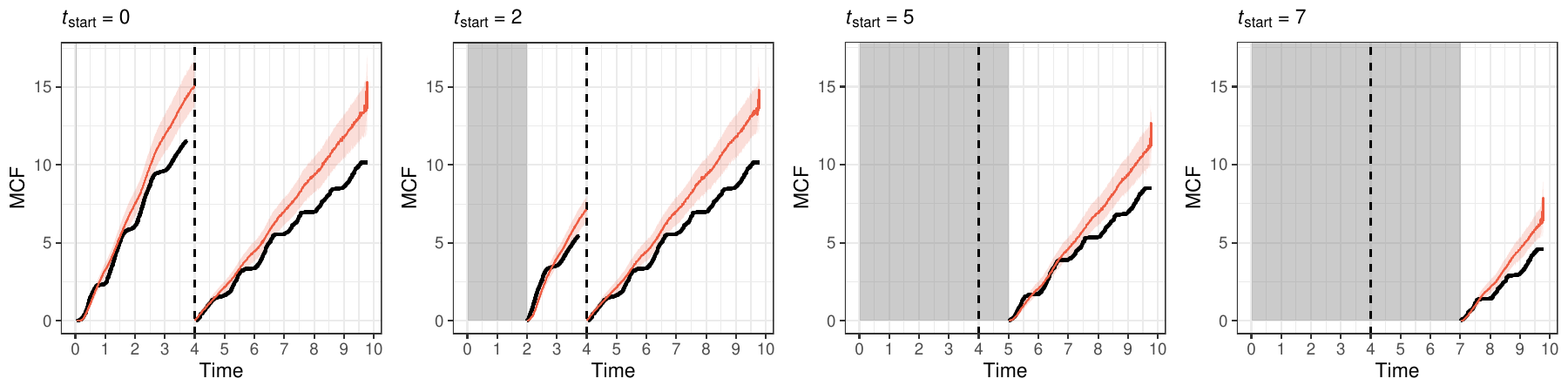}
         \caption{Model assumes an \textbf{additive} treatment effect on the latent process and \textbf{separate} pre- and post-quit treatment parameter sin the hazard submodel.}
         \label{fig:2}
     \end{subfigure}
     \begin{subfigure}[b]{14cm}
         \centering
         \includegraphics[width=14cm]{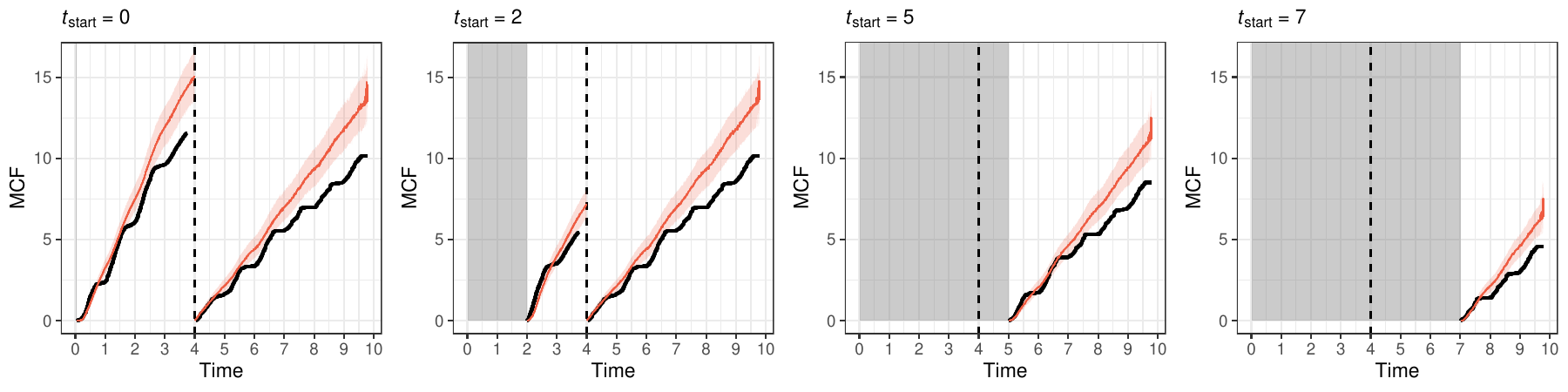}
         \caption{Model assumes a \textbf{drift} treatment effect on the latent process and a \textbf{single} treatment parameter in the hazard submodel.}
         \label{fig:3}
     \end{subfigure}
     \begin{subfigure}[b]{14cm}
         \centering
         \includegraphics[width=14cm]{figures_ouf_jm_tx/ouf_rec_ppc_mcf_v2_tx_effect_drift_delta_tx_8_hazB4_g2_polysub_all_windows.pdf}
         \caption{Model assumes a \textbf{drift} treatment effect on the latent process and \textbf{separate} pre- and post-quit treatment parameter sin the hazard submodel.}
         \label{fig:4}
     \end{subfigure}
        \caption{Median and (5, 95)$^{th}$ percentiles of the mean cumulative function (MCF) for posterior predicted events (red line with shaded ribbon) based on conditional posterior predictions from each fitted joint model.  The black lines show the MCF for observed events.}
        \label{fig:ppc_realdata_h0lognormal}
\end{figure}

We apply the approach described above to the four models fit to the case study data.  We predict posterior events through the 10 days of follow-up, conditioning on 0, 2, 5, and 7 days of follow-up. Recall that the 10 days of follow-up consist of 4 pre-quit days and 6 post-quit days.  Due to computational reasons, rather than computing rolling percentiles to summarize the posterior MCFs, we simply round the time to three decimal places and compute percentiles within each rounded timepoint.  Comparisons of MCFs based on posterior predictions and observed data are shown in Web Figure \ref{fig:ppc_realdata_h0lognormal} for each of the four models considered.  The plots show similar patterns in the MCFs for all of the model: the MCF for posterior predictions and for observed data show close agreement shortly after $t_{start}$, but error accumulates as time passes since $t_{start}$ and so the difference between the predicted and observed MCFs increases. 

\clearpage

\bibliographystyle{imsart-nameyear} 
\bibliography{references}       